\documentstyle[prl,aps,psfig]{revtex}
\begin{document} 
\draft
\title{A Luttinger's theorem revisited}
\author{\sc Behnam Farid}
\address{Max-Planck-Institut f\"ur Festk\"orperforschung,
Heisenbergstra\ss e 1,\\
70569 Stuttgart, Federal Republic of Germany
\thanks{Electronic address: farid@audrey.mpi-stuttgart.mpg.de}\\ 
and Cavendish Laboratory, Department of Physics, 
University of Cambridge,\\
Cambridge CB3 0HE, United Kingdom 
\thanks{Electronic address: bf10000@phy.cam.ac.uk}\\ 
}
\date{{\sl Received 10 November 1998 and accepted 10 January 1999}} 
\maketitle
\begin{abstract}
{\sf
For uniform systems of spin-less fermions in $d$ spatial dimensions
with $d > 1$, interacting through the isotropic two-body potential 
$v({\bf r}-{\bf r}')$, a celebrated theorem due to Luttinger (1961) states 
that under the {\sl assumption} of validity of the many-body perturbation 
theory the self-energy $\Sigma(k;\varepsilon)$, with $0\leq k 
{\mathstrut^{\displaystyle {<} }_{\displaystyle {\sim} }} 3 k_F$ (where 
$k_F$ stands for the Fermi wavenumber), satisfies the following 
universal asymptotic relation as $\varepsilon$ approaches the Fermi 
energy $\varepsilon_F$: ${\rm Im}\Sigma(k;\varepsilon) \sim \mp \alpha_k 
(\varepsilon -\varepsilon_F)^2$, 
$\varepsilon {\mathstrut_{\displaystyle {<} }^{\displaystyle >}}
\varepsilon_F$,
with $\alpha_k \geq 0$. As this is, by definition, specific to 
self-energies of Landau Fermi-liquid systems, treatment of 
non-Fermi-liquid systems are therefore thought to lie outside the 
domain of applicability of the many-body perturbation theory; that, 
for these systems, the many-body perturbation theory should {\sl 
necessarily} break down. We demonstrate that 
${\rm Im}\Sigma(k;\varepsilon)
\sim\mp\alpha_k (\varepsilon-\varepsilon_F)^2$, 
$\varepsilon {\mathstrut_{\displaystyle {<} }^{\displaystyle >}}
\varepsilon_F$,
is {\sl implicit} in Luttinger's proof and that, for $d >1$, in 
principle nothing prohibits a non-Fermi-liquid-type (and, in particular 
Luttinger-liquid-type) $\Sigma(k;\varepsilon)$ from being obtained within
the framework of the many-body perturbation theory. We in addition 
indicate how seemingly innocuous Taylor expansions of the self-energy 
with respect to $k$, $\varepsilon$ or both amount to tacitly assuming 
that the metallic system under consideration is a Fermi liquid, whether 
the self-energy is calculated perturbatively or otherwise. Proofs that 
a certain metallic system is in a Fermi-liquid state, based on such 
expansions, are therefore tautologies. 
}
\end{abstract}

\vskip 1.8cm
\noindent\underline{{\sf Appeared in:} {\sc Philosophical Magazine} 
B, 1999, {\sc Vol.} 79, {\sc No.} 8, 1097 - 1143 }

\noindent
{\sf See:} 
{\tt arXiv:cond-mat/0004444} $\;$ {\&} $\;$
{\tt arXiv:cond-mat/0004451 }

\maketitle

\section{INTRODUCTION}

Landau's (1957) phenomenological theory for the low-lying excited states 
of many-particle systems is based on the {\sl assumption} that these states 
are characterised by a distribution of a dilute gas of quasi-particles 
(QPs), implying that interaction amongst these QPs may, to a good 
approximation, be neglected (Nozi\`eres 1964, Pines and Nozi\`eres 1966,
Abrikosov, Gorkov and Dzyaloshinski 1975)
(for a comprehensive review see Platzman and Wolff (1973)).
\footnote{
\small
To be precise, Landau's Fermi-liquid theory is applicable for temperatures 
$T$ in the range $(T_c, T^*$), where $T^*$ and $T_c$ stand, respectively,  
for the {\sl coherence} and the {\sl transition} temperature; for $T$ below 
$T_c$ the system is in a superconducting state (Kohn and Luttinger 1965). 
For details as well as a comprehensive discussion of instabilities of Fermi 
liquids see (Metzner, Castellani and Di Castro 1998). }
This ideal situation corresponds to the case in which QPs are the {\sl 
exact} one-particle eigenstates of the many-body Hamiltonian, or, 
equivalently, the energies of the QPs are infinitely sharply defined. The 
success of Sommerfeld's independent-particle model (for example Ashcroft 
and Mermin (1981, Ch.~2)) for conduction electrons in simple metals, despite 
the apparent non-negligible strength of the electron-electron Coulomb 
interaction, can be viewed as a most strong pillar of Landau's 
phenomenological theory. Incorporation of the residual two-body 
interaction among the ideal QPs (at least beyond the Hartree-Fock or 
the {\sl static} exchange approximation), which evidently should be 
weaker than the bare Coulomb interaction (insofar as low-energy 
scattering processes are concerned), 
\footnote{
\small
Otherwise the independent-particle model would not be an appropriate
zeroth-order approximation.}
renders the QP states non-stationary. In other words, the actual QPs, if 
such entities can at all be meaningfully defined (\S~III), are merely 
approximate one-particle eigenstates of the many-body Hamiltonian; 
the non-stationary nature of a QP `eigenstate' signifies that its energy, 
in contrast with that of an exact eigenstate, is un-sharp, or the spectral 
function of QPs consists of peaks with finite widths. Thus the true, that 
is interacting, QPs correspond to superpositions of nearly degenerate 
(as well as degenerate) many-body eigenstates, so that an initially 
well-defined QP ceases to be particle-like (in the sense of possessing a 
reasonably well-defined energy) with the passage of time. 

The information with regard to QPs is contained in the single-particle 
Green function $G$. The formal Lehmann (1954) spectral representation of 
this function in the energy domain reveals that energies of the 
single-particle excitations, the QPs, can be identified with the singular 
points (often inappropriately designated as `poles' --- \S~V) along the 
real energy axis ($\varepsilon$-axis) (Fetter and Walecka 1971). Such 
observation is of little practical relevance when systems in the 
thermodynamic limit are concerned. For {\sl macroscopic} systems, the 
continuum of the energy levels 
\footnote{\label{f1}
\small
From dimensional considerations it follows that for a system with linear 
dimension $L$, the separations between the energy levels scale like 
$L^{-2}$ (Landau and Lifshitz 1980, p.~14). } 
gives rise to branch points (which are not {\sl isolated} singularities) 
and branch cuts along the energy axis. This is naturally consistent with 
the above-indicated observation with regard to the broad rather than 
delta-function-like spectral function of QPs. Such a dramatic change in the 
analytic structure of $G(\varepsilon)$ gives rise to a number of 
effects the subtleties of which are often overlooked in the literature. In 
this work we discuss in some detail a number of these, insofar as 
they are relevant to the main objective of our work. One of these effects 
is associated with the fact that whereas a reasonably sharp peak in the 
spectral function may be described by a single pole with some small 
imaginary part, such description is in violation of the fact that
$G(\varepsilon)$ is a perfectly regular function on the entire complex 
energy plane, excluding the real $\varepsilon$-axis. In \S~III we show 
that such complex poles do {\sl not} simply correspond to the analytic 
continuation of $G(\varepsilon)$ into the complex energy plane (i.e. the 
physical Riemann sheet (RS)), rather to the analytic continuation of the 
latter to a RS; we refer to this as a non-physical RS (see Appendix A). 

The main complication, from the restricted viewpoint of the present paper, 
arises from the branch points of the Green function $G(k;\varepsilon)$, in 
particular that at $\varepsilon=\varepsilon_F$ which can be shown to be 
also a branch point of the self-energy, $\Sigma(k;\varepsilon)$. From 
this it follows that, for instance, such expression as (Hugenholtz 1957, 
DuBois 1959b, Luttinger 1961) ${\rm Im}\Sigma(k;\varepsilon) \sim
\mp\alpha_{k}(\varepsilon -\varepsilon_F)^2$, $\varepsilon 
{\mathstrut_{\displaystyle {<} }^{\displaystyle >}} \varepsilon_F$, {\sl 
cannot} be considered as the leading term in the Taylor series expansion 
of the self-energy operator at $\varepsilon=\varepsilon_F$. Thus there 
exists no {\sl a priori} reason for the dispersion of the QP energies
\footnote{\label{f1a}
\small
At places in the present text, such as here, we refer to `QP energies' 
even though QPs may not be well-defined. This is justified by the fact 
that even in these cases the equation for the `QP energies' (i.e. 
Eq.~(\ref{e13} below) does have a solution --- the ill-defined nature of 
QPs in these cases is associated with the fact that the single-particle 
eigenstates corresponding to these energies (see Eq.~(\ref{e10}) below) 
do {\sl not} describe particle-like excitations (for details see in 
particular \S~IV.C.). }
to be a `smooth' function of $k$ in a neighbourhood
\footnote{\label{f1b}
\small
Throughout the present work by `neighbourhood' we refer to an {\sl open} 
non-vanishing interval (e.g. along the real axis) or domain (on the
complex plane) which, however, may be arbitrarily small.}
of $k_F$ (for specification see further on) similar to that of 
non-interacting QPs, namely
\footnote{\label{f2}
\small
We have $\varepsilon_k^0 {:=} \hbar^2 k^2/[2 m_e] \equiv \varepsilon_{k_F}^0 
+ (\hbar^2 k_F/m_e) (k - k_F) + (\hbar^2/[2 m_e]) (k - k_F)^2$. The important 
feature of this energy dispersion is that it is of the general form 
$\varepsilon_k^0 =\varepsilon_{k_F}^0 + \varsigma (k - k_F) + o(k-k_F)$ for 
$k\to k_F$ (see \S~IV.C), where $\varsigma$ is a {\sl finite} constant 
(explicitly, in the case at hand we have $\varsigma\equiv \hbar^2 k_F/m_e$)
and $o(k-k_F)$ stands for a function that in comparison with $(k-k_F)$ is 
vanishingly small as $k\to k_F$. In this respect the property 
$\varepsilon_k^0 \propto k^2$ is to us of relatively minor relevance.
We point out that the coefficient of the {\sl linear} term in the QP energy 
dispersion (here $\varsigma$), if such a coefficient at all exists, has 
the relevance that through it such attributes as the Fermi velocity $v_F$ 
and effective mass $m_e^*$ can be assigned to the corresponding QP (see 
\S~IV.C, specifically Eqs.~(\ref{e19}) and (\ref{e20})). } 
$\varepsilon_k^0 \propto k^2$, with $k {:=} \|{\bf k}\|$. In fact, for the 
special case of `Luttinger liquids' (Haldane 1980, 1981, Luttinger 1963, 
Mattis and Lieb 1965, Dover 1968, Anderson 1997), without hereby specifying 
the dimension of the spatial space to which the corresponding system is 
confined, or marginal Fermi liquids (Varma, {\sl et al.}, 1989, Littlewood 
and Varma 1991, Kotliar, {\sl et al.}, 1991), we explicitly demonstrate 
that $\varepsilon_k$ is {\sl not} differentiable 
at $k=k_F$ (see \S\S~IV.C and VI). As we shall emphasise later in 
this paper (\S~IV.C), QPs in Fermi liquids are special in that their 
energy dispersion in the close vicinity of the Fermi surface is, apart 
from a scaling factor arising from the re-normalisation of the electron 
mass {\sl or} the Fermi velocity, non-interacting-like. Our close 
inspection of the proof of the Luttinger (1961) theorem (\S~IV), the 
contents of which we have spelled out in the Abstract, reveals that this 
`non-interacting-like' assumption with regard to the dispersion of the QP 
energies is implicit in Luttinger's (1961) proof, so that Luttinger's 
theorem in essence amounts to a statement concerning the consistency of 
this assumption with the property ${\rm Im} \Sigma(k;\varepsilon)\sim\mp 
\alpha_k (\varepsilon -\varepsilon_F)^2$, $\varepsilon 
{\mathstrut_{\displaystyle {<} }^{\displaystyle >}} \varepsilon_F$, for 
$\varepsilon\to\varepsilon_F$. In view of our rigorous demonstration with 
regard to non-differentiability of $\varepsilon_k$ at $k=k_F$ for Luttinger 
liquids, or marginal Fermi liquids, which invalidates Luttinger's implicit 
assumption, we are left to conclude that in principle nothing precludes 
existence of a non-Fermi-liquid behaviour for {\sl metallic} systems
\footnote{
\small
Our emphasise throughout this work on the {\sl metallic} nature of the 
systems under consideration corresponds to the fact that non-metallic 
systems are {\sl by definition} non-Fermi liquids.}
in spatial dimensions larger than unity. In particular, contrary to the 
general belief, the break-down of the many-body perturbation theory 
(Luttinger 1961, Anderson 1988, 1989, Varma, {\sl et al.} 1989, Anderson 
1990a,b, 1991, Littlewood and Varma 1991, Kotliar, {\sl et al.} 1991, 
Anderson 1992, 1993, 1997) for a metallic system and the nature of the 
low-lying single-particle excitations in this stand in {\sl no} direct 
relationship.
\footnote{\label{f2a}
\small
We should like to emphasise that often from the contexts of arguments
presented in the literature, in disapproval of the many-body perturbation 
theory, it is not evident what precisely is meant by the `perturbation 
theory'. In the present work we consider a series for the self-energy 
operator in terms of the {\sl interacting} single-particle Green function, 
each term of which is diagrammatically represented by a {\sl skeleton} 
diagram, as constituting a perturbation series. This we do entirely in 
the spirit of Luttinger's (1961) work. We should point out that such a 
series is {\sl not} a power series in the coupling constant of the 
electron-electron interaction and consequently does not conform to the 
orthodox definition for perturbation series. }

For clarity of discussions, we briefly specify the scheme by which we 
classify a {\sl metallic} system as a Fermi liquid. This scheme, which 
to our best knowledge has not been utilised earlier in a similar context, 
has the advantage of being general and independent of the details of the 
metallic system to which it is applied. To this end, we first point out 
that the behaviour ${\rm Im} \Sigma(k;\varepsilon)\sim\mp \alpha_k 
(\varepsilon -\varepsilon_F)^2$, $\varepsilon {\mathstrut_{\displaystyle 
{<} }^{\displaystyle >}} \varepsilon_F$, for $\varepsilon\to\varepsilon_F$, 
which is specific to Fermi liquids, leads to (see \S\S~IV.C, V and 
Appendix C) ${\rm Re}\Sigma(k,\varepsilon) -\Sigma(k,\varepsilon_F) \sim 
\beta_k (\varepsilon -\varepsilon_F)$ for $\varepsilon\to \varepsilon_F$, 
with $\beta_k$ a constant to be specified in Appendix C. The distinguishing 
aspect of $\Sigma(k,\varepsilon)$ whose real and imaginary parts have such 
asymptotic forms is as follows.

\noindent
{\bf Condition A:} $\Sigma(k_F;\varepsilon)$ is a {\sl continuously 
differentiable} function 
\footnote{\label{f2b}
\small
A function $f(x)$ is {\sl continuously differentiable} in the {\sl closed} 
interval $[a,b]$ provided firstly that it is continuous in $[a,b]$ and
secondly that its derivative exists at all points of the {\sl open} interval 
$(a,b)$ {\sl and} coincides at all such points with a function which is 
continuous in $[a,b]$; $f(x)$ is continuously differentiable in a {\sl 
neighbourhood} of $x=x_0$ if $x_0$ is {\sl interior} to a finite interval, 
such as $[a,b]$, in which $f(x)$ is continuously differentiable. }
of $\varepsilon$ in a neighbourhood of $\varepsilon_F$. 

This condition suffices to guarantee a finite discontinuity $Z_{k_F}$ in 
the momentum-distribution function ${\sf n}(k)$ at $k=k_F$ (\S~V). A finite 
$Z_{k_F}$ is generally considered to signify a Fermi liquid. If this merely
evidences a Fermi liquid, then condition (A) would constitute the necessary 
{\sl and} sufficient condition for a metallic system to be one (see \S~V). 
However, condition (A) can be shown (see \S\S~IV.C and V) not to be 
sufficient for the QP energy to possess a `well-defined' dispersion in a 
neighbourhood of $k=k_F$, that is one which is expressible in terms of a 
continuously differentiable function of $k$ in this neighbourhood. For the 
existence of such a dispersion, the following condition is also required
to be satisfied. 

\noindent
{\bf Condition B:}
$\Sigma(k;\varepsilon_F)$ is a {\sl continuously differentiable} function 
of $k$ in a neighbourhood of $k=k_F$. 

We therefore choose to consider a {\sl metallic} system as a Fermi liquid 
provided the corresponding $\Sigma(k;\varepsilon)$ satisfies both 
condition (A) and condition (B); this specifies the notion of `smooth' 
introduced above. According to this, which is physically the most sound 
definition for Fermi liquids,
\footnote{\label{f2c}
\small
The notion of a QP without an associated energy dispersion is devoid 
of physical significance.}
a finite $Z_{k_F}$, although {\sl necessary}, is not {\sl sufficient} 
for a system to be a Fermi liquid. We explicitly demonstrate (\S~VI)
that whereas $\Sigma(k;\varepsilon_F)$ corresponding to a Fermi liquid is 
(by definition) a continuously differentiable function of $k$ in a 
neighbourhood of $k=k_F$, the same {\sl may not} apply to 
$\Sigma(k;\varepsilon)$ when $\varepsilon \not=\varepsilon_F$. We further 
demonstrate (\S~VI) that $\Sigma(k; \varepsilon)$ pertaining to a 
metallic system (whether Fermi liquid or otherwise), which is not 
continuously differentiable with respect to $k$ in a neighbourhood of 
$k=k_0$ (where $k_0$ would possibly be $k_F$) for {\sl some} $\varepsilon 
\not=\varepsilon_F$, is {\sl not} continuously differentiable with 
respect to $k$ in a neighbourhood of $k=k_0$ for {\sl any} $\varepsilon$, 
with the possible exception of $\varepsilon =\varepsilon_F$; this 
exceptional instance is (by definition) reserved for Fermi liquids (see 
above). To be explicit, consider $\alpha_k$ and $\beta_k$ which feature 
in the above expressions for the self-energy of a Fermi liquid (see also 
\S~V). Our above statements imply that these may {\sl not} necessarily be 
continuously differentiable functions of $k$ in a neighbourhood of $k=k_F$; 
possible singularities associated with $\partial\alpha_k/\partial k$ and 
$\partial\beta_k/\partial k$, as $k\to k_F$, are suppressed by 
$(\varepsilon -\varepsilon_F)^2$ and $(\varepsilon-\varepsilon_F)$ 
respectively for $\varepsilon =\varepsilon_F$. In \S~VI we establish 
relationships between the behaviour of $\Sigma(k;\varepsilon)$ for $k$ and 
$\varepsilon$ in neighbourhoods of $k_F$ and $\varepsilon_F$ respectively 
and that of the momentum distribution function for $k$ approaching $k_F$.

Aside from the above-indicated `smoothness' assumption with regard to 
the dispersion of the QP energies which is implicit in Luttinger's (1961) 
proof, there is a second aspect specific to this proof that makes the 
theorem inapplicable to metallic systems of particles interacting through 
the long-range Coulomb interaction function. Luttinger's approach is based 
on the perturbation expansion of the self-energy operator in terms of the 
exact single-particle Green function (the terms in this expansion are 
represented by means of skeleton self-energy Feynman diagrams (Luttinger 
and Ward 1960)) and the {\sl bare} electron-electron interaction function. 
It can be shown (\S~IV.D) that a certain class of terms in this expansion 
are unbounded as a consequence of non-integrable singularities corresponding 
to zero momentum-transfer scattering processes (infrared divergence). 
Therefore, for such systems a formal term-by-term analysis of the 
perturbation series for the self-energy operator can only be meaningful 
when the perturbation expansion is in terms of the dynamically-screened 
interaction function (Hubbard 1957) $W(\varepsilon)$ (for a comprehensive 
discussion see Mattuck (1992, \S~10.4)). In contrast to the {\sl static} 
bare Coulomb interaction $v_c$, $W(\varepsilon)$ is a function with 
singularities along the real energy (i.e. $\varepsilon$) axis (branch 
points, branch cuts, etc. --- similar to the Green function and the 
self-energy operator) and a Lehmann-type spectral representation of this 
function reveals that these singularities coincide with energies of the 
neutral elementary excitations of the system (i.e. energies of excited 
$N$-electron states as measured with respect to the energy of the 
$N$-electron ground state). It follows that an analysis of the self-energy 
operator along the lines of Luttinger (1961) requires {\sl additional} 
knowledge with regard to the dispersion of these neutral excitations. In 
other words, in this case the dispersion of the QP energies and that of 
neutral excitations cannot be considered separately, but must be dealt with 
in conjunction. This is of relevance particularly to uniform {\sl 
two}-dimensional systems of electrons whose spectrum of coherent neutral 
excitations (i.e. plasmons) is gap-less, implying some non-negligible 
amount of interference amongst the neutral and the single-particle, i.e. 
QP, excitations.

Our present work has been motivated by a long-standing discussion in the 
literature with regard to the unusual properties of layered high-$T_c$ 
compounds in their normal states. On the one hand these unusual properties 
have been ascribed to the fact that the normal metallic states of these 
systems not Fermi-liquid states (Anderson 1988, 1989, Varma, {\sl et al.},
1989, Anderson 1990a,b, 1991, Littlewood and Varma 1991, Kotliar, {\sl et
al.}, 1991, Anderson 1992, 1993, 1997) while, on the other hand, according 
to the Luttinger (1961) theorem dealt with in the present work, metallic 
interacting systems in spatial dimensions larger than unity must be Fermi 
liquids. As emphasised by Luttinger (1961), the validity of the Luttinger 
theorem is conditional to that of the many-body perturbation theory for 
the self-energy operator to {\sl all} orders. In this light, it is asserted 
that a non-Fermi-liquid metallic behaviour of a system would necessarily 
imply break-down of the many-body perturbation theory for this system 
(Anderson 1997). These observations give rise to the following questions. 

(1) Can interacting electron systems in spatial dimensions $d$ larger than 
unity, in particular for $d=2$, be non-Fermi liquids? 

(2) does the many-body perturbation theory break down in cases where the 
metallic systems under consideration are non-Fermi liquids?

Concerning question (1), the analyses presented in this work lead us to 
the conclusion that in principle {\sl nothing} stands in the way of 
realisation of non-Fermi-liquid states in spatial dimensions $d$ larger 
than unity. In fact the very strict conditions (A) {\sl and} (B), 
introduced above, imposed on the self-energy of Fermi liquids, suggests 
that non-Fermi liquid metallic systems (in $d>1$) may not be as uncommon 
as generally perceived. We do not touch upon the above question (2) and 
refrain from statements that are likely to be speculative at this stage. 
In this connection we point out that, since the existence of non-Fermi 
liquids (in particular for $d=2$) and the validity of the many-body 
perturbation theory as applied to these are not {\sl a priori} mutually 
exclusive, the question with regard to breakdown of this theory is not as 
urgent as it would be otherwise. This, however, does not diminish the 
relevance of question (2). We should like to emphasise that in calculating 
$\Sigma(k;\varepsilon)$, whether perturbatively or otherwise, account 
has to be taken of the fact that any approximation scheme that involves 
indiscriminate Taylor expansions of functions of energy and momentum
(which expansions implicitly imply {\sl continuous differentiability} of 
these functions in the pertinent regions of energy and momentum) 
{\sl may} inhibit a non-differentiable $\Sigma(k;\varepsilon)$ as
function of both $k$ and $\varepsilon$, and therefore a non-Fermi 
liquid behaviour, from being obtained. 

The above questions, (1) and (2), have been subject of intensive study 
in recent years, the main body of the results thus far obtained pointing 
towards a Fermi-liquid state in two spatial dimensions. In spite of these, 
it has as yet {\sl not} proved possible unequivocally to rule out the
existence of non-Fermi-liquid states in two-dimensional metallic systems. 
This can be ascribed to two main reasons. 

(i) Owing to the complexity of the many-body problem at hand, application 
of reliable theoretical techniques must of necessity be accompanied by 
simplifying approximations, the integrity of which may be a matter of 
dispute. 

(ii) Even relevance of certain calculated quantities to the problem at 
hand has been matter of debate. 

Both these aspects are aptly represented in the following: 
Engelbrecht and Randeria (1990) have found no evidence for the breakdown 
of the many-body perturbation theory and the Fermi-liquid theory in a 
dilute two-dimensional system of fermions interacting through a 
short-range repulsive potential, contradicting the suggestion made by 
Anderson (1990b). It appears, however, that the phase shift as calculated 
by Engelbrecht and Randeria is {\sl not} that which is encountered in the 
arguments by Anderson (see Engelbrecht and Randeria 1991, Anderson 1991). 
\footnote{
\small
For a detailed discussion of the singular effective interaction amongst 
QPs, as encountered in the arguments by Anderson (1991), see Stamp (1993);
see also Houghton, Kwon, and Marston (1994).}

Work by Fujimoto (1990), Fukuyama, Narikiyo and Hasegawa (1991) and Fukuyama, 
Hasegawa and Narikiyo (1991) on the two-dimensional Hubbard model with 
repulsive on-site interaction $U$ within the $t$-matrix approximation 
(which is appropriate to the low-density limit) has indicated that, whereas 
for $k\not=k_F$, ${\rm Im}\Sigma(k,\varepsilon)$ retains the conventional 
Fermi-liquid form in three spatial dimensions (presented above),
\footnote{
\small
Throughout, ${\sim'}$ indicates that the corresponding asymptotic 
relations are correct up to multiplicative constants.}
${\rm Im}\Sigma(k=k_F,\varepsilon) {\sim'} (\varepsilon -\varepsilon_F)^2 
\ln\vert \varepsilon -\varepsilon_F\vert$ as $\varepsilon\to\varepsilon_F$. 
This singular contribution, obtained earlier by Hodges, Smith and Wilkins 
(1971) and Bloom (1975) concerning two-dimensional Fermi systems, is not 
sufficiently strong to render the self-energy non-Fermi-liquid like. 
\footnote{\label{f2d}
\small
It can be shown (see Appendix C) that the associated ${\rm Re}\Sigma(k_F;
\varepsilon)-\Sigma(k_F;\varepsilon_F) {\sim'} (\varepsilon-\varepsilon_F)$ 
so that $\Sigma(k_F;\varepsilon)$ is a continuously differentiable function 
of $\varepsilon$ in a neighbourhood of $\varepsilon =\varepsilon_F$ and 
thus, on account of condition (A) discussed above, gives rise to a finite 
$Z_{k_F}$. }
Serene and Hess (1991), using the conserving `fluctuating-exchange (FLEX) 
approximation' on a finite but large lattice, have also not found evidence 
for a non-Fermi-liquid behaviour in the two-dimensional Hubbard model. In 
a more recent work, Yokoyama and Fukuyama (1997) have arrived at the 
conclusion that in the two-dimensional Hubbard model with repulsive on-site 
interaction, the process of forward scattering gives rise to an anomalous 
contribution to the self-energy and consequently to a vanishing 
quasi-particle weight factor $Z_{k_F}$. This result, Yokoyama and Fukuyama 
(1997) asserted, would be ``a microscopic demonstration of the claim by 
Anderson [(1990a,b), (1993)]''. We (Farid 1999a) have shown that the 
indicated anomalous contribution is {\sl entirely} a consequence of the 
violation of a crucial symmetry in the momentum space (associated with 
the time-reversal symmetry of the ground state of the system under 
consideration) by the particle-particle correlation function employed 
by these workers.

Momentum-space perturbative renormalisation-group calculations by Shankar 
(1991, 1994) have also borne out a Fermi-liquid picture of fermions in two 
spatial dimensions. Shankar (1994) enumerated, however, a number of
possibilities that in principle may render this finding, in its generality, 
invalid (see \S~XI in (Shankar 1994)). Work by Castellani, Di Castro 
(1994), Castellani, Di Castro and Metzner (1994), and Metzner, Castellani 
and Di Castro (1998) modelled on the treatment of one-dimensional 
interacting systems put forward by Dzyaloshinski\v{i} and Larkin (1973),
which exploits the conservation laws and the associated Ward identities, 
and treats $d$, the spatial dimension, as a {\sl real} variable, 
establishes that metallic systems with strong forward scattering (i.e. 
that corresponding to small momentum transfers) are Fermi liquids in $d>1$, 
even though for $d\not=3$ some properties of these systems differ from 
those pertaining to conventional Fermi liquids in $d=3$. Here the 
limited applicability of the formalism to systems with 
dominant forward scattering does not rule out non-Fermi-liquid systems in 
$d > 1$. In a forthcoming work (Farid 1999b) we present a detailed study 
concerning some limiting aspects associated with the technique employed 
in these studies.
 
Systems of particles interacting through long-range repulsive interaction 
functions $v({\bf r}-{\bf r}')$ which for large $\| {\bf r} -{\bf r}'\|$ 
behave like $\sim 1/\|{\bf r}-{\bf r}'\|^{2-d}$, when $1 < d < 2$, and like 
$\sim \ln\|{\bf r}-{\bf r}'\|$, when $d=2$, have been shown to resemble 
one-dimensional Luttinger liquids for $1 < d < 2$ and a ``$Z_{k_F}=0$ 
Fermi liquid'' for $d=2$ (Bares and Wen 1993, Kwon, Houghton and Marston 
1995). We note that in metals the Coulomb interaction function, which 
behaves like $1/\| {\bf r}-{\bf r}'\|$, is screened through the mediation
of the charge polarisation fluctuations (Hubbard 1957); within the framework 
of the random-phase approximation (RPA), the screened interaction function 
in the {\sl static} limit can be shown to behave like $\sim \cos(2 k_F
\|{\bf r}-{\bf r}'\|)/\|{\bf r}-{\bf r}'\|^3$ (Fetter and Walecka 1971, 
pp. 178 and 179; Ashcroft and Mermin 1981, p.~343). It follows that the 
above long-range interactions should have their origin in processes that 
are not related to the electrostatic electron-electron interaction. The 
only known interaction that remains long-ranged in metals corresponds to 
that of electronic currents mediated by the exchange of transverse 
photons (Holstein, Norton, and Pincus 1973, Reizer 1989, 1991). Lack 
of the {\sl static} screening of this interaction leads to the asymptotic 
behaviour of the self-energy $\Sigma(k;\varepsilon_k)$ (on-the-mass-shell
self-energy),
\footnote{
\small
For the equation satisfied by the QP energy dispersion $\varepsilon_k$ see 
Eq.~(\ref{e17}) below.}
as $k\to k_F$ (or $\varepsilon_k\to\varepsilon_F$), to be that 
characteristic of the marginal Fermi liquids (Varma, {\sl et al.}, 1989, 
Littlewood and Varma 1991, Kotliar, {\sl et al.} 1991); on the other hand, 
for the fixed $k\not=k_F$, $\Sigma(k;\varepsilon)$ behaves Fermi-liquid 
like as $\varepsilon\to\varepsilon_F$ (Holstein, Norton, and Pincus 1973). 
We note that the $\Sigma(k;\varepsilon)$ considered here does {\sl not} 
involve the effects of the Coulomb interaction beyond a mean-field level; 
it merely describes the self-energy of the current-current interaction 
whose bare coupling constant is proportional to the square of the ratio of 
the Fermi velocity to the light velocity in vacuum ($\sim 10^{-4}$) and 
therefore is not of dominating influence except in extremely pure metals 
and at very low temperatures. We shall not enter into further details 
concerning this subject here
\footnote{
\small
In particular we do not touch upon the subject of electrons interacting 
with a gauge field, such as is the case in two-dimensional electron
systems exposed to external magnetic field (the gauge field here being 
the statistical Chern-Simons field), where {\sl in principle} at fractional
Landau-level filling factors $\nu$ with even denominators (specifically 
at and close to $\nu=1/2$) the states are metallic but may be non-Fermi 
liquids (Kalmeyer and Zhang 1992, Halperin, Lee and Read 1993). }  
and refer the reader to the cited literature.
\footnote{
\small
For a concise review see (Tsvelik 1995, Ch.~12). }

The organisation of this work is as follows. In \S~II we present the 
Lehmann spectral representation for the single-particle Green function 
and clarify certain elements that are of particular relevance to our 
analysis of the Luttinger (1961) theorem. In \S~III we rigorously define 
the notion of QP and derive equations from which QP energies and wavefunctions 
can be obtained. Here we indicate the necessary steps to be undertaken 
for obtaining complex-valued QP energies. We devote \S~IV to the main 
objective of our work, namely a detailed analysis of the Luttinger theorem. 
In \S~V we briefly review and discuss a theorem due to Migdal (1957) in 
light of our findings in \S~IV. In \S~VI we compare Fermi liquids with 
non-Fermi liquids in terms of the `smoothness' properties of $\Sigma(k;
\varepsilon)$, as a function of both $k$ and $\varepsilon$. In \S~VII we 
conclude this work by a brief discussion and a review of our main results. 
In Appendix A we introduce and discuss a number of mathematical concepts 
that we repeatedly encounter in the present work. Here we further attempt 
to clarify the physical relevance of the introduced notions by means of 
some simple examples. In Appendix B we derive the leading asymptotic 
contribution to the self-energy operator $\Sigma(\varepsilon)$ at large 
$\vert\varepsilon\vert$. For this, the next-to-the-leading-order asymptotic 
contribution to the single-particle Green function needs be calculated;
in the same Appendix we present this contribution as well as some details 
underlying its derivation. In Appendices C and D we consider the 
next-to-the-leading-order asymptotic term to the self-energy pertaining 
to Fermi- and marginal Fermi-liquids (Appendix C) and the Luttinger 
liquid (Appendix D). Here we separately deal with the cases corresponding
to $k=k_F$ and $k\not=k_F$. In Appendix D we explicitly demonstrate
that the self-energy of the one-dimensional Luttinger model is
not continuously differentiable at the Fermi points, in full conformity
with a general result established in Appendix C. {\sl In this work we 
identify electrons with spin-less fermions.} 

\section{The single-particle Green function}

Here we explicitly deal with the single-particle Green function. In 
addition to an exposition of the formal significance of the singularities 
of this function, we discuss three specific and distinct energies $\mu$, 
$\mu_N$ and $\mu_{N+1}$, that are invariably (but un-justifiably) 
identified in the literature concerning the free-electron system. 

Consider the following Lehmann (1954) representation (Fetter and Walecka 
1971) for the (`physical') single-particle Green function in the 
coordinate-free representation 
\begin{equation}
\label{e1}
G(\varepsilon) = \hbar \sum_s \Lambda_s \Lambda_s^{\dag}
\Big\{ {{\Theta(\mu-\varepsilon_s)}\over
\varepsilon - \varepsilon_s - i\eta}
+ {{\Theta(\varepsilon_s-\mu)}\over \varepsilon
-\varepsilon_s + i\eta}\Big\},\;\;\; \eta\downarrow 0,
\end{equation}
where
\begin{eqnarray}
\label{e2}
\Lambda_s {:=} \left\{ \begin{array}{ll}
\langle\Psi_{N-1,s}\vert\widehat{\psi}\vert\Psi_{N,0}\rangle,
&\;\;\; {\rm when}\;\; \varepsilon_s < \mu, \\
\langle\Psi_{N,0}\vert\widehat{\psi}\vert\Psi_{N+1,s}\rangle,
&\;\;\; {\rm when}\;\; \varepsilon_s > \mu
\end{array} \right.
\end{eqnarray}
denotes a `Lehmann amplitude', and
\begin{eqnarray}
\label{e3}
\varepsilon_s {:=} \left\{ \begin{array}{ll}
E_{N,0} - E_{N-1,s},
&\;\;\; {\rm when}\;\; \varepsilon_s < \mu, \\
E_{N+1,s} - E_{N,0},
&\;\;\; {\rm when}\;\; \varepsilon_s > \mu.
\end{array} \right.
\end{eqnarray}
Above $\vert\Psi_{M,s}\rangle$ stands for the {\sl exact} $M$-particle 
eigenstate of the interacting system and $E_{M,s}$ for the corresponding 
eigenenergy; $s$ denotes the set of all quantum numbers that uniquely 
specify $\vert\Psi_{M,s}\rangle$, (we choose $s=0$ to signify the ground 
state which we assume to be non-degenerate); $\widehat{\psi}$ is the 
annihilation field operator.
\footnote{\label{f3}
\small
Throughout this work, $N$ denotes the actual number of the electrons 
in the system.}
Here the `chemical potential' $\mu$ is {\sl any} real value which 
satisfies $\mu_N \leq \mu \leq \mu_{N+1}$, where $\mu_N {:=} E_{N,0} - 
E_{N-1,0}$ and $\mu_{N+1} {:=} E_{N+1,0} - E_{N,0}$. That such a $\mu$ 
should exist follows from the requirement of stability of the ground 
state, meaning that $\varepsilon_g {:=} (E_{N+1,0}-E_{N,0})-(E_{N,0} 
-E_{N-1,0})$ $\equiv \mu_{N+1}-\mu_N$ be {\sl non-negative}. For a 
system with uniform electron distribution, which is thus a system
in the thermodynamic limit, it holds $\mu_{N+1} =\mu_{N} + {\cal O}
(N^{-p})$, with $p> 0$.
\footnote{\label{f4}
\small
The value $p=1$ as given in, e.g., Fetter and Walecka (1971, p.~75), 
is incorrect.}
Therefore with regard to uniform systems, we are considering a case in 
which $(\mu_N,\mu_{N+1})$ is a {\sl finite} infinitesimal open interval. 
In \S~IV.C we shall see that $\mu_N \equiv \varepsilon_F$, the Fermi 
energy of the system of $N$ electrons.

It can be easily verified that (see Appendix A and in particular \S~A.3,
for our notational conventions)
\begin{equation}
\label{e4}
\widetilde{G}(z) {:=}
\hbar \sum_s {{\Lambda_s \Lambda_s^{\dag}}\over z -\varepsilon_s}
\end{equation}
is the analytic continuation (Whittaker and Watson 1927, pp. 96-98,
Titchmarsh 1939, pp. 138-164) of the (`physical') single-particle
Green function $G(\varepsilon)$ into the complex $z$-plane, that it 
is $\widetilde{\widetilde{G}}(z)$ on the physical RS (see Appendix A); 
$G(\varepsilon)$ is obtained from $\widetilde{G}(z)$ according to the 
following prescription
\footnote{
\small
Below $\widetilde{f}(z)$ ($f(\varepsilon)$) is a representative for any 
function of $z$ ($\varepsilon$) that we encounter in the present work.}
\footnote{\label{f5}
\small
It should be noted that the prescription given in Eq.~(\ref{e5}) is 
connected to our convention with regard to the time-Fourier transform 
of the time-dependent functions, namely $f(\varepsilon) {:=} 
\int_{-\infty}^{\infty} {\rm d}t\; F(t) \exp(+i\varepsilon t/\hbar)$. Had 
we chosen $\exp(-i\varepsilon t/\hbar)$ rather than 
$\exp(+i\varepsilon t/\hbar)$, `$\pm$' would have to be `$\mp$'.}
\begin{equation}
\label{e5}
f(\varepsilon) {:=} \lim_{\eta\downarrow 0}
\widetilde{f}(\varepsilon\pm i\eta),\;\;\;\mbox{\rm for}\;\;
\varepsilon {\mathstrut_{\displaystyle {<} }^{\displaystyle >}}\mu.
\end{equation}

For a {\sl finite} system, the Lehmann representation indicates that
$\widetilde{G}(\varepsilon)\equiv G(\varepsilon)$ has {\sl poles}
\footnote{\label{f6}
\small
There is a subtlety involved here. For an open stable system, the spectrum 
consists of both a discrete and a continuous part. The former part 
corresponds to bound states, and the latter to scattering states; the 
continuous part of the spectrum gives rise to branch-cut discontinuities 
in the energy-dependent correlation functions of this system. Further, 
even though in such a system the number of particles is finite, the set 
of discrete one-particle excitation energies has an accumulation point 
which, as indicated in \S~A.1, is {\sl not} an isolated singularity of 
$\widetilde{G}(z)$. For closed and finite systems (that is those with 
impenetrable boundaries), the completeness of the one-particle eigenstates 
implies that the set of one-particle eigenenergies $\{\varepsilon_s\}$ 
possesses at least one accumulation point (this follows from the 
Bolzano-Weierstrass theorem --- see Whittaker and Watson (1927, pp. 12 
and 13)). Therefore, the non-isolated nature of the singular points of, 
for example, $\widetilde{G} (z)$ is {\sl not exclusively} a peculiarity 
of systems in the thermodynamic limit. }
at $\varepsilon = \varepsilon_s$ for all $s$ (see Eq.~(\ref{e3})). In the 
thermodynamic limit, the singular points of $\widetilde{G}(z)$ correspond 
to branch points (isolated singularities are not {\sl a priori} excluded) 
and continua of branch cuts covering (parts of) the {\sl real} energy axis 
(see \S~A.4). For finite systems as well as those in the thermodynamic
limit, $\widetilde{G}(z)$ is analytic for {\sl all} $z$ with ${\rm Im}(z)
\not=0$.

In \S~III we shall encounter the inverse operator $\widetilde{G}^{-1}(z)$. 
In view of this, we point out that from the representation in Eq.~(\ref{e4}) 
it can be easily demonstrated that $\widetilde{G}(z)$ has {\sl no} complex 
zeros (or, better, {\sl zero eigenvalues}) on the $z$-plane (Luttinger 1961). 
Thus $\widetilde{G}^{-1}(z)$, similar to $\widetilde{G}(z)$, is analytic 
for {\sl all} $z$ with ${\rm Im}(z) \not=0$.

\section{Quasi-particles; their ``energies'' and wavefunctions}

As we have mentioned in \S~I, to quasi-particles correspond one-particle 
wavefunctions which are eigenfunctions of some one-particle-like 
Schr\"odinger equation corresponding to an energy-dependent non-Hermitian 
one-body Hamiltonian, the potential-energy part of which consists of the 
self-energy operator 
\footnote{\label{f7}
\small
In spite of the fact that according to our present convention 
$\Sigma(k;\varepsilon)$ has dimension ${\rm s}^{-1}$, i.e. inverse 
second, we refer to it as self-{\sl energy}.}
and of possibly a non-trivial contribution due to an external potential 
(such as an ionic potential in a solid). We devote this Section to 
derivation as well as to the discussion of the equations for QP energies 
and wavefunctions. Some specific properties of the equation for the QP 
energies are of considerable relevance to our discussion of the Luttinger 
(1961) theorem in \S~IV.

In the representation-free form and for the complex energy $z$, from 
the Dyson equation $\widetilde{G}(z)=\widetilde{G}_0(z)+\widetilde{G}_0(z) 
\widetilde{\Sigma}(z)\widetilde{G}(z)$, with $\widetilde{G}_0(z)$ 
the single-particle Green function pertaining to the non-interacting 
system, one obtains
\begin{equation}
\label{e6}
\widetilde{G}^{-1}(z)={1\over\hbar}\{z I -\widetilde{\cal H}(z)\}
\end{equation}
where $I$ denotes the unit operator in the single-particle Hilbert
space and 
\begin{equation}
\label{e7}
\widetilde{\cal H}(z) {:=} H_0 + \hbar \widetilde{\Sigma}(z).
\end{equation}
Any singular point of $\widetilde{G}(z)$ is a solution of 
$\det\big(\widetilde{G}^{-1}(z)\big)=0$, or, making use of 
Eq.~(\ref{e6}), of
\begin{equation}
\label{e8}
\det\big(z I - \widetilde{\cal H}(z)\big) = 0.
\end{equation}
This is reminiscent of the equation for the energies of the `ideal' (i.e.
non-interacting) QPs, namely $\det\big(z I - H_0\big)=0$; the difference 
between the two arises from the difference between $\widetilde{\cal H}(z)$ 
and $H_0$ which according to Eq.~(\ref{e7}) is equal to $\hbar
\widetilde{\Sigma}(z)$. The latter, a function of the complex energy 
parameter $z$, is a non-Hermitian operator; even for $z\to\varepsilon \pm 
i\eta$, $\eta\downarrow 0$, for $\varepsilon {\mathstrut_{\displaystyle 
{<} }^{\displaystyle >}} \varepsilon_F$, leading to $\widetilde{\Sigma}(z)
\to\Sigma(\varepsilon)$ (see Eq.~(\ref{e5}) above), $\widetilde{\cal H}(z)
\equiv {\cal H}(\varepsilon)$ is {\sl not} in general Hermitian. This 
implies, among others, that, unless $\widetilde{{\cal H}}(z)$ is Hermitian, 
the spectral decomposition of $\widetilde{\cal H}(z)$ involves the {\sl 
distinct} sets of left and right eigenvectors, $\{\widetilde{\phi}_{\ell}
(z)\}$ and $\{\widetilde{\psi}_{\ell}(z)\}$ respectively. Since the sets 
of eigenvalues pertaining to the latter two sets are up to ordering 
identical (Morse and Feshbach 1953, p. 885), by choosing the appropriate 
ordering, the spectral (or {\sl bi-orthonormal} (Morse and Feshbach 1953, 
pp. 884-886, Layzer 1963) --- see later) representation of 
$\widetilde{\cal H}(z)$ is as follows
\begin{equation}
\label{e9}
\widetilde{\cal H}(z) = \sum_{\ell} \widetilde{E}_{\ell}(z)
\widetilde{\psi}_{\ell}(z) \widetilde{\phi}_{\ell}^{\dag}(z),
\end{equation}
where $\widetilde{E}_{\ell}(z)$ denotes the common eigenvalue
corresponding to $\widetilde{\phi}_{\ell}(z)$ and 
$\widetilde{\psi}_{\ell}(z)$:
\begin{equation}
\label{e10}
\widetilde{\phi}_{\ell}^{\dag}(z)\widetilde{\cal H}(z)
= \widetilde{E}_{\ell}(z) \widetilde{\phi}_{\ell}^{\dag}(z);\;\;\;
\widetilde{\cal H}(z) \widetilde{\psi}_{\ell}(z)
= \widetilde{E}_{\ell}(z) \widetilde{\psi}_{\ell}(z).
\end{equation}
For $\widetilde{E}_{\ell}(z) \not= \widetilde{E}_{\ell'}(z)$ it holds 
$\langle\widetilde{\phi}_{\ell}(z),\widetilde{\psi}_{\ell'}(z)\rangle = 
\delta_{\ell,\ell'}$ (Morse and Feshbach 1953, p. 885). In the case 
of degeneracy, that is $\widetilde{E}_{\ell}(z) = \widetilde{E}_{\ell'}(z)$ 
for $\ell\not=\ell'$, the degenerate left and right eigenvectors can be 
made orthogonal through a Gram-Schmidt orthogonalisation procedure. When 
the left and right eigenfunctions are simultaneously bases of the unitary 
irreducible representations of the symmetry group of the QP Schr\"odinger 
equation, Eq.~(\ref{e10}), the degenerate eigenvectors are automatically 
orthogonal as long as they belong to different unitary irreducible 
representations (Cornwell 1984, pp. 81-83). 
\footnote{\label{f8}
\small
Consider, for instance, the uniform-electron system with which we deal 
in detail in the present work. The symmetry group of this system is the 
continuous translation group (which is Abelian) and therefore the 
corresponding bases for the (one-dimensional) unitary irreducible 
representations are specified by wave-vectors ${\bf k}$. The time-reversal 
symmetry of the problem implies degeneracy of eigenfunctions at ${\bf k}$ 
and $-{\bf k}$ (``Kramers' degeneracy'' (Landau and Lifshitz 1977, pp. 
223-226)). Since for ${\bf k} \not={\bf 0}$, ${\bf k}\not= -{\bf k}$, the 
eigenfunctions corresponding to ${\bf k}$ and $-{\bf k}$ are therefore 
automatically orthogonal. Note that, since our `electrons' are spin-less, 
for the application of the Kramers theorem we {\sl cannot} rely on the 
condition that the total spin of the system is half-integer. However, we 
can rely on the fact that {\sl all} irreducible representations of the 
translation group corresponding to ${\bf k}\not= {\bf 0}$ are {\sl 
essentially} complex. }
Further, $\{\widetilde{\phi}_{\ell}(z)\}$ and $\{\widetilde{\psi}_{\ell}
(z)\}$ satisfy the completeness relation (Morse and Feshbach 1953, p. 886)
\footnote{\label{f9}
\small
The left-hand side of this relation is the so-called {\sl idemfactor}
of the one-particle Hilbert space.}
\begin{equation}
\label{e11}
\sum_{\ell} \widetilde{\psi}_{\ell}(z) 
\widetilde{\phi}_{\ell}^{\dag}(z) = I,
\end{equation}
or $\sum_{\ell} \widetilde{\psi}_{\ell}({\bf r};z)
\widetilde{\phi}_{\ell}^*({\bf r}';z) = \delta({\bf r}-{\bf r}')$.

Because $\widetilde{\cal H}^{\dag}(z) = \widetilde{\cal H}(z^*)$, for 
${\rm Im}(z)\not=0$, a property that follows from 
$\widetilde{\Sigma}^{\dag}(z)=\widetilde{\Sigma}(z^*)$ (Appendix B in 
DuBois (1959a), Luttinger 1961), it can be readily shown that, for 
${\rm Im}(z)\not=0$,
\begin{equation}
\label{e12}
\widetilde{\phi}_{\ell}(z) \equiv \widetilde{\psi}_{\ell}(z^*),\;\;\;
\widetilde{\psi}_{\ell}(z) \equiv \widetilde{\phi}_{\ell}(z^*),\;\;\;
\widetilde{E}_{\ell}(z^*)=\widetilde{E}_{\ell}^*(z).
\end{equation}
As we shall see later, the last result
\footnote{\label{f10}
\small
This result is the expression of the Riemann-Schwarz reflection principle
(Titchmarsh 1939, p. 155) which follows from the analyticity of 
$\widetilde{\Sigma}(z)$ over the entire complex $z$-plane, with the 
exception of the real $\varepsilon$-axis, and the fact that over the finite 
(even though infinitesimal) {\sl open} interval $(\mu_N,\mu_{N+1})$, 
$\Sigma(\varepsilon)$, or $G(\varepsilon)$, is real valued.}
is of particular significance to our considerations. By defining 
$E_{\ell}(\varepsilon)$ in accordance with Eq.~(\ref{e5}), for 
${\rm Im}E_{\ell}(\varepsilon)$ the following must hold
\begin{equation}
\label{e12a}
{\rm Im}E_{\ell}(\varepsilon)\;
{\mathstrut_{\displaystyle {\geq} }^{\displaystyle \leq}}\; 0,
\;\; {\rm for}\;\;\;\;
\varepsilon\; {\mathstrut_{\displaystyle {<} }^{\displaystyle >}}\; \mu.
\end{equation}
Violation of these inequalities signifies breakdown of causality, or 
instability of the ground state due to its collapse into a lower-energy 
state; to appreciate this, note that (combine Eqs.~(\ref{e6}) and 
(\ref{e11})) $\widetilde{G}(z) =\hbar \sum_{\ell} \widetilde{\psi}_{\ell}(z) 
\widetilde{\phi}_{\ell}^{\dag}(z)/\big(z-\widetilde{E}_{\ell}(z)
\big)$ and compare this with the Lehmann representation in
Eq.~(\ref{e1}) (see also Eq.~(\ref{e3})).

From the spectral representation for $\widetilde{\cal H}(z)$ in 
Eq.~(\ref{e9}) above, together with Eq.~(\ref{e11}), the equivalence of
Eq.~(\ref{e8}) with the set of equations
\begin{equation}
\label{e13}
\widetilde{E}_{\ell}(z) = z, \;\;\;\forall \ell,
\end{equation}
is readily established.
As we have mentioned in \S~II, singular points of $G(\varepsilon)$, 
according to the Lehmann representation, coincide with the real-valued 
quantities $\varepsilon_s$ (see Eq.~(\ref{e3})). Since Eq.~(\ref{e13}) is 
the equation for these singular points, it may be expected that by taking 
the appropriate limits $z\to \varepsilon \pm i\eta$, with $\eta \downarrow 
0$ (see Eq.~(\ref{e5})), Eq.~(\ref{e13}) should transform into an equation 
with real-valued solutions $\varepsilon_s$; owing to the analyticity of 
$\widetilde{G}^{-1}(z)$ for ${\rm Im}(z)\not=0$ (see \S~II), it follows 
that Eq.~(\ref{e13}) cannot be satisfied for any $z$ with ${\rm Im}(z)
\not=0$ (for a physical interpretation of this statement see the following 
paragraph). As we shall demonstrate below, in the thermodynamic limit the 
number of real-valued solutions of Eq.~(\ref{e13}), with $z\to \varepsilon 
\pm i\eta$ (see Eq.~(\ref{e5})), is limited. Further, in the same limit, 
these real-valued solutions are in general not poles, in contrast with 
what the Lehmann representation in Eq.~(\ref{e1}) (or Eq.~(\ref{e4})) may 
suggest. This is not a contradiction, since as we indicate in \S~A.4, 
analytic properties of correlation functions in the thermodynamic limit 
are fundamentally different from those of finite systems. 

It is often mentioned that finite lifetimes of QPs is reflected 
in the energies of these being complex valued. This statement, without 
further qualification, is misleading. To clarify this, consider 
Eq.~(\ref{e13}) and suppose that it were satisfied by $z=z_0$ with ${\rm 
Im}(z_0)\not=0$. Owing to the reflection property of $\widetilde{E}_{\ell}
(z)$ presented in Eq.~(\ref{e12}), it follows that $z=z_0^*$ must also 
satisfy Eq.~(\ref{e13}). This {\sl cannot} be the case, since depending on 
whether ${\rm Re}(z_0)$ is less than or larger than $\mu$, one of the two 
solutions $z_0$ and $z_0^*$ signals violation of the principle of 
causality. In this connection, recall that the single-particle Green 
function has been defined in terms of the {\sl time-ordered} product of two 
field operators in the Heisenberg representation (Fetter and Walecka 1971). 
Consequently, no $z_0$ can satisfy Eq.~(\ref{e13}) unless ${\rm Im}(z_0)=0$. 
Thus for a given $\ell$, Eq.~(\ref{e13}) has either a real-valued solution 
{\sl or} no solution at all; the possible real-valued solutions (see \S~A.1) 
are, in general, {\sl not} isolated. To obtain complex-valued 
quasi-particle energies, the appropriate equation to be solved is the 
following
\begin{equation}
\label{e14}
\widetilde{\widetilde{E}}_{\ell}(z) = z,\;\;\;\forall \ell,
\end{equation}
where $\widetilde{\widetilde{E}}_{\ell}(z)$ represent {\sl the} unique
analytic function of which $\widetilde{E}_{\ell}(z)$ is one specific 
branch (see \S~A.3). The analytic function $\widetilde{\widetilde{E}}_{
\ell}(z)$ (i.e. any of its branches; there may be infinity of these, 
depending on the nature of the branch points of $\widetilde{E}_{\ell}(z)$) 
can in principle be calculated from the knowledge of $\widetilde{E}_{\ell}
(z)$. To this end, one starts from $\widetilde{E}_{\ell}(z)$ and analytically 
continues it across its branch cuts into a RS `above' or `below' the 
physical RS (see examples in \S\S~A.3 and A.4). This process can be 
repeated on any of the RSs and in this way one recovers all the possible 
branches of $\widetilde{\widetilde{E}}(z)$. From the specification 
presented in Eq.~(\ref{e5}), it is readily observed that, to obtain 
the physically most relevant complex-valued solutions of Eq.~(\ref{e14}), 
one has to analytically continue $\widetilde{E}_{\ell}(z)$ from the first 
(third) quadrant of the physical RS into the fourth (second) quadrant of 
a non-physical RS when ${\rm Re}(z) >\mu$ (${\rm Re}(z) < \mu$). We note 
that solution $z_0$ of $\widetilde{\widetilde{E}}_{\ell}(z)=z$ with 
${\rm Re}(z_0) > \mu$ corresponds to a QP that propagates and attenuates 
forwards in time (i.e. it concerns a `particle-like' QP) and that with 
${\rm Re}(z_0) < \mu$ to one that propagates and attenuates backwards in 
time (a `hole-like' QP).

\section{Discussion of the Luttinger theorem}

We are now in a position to deal with the central objective of the 
present work. In doing so we distinguish three major steps in the proof 
by Luttinger (1961). We shall separately describe each of these steps 
and, when necessary, comment on the details. We conclude that Luttinger's
theorem does {\sl not} exclude the existence of non-Fermi-liquid metallic
states for systems in spatial dimensions larger than unity. Rather, it 
merely establishes that Landau's Fermi-liquid theory for metals is 
consistent with the principles of the many-body theory of interacting 
systems.

\subsection{Preliminaries}

Before proceeding with the proof of the Luttinger (1961) theorem, several 
remarks are in order.

First, our considerations in \S~II have made evident that $\mu_N$ and 
$\mu_{N+1}$ are two branch points of $\widetilde{G}(z)$ and
$\widetilde{\Sigma}(z)$ and consequently of $\widetilde{E}_{\ell}(z)$, 
for all $\ell$. For simplicity of notation, let $\widetilde{f}(z)$ denote 
any of these functions of $z$. The fact that $\mu_N$ and $\mu_{N+1}$ are 
branch points of $\widetilde{f}(z)$ implies that such expansion as 
${\rm Im}\Sigma(k;\varepsilon)\sim\mp \alpha_{k} (\varepsilon 
-\varepsilon_F)^2$, $\varepsilon {\mathstrut_{\displaystyle {<} 
}^{\displaystyle >}} \varepsilon_F$, {\sl cannot} correspond to the 
leading-order term in a Taylor expansion of $\widetilde{\Sigma}(k;z)$ 
when $z=\varepsilon_F$ is identified with either $\mu_N$ or $\mu_{N+1}$.
Rather, it corresponds to the leading-order term in an asymptotic 
expansion (Whittaker and Watson 1927, Ch. VIII, Copson 1965, Dingle 1973, 
Lauwerier 1977) of ${\rm Im}\Sigma(k;\varepsilon)$ for $\varepsilon\to
\varepsilon_F$ (see \S~A.2). 

In view of the fact that for the uniform-electron system under 
consideration, the interval $(\mu_N,\mu_{N+1})$ is infinitesimally small, 
it is tempting to identify $\varepsilon_F$, $\mu_N$ and $\mu_{N+1}$, 
as is commonly done in the literature. Such an identification is not 
justified. Our discussion of a Migdal's (1957) theorem in \S~V further 
clarifies this statement. We note that, strictly speaking, $\varepsilon_F$ 
is identical with $\mu_N$. Thus the precise statement of the Luttinger 
theorem under consideration is as follows.
\footnote{\label{f11}
\small
In the work by Luttinger (1961), there is no explicit mention of the 
value or range of values of $k$ for which ${\rm Im}\Sigma(k;\varepsilon)
\sim\mp\alpha_k (\varepsilon-\varepsilon_F)^2$, $\varepsilon 
{\mathstrut_{\displaystyle {<} }^{\displaystyle >}} \varepsilon_F$,
is valid. This is, however, implicit in Eqs.~(59)-(61) of Luttinger's 
work: Eq.~(61) implies $u$ to be small in magnitude ($u{:=}(\mu-x)$; in 
our notation, $u{:=} (\varepsilon_F-\varepsilon)$), so that by Eq.~(60) 
one deduces $t_1$, $t_2$ and $t_3$ to be small, requiring by Eq.~(59) 
that $k_1, k_2, k_3\approx k_F$. From ${\bf k}_3 = {\bf k}_1 +{\bf k}_2 - 
{\bf k}$, using a geometrical construction, it is easily deduced that 
$0 \leq k {\mathstrut_{\displaystyle {\sim}}^{\displaystyle <}} 3 k_F$.} 
${\rm Im}\Sigma(k;\varepsilon)\equiv 0$ for $\varepsilon\in (\mu_N,
\mu_{N+1})$; ${\rm Im}\widetilde{\Sigma}(k;\varepsilon+i\eta)$ $\equiv 
{\rm Im}\Sigma(k;\varepsilon)$ $\sim -\alpha_k (\varepsilon -\mu_{N+1})^2$ 
for $\varepsilon\geq \mu_{N+1}$ and ${\rm Im}\widetilde{\Sigma}
(k;\varepsilon-i\eta)$ $\equiv {\rm Im}\Sigma(k;\varepsilon)$ $\sim 
\alpha_k (\varepsilon -\mu_N)^2$ for $\varepsilon\leq \mu_N$.

Second, the above asymptotic expansions are not {\sl uniform} (see \S~A.2), 
as is evident from the sign of ${\rm Im}\Sigma(k;z)$ which depends on the 
sign of ${\rm Im}(z)$ (or on $\arg(z)$). Thus $z=\mu_N$ and $z=\mu_{N+1}$ 
are indeed branch points.

Third, analysis of the next-to-leading order terms in the asymptotic 
expansions of $\widetilde{\Sigma}(k;z)$ around $z=\mu_N$ and $z=
\mu_{N+1}$ reveal that these terms involve the logarithm function, 
implying that both $\mu_N$ and $\mu_{N+1}$ are branch points of {\sl 
infinite} order, that is to say $\widetilde{\widetilde{f}}(z)$ (see \S~A.3) 
consists of infinitely many branches. Already the fact that $\mu_N$ and 
$\mu_{N+1}$ are not regular singularities of $\widetilde{f}(z)$ suggests
that care should be exercised in defining the QP weights, in particular 
$Z_{k_F}$ which features in the Migdal (1957) theorem and which determines 
the value of the discontinuity in the momentum distribution functions 
at the Fermi momentum $\hbar k_F$. We shall return to this point in \S~V.

In spite of our first remark above, for conformity with the existing 
texts in what follows we shall use the conventional notation (such as 
that in our Abstract) and proceed with the demonstration of the implicit 
assumption involved in the proof of the Luttinger (1961) theorem. To this 
end, we first summarise the main aspects of Luttinger's (1961) proof. 
This proof is based on the assumption of validity of the many-body 
perturbation theory and further rests on the analyses of the terms in an 
{\sl implicit} perturbation expansion of the self-energy operator; the 
{\sl implicit} nature of this expansion is associated with the fact that 
contributions in this are in terms of the single-particle Green 
function pertaining to the {\sl interacting} system which is, in turn, a 
functional of the self-energy operator itself. Diagrammatically, the 
terms in this perturbation series are represented by the proper {\sl 
skeleton} self-energy diagrams (Luttinger and Ward 1960), that is those 
connected self-energy diagrams that do not accommodate any simpler
self-energy parts in them. 

In the proof by Luttinger (1961), the skeleton diagrams are in terms of 
the isotropic bare electron-electron interaction function $v({\bf r}
-{\bf r}')$ and this, as we shall discuss in \S~IV.D, implies that 
Luttinger's approach is {\sl not} directly applicable to cases where 
$v({\bf r}-{\bf r}')$ is the long-range Coulomb interaction function. 

Since the many-body perturbation theory underlies the proof by Luttinger
(1961), the Luttinger theorem is valid insofar as this perturbation theory 
is valid. This aspect leads to the natural conclusion that metallic 
non-Fermi-liquid systems in spatial dimensions $d$ larger than unity
(see footnote \ref{f15} further on) fall outside the domain of 
applicability of the many-body perturbation theory, or that this theory 
should break down when applied to such systems. As has been emphasised 
earlier (Anderson 1993), validity or otherwise of a many-body perturbation 
series cannot be decided solely on the basis of its convergence or 
divergence, respectively. Despite convergence, the calculated limit may 
be unphysical.
\footnote{\label{f12}
\small
The following example due to Simon (1970) should be clarifying: Consider 
the hydrogen-like Hamiltonian in three dimensions (here we use the Hartree
atomic units): ${\cal H} {:=} -\nabla^2/2 - \lambda/r$, $r\geq 0$. The 
discrete energies of this Hamiltonian for $\lambda > 0$ are ${\cal 
E}_n(\lambda) = -\lambda^2/[2 n^2]$, $n=1,2,\dots$. For $\lambda\leq 0$ 
there are no discrete levels (no bound states). However, ${\cal 
E}_n(\lambda)$ being a finite-order polynomial (technically, an {\sl 
entire} function) of $\lambda$, the Rayleigh-Schr\"odinger perturbation 
expansion for, say, the ground-state energy (i.e. ${\cal E}_1(\lambda)$), 
with $-\lambda/r$ playing the role of the `perturbation', yields 
{\sl exactly} $-\lambda^2/2$, irrespective of whether $\lambda > 0$ or 
$\lambda\leq 0$ and independent of the magnitude of $\vert\lambda\vert$; 
the coefficients of $\lambda^m$, for $m=0,1$ and $m > 2$, in this 
expansion are identically vanishing. Evidently, the convergence of this 
perturbation series for $\lambda \leq 0$ is a `bogus' convergence.}

\subsection{Explicit analysis of the Luttinger theorem}

After indicating that the perturbation contributions in the 
above-indicated expansion for $\Sigma(k;\varepsilon)$ in terms of the 
skeleton self-energy diagrams below second order in the electron-electron 
interaction $v$ are $\varepsilon$-independent, Luttinger (1961) considered
in detail the second-order self-energy diagram that we have depicted 
in Fig.~1 (for facilitating direct comparison with the results in the 
work by Luttinger, below we adopt the notation that he employed as 
closely as possible). In doing so, Luttinger first evaluated the 
imaginary part of the self-energy operator corresponding to this diagram, 
with the single-particle Green function pertaining to the {\sl interacting} 
system, that is $G(\varepsilon)$ (represented by full line in Fig.~1), 
replaced by that pertaining to the {\sl non-interacting} system, i.e. 
$G_0(\varepsilon)$.
\footnote{\label{f13}
\small
Explicitly, we have $G_0(k;\varepsilon) \equiv\hbar\big\{\Theta(k_F - k)
/(\varepsilon -\varepsilon_k^0-i\eta)+\Theta(k - k_F)/(\varepsilon 
-\varepsilon_k^0 + i\eta)\big\}$, $\eta\downarrow 0$; $\varepsilon_k^0$ 
is defined in Eq.~(\ref{e15}).} From the well-known rules for transcribing 
Feynman diagrams into mathematical expressions in the momentum 
representation (Fetter and Walecka 1971, pp. 100-105), it is readily seen 
that the expression corresponding to the diagram in Fig.~1 involves three 
wave-vector integrations, over ${\bf k}_1$, ${\bf k}_2$ and ${\bf k}_3$. 
Following the fact that for the matrix elements of the electron-electron 
interaction, $({\bf k},{\bf k}_3\vert v\vert {\bf k}_1,{\bf k}_2)$ 
$= \overline{v}({\bf k}-{\bf k}_1) \delta_{{\bf k} -{\bf k}_1,{\bf k}_2
-{\bf k}_3}$ holds, where ${\bf k}$ stands for the {\sl external} 
wave-vector (i.e. the wave-vector associated with $k {:=}\|{\bf k}\|$ 
in the argument of $\Sigma(k;\varepsilon)$), it would be tempting to 
eliminate one of the wave-vector integrals. This elimination would make 
the subsequent algebra extremely cumbersome. Rather, Luttinger (1961) 
transforms the three wave-vector integrals into three one-dimensional 
integrals over energies $\varepsilon_{k_1}^0$, $\varepsilon_{k_2}^0$ 
and $\varepsilon_{k_3}^0$ of the non-interacting electrons, where 
\begin{equation}
\label{e15}
\varepsilon_k^0 {:=} {\hbar^2\over 2 m_e} k^2, 
\end{equation}
with $m_e$ the bare-electron mass. 
\footnote{\label{f14}
\small
The approach by Luttinger (1961) aims at obtaining the leading-order term 
in $(\varepsilon -\varepsilon_F)$ (in Luttinger's notation, $u$), so that 
other integrals which merely determine numerical pre-factors are not 
explicitly dealt with by him. For clarity, consider a three-dimensional 
non-interacting system with $\varepsilon_k^0$ the one-particle energy
dispersion presented in Eq.~(\ref{e15}). In the spherical-polar coordinate 
system $(k,\phi,\theta)$ we have ${\rm d}^3k = k^2 \sin(\theta)\, 
{\rm d}k\, {\rm d}\phi\, {\rm d}\theta$ $\equiv {1\over 2} 
(2 m_e/\hbar^2)^{3/2} \sqrt{\varepsilon_k^0} \sin(\theta)\, 
{\rm d}\varepsilon_k^0\, {\rm d}\phi\, {\rm d}\theta$. In the treatment 
by Luttinger (1961), ${\rm d}^3k \propto {\rm d}\varepsilon_k^0$ plays 
an all-important role.} 

A most crucial ingredient in Luttinger's proof consists of the recognition 
that, although the momentum conservation, enforced by the above $\delta_{
{\bf k}-{\bf k}_1, {\bf k}_2-{\bf k}_3}$, uniquely determines one of the 
three wave-vectors in terms of the other two, say ${\bf k}_3$ in terms of 
${\bf k}_1$ and ${\bf k}_2$, nonetheless, barring one-dimensional systems
(Luttinger 1961, footnote 5), $\varepsilon_{k_1}^0$, $\varepsilon_{k_2}^0$ 
and $\varepsilon_{k_3}^0$ can be independently and continuously varied 
over some {\sl finite} range of values.
\footnote{\label{f15}
\small
This is the instance where one-dimensional interacting systems are singled 
out as being in some fundamental way different from higher-dimensional 
interacting systems. The Luttinger model, which turns out to be akin to
a large class of one-dimensional models (Haldane 1980, 1981), is however
introduced in 1963 by Luttinger (1963) and was first correctly treated 
by Mattis and Lieb (1965). An earlier model to which the Luttinger model 
is similar, is due to Tomonaga (1950). For three different treatments 
of the Tomonaga-Luttinger models see Dzyaloshinski\v{i} and Larkin (1973), 
Luther and Peschel (1974), and Everts and Schulz (1974). For reviews 
concerning one-dimensional systems of interacting fermions see S\'olyom 
(1979), Mahan (1981, \S~4.4), Voit (1994), Sch\"onhammer (1997) and
Schulz, Cuniberti and Pieri (1998). }
This can be seen as follows: using the above dispersion for 
$\varepsilon_k^0$ as well as ${\bf k}_3 = {\bf k}_1 + {\bf k}_2 - 
{\bf k}$, it is readily seen that
\begin{equation}
\label{e16}
\varepsilon_{k_3}^0 = \varepsilon_k^0 + \varepsilon_{k_1}^0 
+ \varepsilon_{k_2}^0 + {\hbar^2\over m_e}
\big\{ {\bf k}_1\cdot {\bf k}_2 - {\bf k}\cdot {\bf k}_1
- {\bf k}\cdot {\bf k}_2\big\}.
\end{equation}
For fixed values of $\varepsilon_{k_1}^0$ and $\varepsilon_{k_2}^0$, and 
in spite of ${\bf k}_3 = {\bf k}_1 + {\bf k}_2 - {\bf k}$, for systems 
in spatial dimensions $d > 1$, the {\sl orientational} freedom of vectors 
${\bf k}_1$ and ${\bf k}_2$ with respect to each other as well as ${\bf k}$ 
allows for {\sl continuous} finite variations in $\varepsilon_{k_3}^0$. A 
crucial element in our arguments that follow, is derived from the fact 
that freedom in the independent variations of $\varepsilon_{k_3}^0$ is a 
consequence of the expression in Eq.~(\ref{e16}), which, in turn, follows 
from the {\sl specific} form of the energy dispersion pertaining to {\sl 
non-interacting} electrons in Eq.~(\ref{e15}). As a matter of course the 
latter form is {\sl not} unique in that one can construct energy dispersions 
$\varepsilon_k^0$ different from that in Eq.~(\ref{e15}) that equally give 
rise to the possibility of independent and continuous variations of 
$\varepsilon_{k_j}^0$, $j=1,2,3$, over some non-vanishing region. At the 
same time, it is not difficult to put forward energy dispersions that do 
{\sl not} allow for such an independent continuous variation over a finite 
domain. We shall return to this point in \S~IV.C.

It is from the observation with regard to the possibility of independent 
variations of the three energies $\varepsilon_{k_j}^0$, $j=1,2,3$, that 
Luttinger (1961) has established the leading-order contribution to 
${\rm Im} \Sigma(k;\varepsilon)$ of the second-order diagram in Fig.~1 
to be proportional to $(\varepsilon -\varepsilon_F^0)^2$ as $\varepsilon
\to\varepsilon_F^0$. 
\footnote{
\small
We note in passing that `${u^2}$' in Eq.~(60) of Luttinger (1960) is in 
error; the correct pre-factor is unity. }

The second step and, from the viewpoint of our present work, the most 
crucial step in Luttinger's (1961) proof consists in establishing that, 
upon employing $G(\varepsilon)$ rather than the $G_0(\varepsilon)$ of the 
first step, the leading-order contribution to ${\rm Im}\Sigma(k;\varepsilon)$ 
pertaining to the second-order self-energy diagram in Fig.~1 is also 
proportional to $(\varepsilon -\varepsilon_F)^2$, for $\varepsilon\to
\varepsilon_F$. This conclusion is of fundamental importance to the third 
part in Luttinger's proof, where, after having established the latter 
property, namely that the leading-order contribution to ${\rm Im}\Sigma(k;
\varepsilon)$, as $\varepsilon\to\varepsilon_F$, is unaffected by 
evaluating the contributions of the {\sl skeleton} self-energy diagrams 
in terms of $G_0(\varepsilon)$ rather than $G(\varepsilon)$, Luttinger 
demonstrated that to {\sl all} orders in the perturbation theory 
${\rm Im}\Sigma(k;\varepsilon) {\sim'} (\varepsilon-\varepsilon_F)^2$ 
or explicitly, that any skeleton self-energy diagram (in terms of 
$G_0(\varepsilon)$) of second and higher order in the {\sl bare} 
electron-electron interaction $v$ has a contribution to ${\rm Im}
\Sigma(k;\varepsilon)$ proportional to $(\varepsilon-\varepsilon_F^0)^{2m}$, 
where $m\geq 1$, for $\varepsilon \to\varepsilon_F^0$.
\footnote{\label{f15a}
\small
It is known (Hodges, Smith and Wilkins 1971, Bloom 1975, Fujimoto 1990, 
Fukuyama, Narikiyo and Hasegawa 1991, Fukuyama, Hasegawa and Narikiyo 
1991) that, in isotropic two-dimensional interacting systems, ${\rm Im}
\Sigma(k=k_F;\varepsilon) {\sim'} (\varepsilon-\varepsilon_F)^2
\ln\vert\varepsilon-\varepsilon_F\vert$ as $\varepsilon\to \varepsilon_F$. 
Since the leading $\varepsilon$-dependent contribution to ${\rm Re}
\Sigma(k=k_F;\varepsilon)$ associated with this ${\rm Im}\Sigma(k=k_F;
\varepsilon)$ is proportional to $(\varepsilon-\varepsilon_F)$ (see 
Appendix C), we have that $\Sigma(k_F;\varepsilon)$ is a continuously
differentiable function of $\varepsilon$ in a neighbourhood of $\varepsilon
=\varepsilon_F$. In view of our statements in \S~I (see also \S~IV.C), 
the above logarithmic contribution does {\sl not} turn the system into a 
non-Fermi liquid. }

In this second part of his proof, Luttinger (1961) proceeded as follows. 
On evaluating the contribution of the diagram in Fig.~1 to ${\rm Im}
\Sigma(k;\varepsilon)$ in terms of $G(\varepsilon)$, he made a 
simplifying approximation, which, roughly speaking, corresponds to 
replacing a Lorentzian by a Dirac $\delta$-function. 
\footnote{\label{f16}
\small
This approximation neglects the finite lifetimes of the QPs. From ${\rm 
Im}\Sigma(k;\varepsilon_F)\equiv 0$ (note in passing that this together 
with $\varepsilon_k^0 + \hbar\Sigma(k;\varepsilon_F)=\varepsilon_F$ define 
the Fermi surface --- see footnote \ref{f21} below) and the requirement of 
{\sl continuity} of ${\rm Im} \Sigma(k;\varepsilon)$ for $\varepsilon\to
\varepsilon_F$, we have ${\rm Im}\Sigma(k;\varepsilon)\to 0$ for 
$\varepsilon \to\varepsilon_F$. Consequently, for $\varepsilon\to 
\varepsilon_F$ life-time effects are {\sl not} of relevance to the {\sl 
leading}-order contribution to ${\rm Im}\Sigma(k;\varepsilon)$ as 
$\varepsilon \to\varepsilon_F$. The mentioned `Lorentzian' has its origin 
in the finite lifetimes of the QPs. For a summary of various ways in which 
${\rm Im}\Sigma(k;\varepsilon)$ can vanish as $\varepsilon\to\varepsilon_F$, 
see the text following Eq.~(\ref{e30}) in \S~V. }
Upon this, Luttinger arrived at an expression which has the same formal 
structure as the one discussed above. Similar to the second-order 
expression for ${\rm Im}\Sigma(k;\varepsilon)$ in terms of $G_0$, the 
simplified expression in terms of $G$ involves three wave-vector integrals 
over ${\bf k}_1$, ${\bf k}_2$ and ${\bf k}_3$. Now, at this place, Luttinger 
took a step which, as will become evident below, implicitly contains the 
property to be proven. Here Luttinger transforms the three wave-vector 
integrals into three energy integrals, the energy dispersions being 
$\varepsilon_{k_1}$, $\varepsilon_{k_2}$ and $\varepsilon_{k_3}$, to be
compared with $\varepsilon_{k_1}^0$, $\varepsilon_{k_2}^0$ and 
$\varepsilon_{k_3}^0$ considered above (see \S~IV.C). Note that the 
dispersions $\varepsilon_{k_j}$, $j=1,2,3$, correspond to the interacting 
(i.e. true) QPs and satisfy Eq.~(\ref{e13}) or Eq.~(\ref{e17}) below.
\footnote{\label{f17}
\small
In Luttinger's (1961) work there is no symbolic distinction between, 
for example, what we have denoted as $\widetilde{\Sigma}(k;z)$ and 
$\Sigma(k;z)$. Further, Luttinger does not discuss whether Eq.~(\ref{e17}) 
can be exactly satisfied or not (for details see \S~IV.C).}
According to Luttinger, {\sl all} the necessary mathematical steps that 
are to be taken in order to obtain, from the last-indicated integral, the 
leading-order contribution to ${\rm Im}\Sigma(k;\varepsilon)$, as 
$\varepsilon\to\varepsilon_F$, are identical with those taken in dealing 
with ${\rm Im}\Sigma(k;\varepsilon)$ evaluated in terms of $G_0$. {\sl 
This is an unjustified statement.} The dependence on $k$ of $\varepsilon_k$ 
being unknown, there is no {\sl a priori} reason why $\varepsilon_{k_3}$ 
can be varied continuously and independently of $\varepsilon_{k_1}$ and 
$\varepsilon_{k_2}$ over a finite range. 
\footnote{
\small
The contribution of the diagram in Fig.~1 to ${\rm Im}\Sigma(k;
\varepsilon)$ for one-dimensional metals, according to Luttinger (1961,
p.~946, footnote~5) amounts to ${\sim'} (\varepsilon-\varepsilon_F)$, 
which we know to be characteristic of marginal-Fermi liquids (see 
Appendix~C) rather than one-dimensional Luttinger liquids (see Appendix~D) 
over a physically-relevant range of values for the anomalous dimension 
$\alpha\equiv 2\gamma_0$, that is $0 < \alpha < 1$. At the root of this 
incorrect inference by Luttinger (1961) lies the fact that in 
one-dimensional metals, $\varepsilon_k$, contrary to Luttinger's {\sl 
implicit} assumption, is {\sl not} a continuously-differentiable function 
of $k$ and therefore ${\rm d}^1k \propto {\rm d}\varepsilon_k$ does {\sl 
not} apply, while ${\rm d}^1k \propto {\rm d}\varepsilon_k^0$ is valid. 
See \S~IV.C. }
If the effect of electron-electron interaction on the dispersion of the QP 
energies close to the Fermi surface were only to re-normalise the QP mass, 
then of course for $k$ in a neighbourhood of $k=k_F$ one invariably had 
$\varepsilon_k \sim\hbar^2 k^2/(2 m_e^*)$, with $m_e^*$ an electron's 
renormalised mass (Pines and Nozi\`eres 1966, Abrikosov, Gorkov and 
Dzyaloshinski 1975). In such an event, for $k$ in a neighbourhood of $k
=k_F$, Eq.~(\ref{e15}) would hold for $\varepsilon_k$ with $m_e$ replaced 
by $m_e^*$, and thus the just-mentioned assertion by Luttinger (1961) were 
correct. However, as we shall demonstrate in \S~IV.C, $\varepsilon_k \sim 
\hbar^2 k^2/(2 m_e^*)$ can only apply if $\Sigma(k;\varepsilon)$ is of the 
Fermi-liquid type, thus establishing that Luttinger's proof amounts to a 
demonstration of consistency of the Fermi-liquid state with the many-body 
theory of interacting metals. This demonstration does not rule out 
existence of states different from the Fermi-liquid state.

\subsection{On the dispersion of the QP energies}

Here we analyse some general features of the QP energies pertinent to a 
uniform-electron system. 

Consider Eq.~(\ref{e13}) specialised to a uniform system. Any possible
{\sl real}-valued QP energy $\varepsilon_k$ is a solution (which for an 
arbitrary $k$ may not exist; see \S~V) of the following equation (see 
also Eq.~(\ref{e25}) below)
\begin{equation}
\label{e17}
\varepsilon_k = \varepsilon_k^0 + \hbar\Sigma(k;\varepsilon_k).
\end{equation}
The fact that $\widetilde{\Sigma}(k;z)$ has a branch point at $z=\mu_N
\equiv\varepsilon_F {:=}\varepsilon_{k_F}$ (\S~IV.A), implies that the 
following equation is well satisfied by the {\sl real}-valued energy 
$\varepsilon_F$ (below $\varepsilon_F^0 {:=} \varepsilon_{k_F}^0$):
\footnote{\label{f18}
\small
Isotropy of the system under consideration implies $k_F$ to be
independent of the electron-electron interaction. We note in
passing that this is a corollary to another celebrated theorem
due to Luttinger (1960); see also Luttinger and Ward (1960). }
\begin{equation}
\label{e18}
\varepsilon_F = \varepsilon_F^0 + \hbar\Sigma(k_F;\varepsilon_F).
\end{equation}
{\sl Assuming} $\Sigma(k;\varepsilon)$ to be a continuously differentiable
function of $k$ and $\varepsilon$ in some neighbourhoods of $k=k_F$ and 
$\varepsilon=\varepsilon_F$, respectively, from Eq.~(\ref{e17}) together 
with the manifest differentiability of $\varepsilon_k^0$ in Eq.~(\ref{e15}) 
with respect to $k$ it follows that $\varepsilon_k$ is similarly a 
continuously differentiable function of $k$ in a neighbourhood of $k=k_F$. 
Consequently,
\footnote{\label{f19}
\small
As indicated earlier in this work, $o$ in $f(x) = o\big(g(x)\big)$ 
signifies $f(x)/g(x)\to 0$ for $x\to 0$.}
\begin{equation}
\label{e19}
\varepsilon_k = \varepsilon_F + \hbar v_F (k - k_F) + o\big(k-k_F\big),
\end{equation}
where
\begin{equation}
\label{e20}
v_F {:=} {1\over \hbar}\left. {{\partial\varepsilon_k}\over
\partial k}\right|_{k=k_F} \equiv {{\hbar k_F}\over m_e^*}
\end{equation}
stands for the Fermi velocity. From Eq.~(\ref{e17}) it can be shown that
\begin{equation}
\label{e21}
\left. {{\partial \varepsilon_k}\over 
\partial k}\right|_{k=k_F}
= Z_{k_F} \left. {{\partial\{\varepsilon_k^0 
+\hbar\Sigma(k;\varepsilon_F)\}} \over\partial k}\right|_{k=k_F},
\end{equation}
where
\begin{equation}
\label{e22}
Z_{k_F} {:=} \Big( 1 -\hbar\left. 
{{\partial \Sigma(k_F;\varepsilon)}\over\partial\varepsilon}
\right|_{\varepsilon=\varepsilon_F}\Big)^{-1}
\end{equation}
is, according to a Migdal's (1957) theorem (see \S~V), exactly the amount 
of discontinuity in the momentum distribution function ${\sf n}(p)$ at the 
Fermi momentum $p_F \equiv \hbar k_F$. Eqs.~(\ref{e19}) and (\ref{e20}) 
are readily seen to imply that for $k\to k_F$, $\varepsilon_k \sim \hbar^2 
k^2/(2 m_e^*)$. The above steps make explicit how the latter dispersion 
relation is rooted in the assumption of continuous differentiability 
of $\Sigma(k;\varepsilon_k)$ in a neighbourhood of $k = k_F$. For Fermi 
liquids $0 < Z_{k_F} \leq 1$ holds,
\footnote{
\small
Recall that $0 < Z_{k_F} \leq 1$ although necessary, is {\sl not}
sufficient for rendering a metallic system a Fermi liquid (see \S~I, 
conditions (A) and (B)).}
where $Z_{k_F}=1$ corresponds to the case of non-interacting QPs. The 
condition $Z_{k_F}=0$, which is a signature of marginal- and
Luttinger-liquid systems, corresponds, according to Eq.~(\ref{e22}), to 
the case where $\partial\Sigma(k_F;\varepsilon)/\partial\varepsilon$ is 
unbounded at $\varepsilon=\varepsilon_F$. In cases where $Z_{k_F}$ is 
vanishing, Eq.~(\ref{e21}) {\sl formally} implies that, for the Fermi 
velocity $v_F$ to be finite, $\partial\Sigma(k;\varepsilon_F)/\partial k$ 
is also to be unbounded at $k=k_F$. Although {\sl formally} Eq.~(\ref{e19}) 
can be maintained with $v_F = -\hbar^{-1}\partial\Sigma(k;\varepsilon_F)/
\partial k\vert_{k=k_F}/\partial\Sigma(k_F;\varepsilon)/\partial\varepsilon 
\vert_{\varepsilon=\varepsilon_F}$, it should be evident, however, that in 
the case at hand the right-hand side of Eq.~(\ref{e19}) neglects the fact 
that $\Sigma(k;\varepsilon)$ is singular at both $k=k_F$ and $\varepsilon 
=\varepsilon_F$ with the singularities showing up already in the first 
derivatives, thus invalidating this formal definition of the Fermi 
velocity. To appreciate the significance of this observation, note that 
the key element in obtaining Eq.~(\ref{e19}) has been the {\sl assumption} 
of continuous differentiability of $\Sigma(k;\varepsilon)$ in some 
neighbourhoods of $k=k_F$ {\sl and} $\varepsilon=\varepsilon_F$, whereby 
it has been possible to write
\footnote{\label{f20}
\small
Compare with Eq.~(\ref{e19}). See also footnotes \ref{f2} and \ref{f19}. }
$\Sigma(k;\varepsilon_F)\sim\Sigma(k_F;\varepsilon_F) +\beta (k - k_F) + 
o(k - k_F)$ {\sl and} $\Sigma(k_F;\varepsilon) \sim \Sigma(k_F;
\varepsilon_F) +\gamma (\varepsilon - \varepsilon_F) + o(\varepsilon
-\varepsilon_F)$, with $\beta$ and $\gamma$ {\sl bounded}, in some 
neighbourhoods of $k=k_F$ and $\varepsilon =\varepsilon_F$, respectively. 
These asymptotic relations signify the facts that $\{\Sigma(k;\varepsilon_F) 
-\Sigma(k_F;\varepsilon_F)\}/(k-k_F)$ and $\{\Sigma(k_F;\varepsilon) 
-\Sigma(k_F;\varepsilon_F)\}/(\varepsilon -\varepsilon_F)$ are {\sl 
vanishing} for $k\to k_F$ and $\varepsilon\to \varepsilon_F$, respectively. 
With, for instance, $\Sigma(k;\varepsilon_F)\sim \Sigma(k_F;\varepsilon_F) 
+\beta \vert k - k_F\vert^{\gamma_0}$, $0<\gamma_0 < 1$ (in the present 
case $\alpha {:=} 2\gamma_0$ is referred to as the {\sl anomalous dimension}; 
see Appendix~D, in particular paragraph following Eq.~(\ref{ed30})), for 
$k\to k_F$, or $\Sigma(k_F;\varepsilon) \sim \Sigma(k_F;\varepsilon_F) 
+\gamma (\varepsilon-\varepsilon_F) \ln\vert\varepsilon -\varepsilon_F\vert$, 
for $\varepsilon\to \varepsilon_F$, which have {\sl not} been accounted for 
by Eq.~(\ref{e19}), $\{\Sigma(k;\varepsilon_F) -\Sigma(k_F;\varepsilon_F)\}/
(k-k_F)$ and $\{\Sigma(k_F;\varepsilon) -\Sigma(k_F;\varepsilon_F)\}/
(\varepsilon -\varepsilon_F)$ clearly diverge for $k\to k_F$ and 
$\varepsilon\to \varepsilon_F$, respectively, and consequently both 
Eq.~(\ref{e19}) and the notion of Fermi velocity of QPs in the vicinity 
of the Fermi surface become altogether meaningless.
\footnote{\label{f21}
\small
We note in passing that for metallic systems the locus of the 
${\bf k}$-points satisfying $\varepsilon_k^0 + \hbar\Sigma(k;
\varepsilon_F) = \varepsilon_F$ (compare with Eq.~(\ref{e17})) is the 
Fermi surface, with $\varepsilon_k^0 + \hbar\Sigma(k;\varepsilon_F) 
< \varepsilon_F$ ($> \varepsilon_F$) defining the interior (exterior) 
of the Fermi sea (Galitskii and Migdal 1958, Luttinger 1960, Eqs.~(6) 
and (94)). Clearly, the existence of a Fermi surface is {\sl in}dependent 
of the value of $Z_{k_F}$, as the latter is determined by the {\sl 
derivative} with respect to $\varepsilon$ of $\Sigma(k_F;\varepsilon)$; 
see Eq.~(\ref{e22}). In particular $Z_{k_F}=0$ does {\sl not} rule out a 
Fermi surface. We point out that ${\rm Im}\Sigma(k;\varepsilon_F) \equiv 
0$, for {\sl all} $k$ (see footnote \ref{f16} above), and that in an 
isotropic system the Fermi surface can consist of disconnected concentric 
surfaces; since $\varepsilon_k^0$ is a monotonically increasing function 
of $k$, such Fermi surfaces can exist (in isotropic systems) {\sl only} 
in consequence of the electron-electron interaction: for $\varepsilon_k^0 
+ \hbar\Sigma(k;\varepsilon_F) = \varepsilon_F$ to have more than one 
solution, it is necessary that $\hbar\Sigma(k;\varepsilon_F)$ can counter 
$\varepsilon_k^0$. In the present work we explicitly deal with a single 
Fermi surface of radius $k_F$. }

The following two observations should be clarifying: first, any positive 
{\sl non-integer} $\gamma_0$ in the above expression would signal a 
non-analytic singularity in the $k$-dependence of $\Sigma(k;\varepsilon)$, 
even when $\gamma_0 > 1$; second, even if one puts aside the possibility
of non-differentiability with respect to $k$ of $\Sigma(k;\varepsilon)$
in a neighbourhood of $k=k_F$, it is certain that $\widetilde{\Sigma}
(k;z)$ possesses an essential singularity at $z=\varepsilon_F\equiv 
\mu_N$. For instance, Eq.~(\ref{e19}) with $v_F$ defined in accordance 
with Eqs.~(\ref{e20}) and (\ref{e21}) takes {\sl no} account of the 
possibility of, for example, ${\rm Im}\Sigma(k;\varepsilon) {\sim'} 
(\varepsilon -\varepsilon_F)$ whose corresponding ${\rm Re}\Sigma(k;
\varepsilon)-\Sigma(k;\varepsilon_F) {\sim'} (\varepsilon -\varepsilon_F) 
\ln\vert\varepsilon-\varepsilon_F\vert$ as $\varepsilon\to \varepsilon_F$ 
(see text following Eq.~(\ref{e30}) below as well as Appendix~C) which 
is specific to `marginal' Fermi liquids (Varma, {\sl et al.} 1989, 
Littlewood and Varma 1991, Kotliar, {\sl et al.} 1991).

Having demonstrated the difficulties that arise as a consequence of 
the fact that $\Sigma(k;\varepsilon)$ (and, owing to Eq.~(\ref{e17}), 
$\varepsilon_k$) is not {\sl a priori} `smooth' (in our above-indicated 
sense), it becomes evident that the implicit assumption in the proof 
of the Luttinger (1961) theorem, as though $\Sigma(k;\varepsilon)$ were 
invariably `smooth', has {\sl no} theoretical justification. In particular, 
we should like to emphasise that our above exposition makes evident that 
no theoretical treatments that aim to expose the behaviour of 
$\Sigma(k;\varepsilon)$, for $\varepsilon \to\varepsilon_F$, whether 
$\Sigma(k;\varepsilon)$ is evaluated by perturbative or non-perturbative 
techniques, should rely on the expansion in Eq.~(\ref{e19}), since 
{\sl any} consistent treatment that relies on this expansion is bound 
to arriving at the conclusion that $\Sigma(k;\varepsilon)$ is
Fermi-liquid like (validity of the expansion in Eq.~(\ref{e19}) is 
necessary and sufficient for the satisfaction of conditions (A) and 
(B) in \S~I); a different conclusion must of necessity signal some 
inconsistency (or inconsistencies) in the treatment, including plain 
algebraic errors.

One of the main conclusions that may be drawn from our above analyses 
is that {\sl a non-Fermi-liquid-type $\Sigma(k;\varepsilon)$ does not 
necessarily signal breakdown of the many-body perturbation theory.} 
Consequently, we observe that {\sl non-Fermi-liquid behaviour can in 
principle be manifested in any spatial dimension higher than $d=1$}.

\subsection{A specific feature}

Here we comment on one of the specific features in the treatment by 
Luttinger (1961) which brings out some additional aspects concerning the 
above-indicated implicit assumption. The diagram in Fig.~1 has the property 
that it involves a so-called `polarisation part', that is a part which can 
be disconnected from the self-energy diagram through removing two 
interaction lines (indicated by broken lines). Momentum conservation 
implies that to both of these interaction lines the same momentum-transfer 
vector is associated. Consequently, any polarisation part in a self-energy 
diagram contributes at least a {\sl square} of the electron-electron 
interaction to the respective integrand. As a consequence, the contribution 
of the diagram in Fig.~1 is divergent when the electron-electron interaction 
is the bare Coulomb interaction (for a comprehensive treatment see Mattuck 
(1992, \S~10.4)) for which $v({\bf r}-{\bf r}') \propto 1/\|{\bf r}
-{\bf r}'\|$ holds and thus $\overline{v}(k)\propto 1/k^2$ in three spatial 
dimensions. Thus Luttinger's analysis is not {\sl directly} applicable 
to systems of particles interacting via the Coulomb interaction.
\footnote{\label{f23}
\small
Here we are referring to metallic systems. For non-metallic ground states, 
this problem of divergent self-energy contributions does not arise. Since
non-metallic systems are by definition non-Fermi liquids, they do not 
concern us here.}
However, as is well known, the contribution of the set of all divergent 
polarisation diagrams (the random-phase approximation, RPA, bubble-like 
diagrams) can be exactly calculated (as these diagrams give rise to a 
geometric series) and one observes that screening effects in the static 
limit remove the singularity of the bare Coulomb interaction function 
$\overline{v}(k)$ at $k=0$ (see, e.g., Fetter and Walecka 1971, pp. 178 
and 179). In analysing the leading-order term in ${\rm Im}\Sigma(k;
\varepsilon)$ as $\varepsilon \to\varepsilon_F$, the 
$\varepsilon$-dependence of the dynamically-screened interaction function 
(Hubbard 1957) $W(\varepsilon)$ necessitates knowledge not only of the 
energy dispersion of the QP excitations, i.e. $\varepsilon_k$, but also 
of that of the bosonic neutral excitations, i.e. ${\sf e}_k$; here for 
the energies of the `neutral' excitations we have ${\sf e}_s {:=} E_{N,s}
-E_{N,0} \geq 0$ (Fetter and Walecka 1971). For an interacting system, 
the exact dispersion of ${\sf e}_k$ is unknown, similar to that of 
$\varepsilon_k$; ${\sf e}_k$ can be in principle determined from the 
dynamical density-density correlation function $\chi(k;\varepsilon)$;
for ${\sf e}_k$ this function plays a similar role as $G(k;\varepsilon)$ 
does for $\varepsilon_k$. It follows that for metallic Coulomb systems, 
the behaviour of ${\rm Im}\Sigma(k;\varepsilon)$ {\sl cannot} be solely 
determined from the perturbation series expansion for the self-energy 
operator; one needs in addition the perturbation series expansion for 
$\chi(k;\varepsilon)$.

From the above arguments we conclude that, even if the implicit assumption 
by Luttinger (1961), discussed in \S\S~IV.B and IV.C, were correct, 
Luttinger's asymptotic expression for ${\rm Im}\Sigma(k;\varepsilon)$ as 
presented in the Abstract, would {\sl not} be justifiably applicable to 
self-energies of metallic systems of particles interacting via the 
Coulomb interaction function (see \S~I for references to studies on 
systems of particles interacting through long-range repulsive interaction
functions).

\section{A brief discussion of a Migdal's theorem}

In \S~IV we have mentioned that the quantity $Z_{k_F}$, which coincides
with weight of the Landau quasi-particles on the Fermi surface, is also 
equal to the amount of discontinuity of the momentum-distribution function 
${\sf n}(p)$ at $p_F = \hbar k_F$. This statement constitutes a theorem
due to Migdal (1957) (Luttinger 1960). Here we briefly discuss this 
theorem in light of our above considerations. In what follows we deal 
with ${\sf n}(k)$ and use the commonly-employed designation 
`momentum-distribution function', despite the fact that $k =p/\hbar$ is 
not momentum. For ${\sf n}(k)$ we have
\begin{equation}
\label{e23}
{\sf n}(k) {:=} 
\langle\Psi_{N,0}\vert \widehat{a}_k^{\dag}
\widehat{a}_k\vert\Psi_{N,0}\rangle
\equiv {1\over\hbar} \int_{\cal C}
{{{\rm d} z}\over 2\pi i}\; \widetilde{G}(k;z),
\end{equation}
where $\widehat{a}_k^{\dag}$ and $\widehat{a}_k$ stand for creation and 
annihilation operators, respectively. The contour ${\cal C}$ of integration 
is depicted in Fig.~2. To prevent confusion, we point out that the 
right-hand side of Eq.~(\ref{e23}) differs by a minus sign from the 
corresponding expression in Migdal's (1957) work. This is because 
the single-particle Green function adopted here is defined according 
to the modern convention and is equal to minus the Green function employed 
by Migdal (1957). Further, our use of the symbol $\widetilde{G}(k;z)$ (rather 
than $G(k;\varepsilon)$ which is the common notation) is in accordance with 
our present conventions (see \S~A.3) and has the advantage that our 
following discussions will not suffer from mathematical ambiguities. One 
of these ambiguities can be found in the original work by Migdal (1957). 
Migdal stated namely that, for $k$ infinitesimally less than $k_F$ 
(denoted by $k_F^-$), $\widetilde{G}(k;z)$ would have a {\sl pole}, with an 
infinitesimal imaginary part (apparently due to ${\rm Im}\Sigma(k;
\varepsilon)$ as indicated in the Abstract), enclosed by ${\cal C}$. Migdal 
(1957) further asserted that, through increasing $k$ from $k_F^-$ to 
$k_F^+$ ($k_F^+$ denotes a value infinitesimally larger than $k_F$), the 
imaginary part of the mentioned {\sl pole} would change sign, upon which 
this {\sl pole} would leave the interior of ${\cal C}$. Since the regular 
or {\sl incoherent} part of $\widetilde{G}(k;z)$ does not contribute to 
the {\sl difference} ${\sf n}(k_F^-) - {\sf n}(k_F^+)$, from Eq.~(\ref{e23}) 
it would follow that $Z_{k_F}$ would be the residue of an isolated {\sl 
pole} of $\widetilde{G}(k_F^-;z)$ at $z=\varepsilon_F$. The Migdal theorem 
(Migdal 1957, Luttinger 1960) thus states
\begin{equation}
\label{e24}
{\sf n}(k_F^-) - {\sf n}(k_F^+) = Z_{k_F}.
\end{equation}

From the Dyson equation we have
\begin{equation}
\label{e25}
\widetilde{G}(k;z) = {\hbar\over z - \widetilde{E}_k(z)},
\;\;\;\;\mbox{\rm where}\;\;\;\;
\widetilde{E}_k(z) {:=} 
\varepsilon_k^0 +\hbar\widetilde{\Sigma}(k;z).
\end{equation}
{\sl Poles} of $\widetilde{G}(k;z)$ are thus seen to be the {\sl 
isolated} (see \S~A.1) solutions of 
\begin{equation}
\label{e26}
\widetilde{E}_k(z) = z,
\end{equation}
which is equivalent to Eq.~(\ref{e13}) specialised to the uniform 
isotropic system under consideration. According to our conventions, 
$k_F^-$ has to be identified with $k_F$ and we need only to have $k_F^+$ 
for indicating a $k$ infinitesimally larger than $k_F$. As we have 
mentioned in \S~IV.C, indeed $E_{k_F}(\varepsilon_F) = \varepsilon_F$ 
must be satisfied. However, since $\widetilde{\Sigma}(k_F;z)$ has a {\sl 
branch point} at $z=\varepsilon_F$, it follows that $\widetilde{G}(k_F;z)$ 
{\sl cannot} have a pole at $z=\varepsilon_F$, since a branch point is {\sl 
not} an isolated singularity (see \S~A.1). Moreover, as we have discussed 
in \S~III, Eq.~(\ref{e13}), or its equivalent, Eq.~(\ref{e26}), {\sl 
cannot} have a complex-valued solution, so that there cannot be any 
question of a `pole' of $\widetilde{G}(k;z)$ changing the sign of its 
imaginary part. The question, therefore, arises as to how, despite the 
fact that $\widetilde{G} (k_F;z)$, or $G(k_F;\varepsilon)$ in Migdal's 
notation, has {\sl no} pole at $z=\varepsilon_F$, the above-introduced 
Migdal (1957) theorem could hold.

To answer the above question, we note that since, for $k=k_F$ and $z\to 
\varepsilon_F$, $z-\widetilde{E}_k(z)$ approaches zero (by continuity 
and the fact that Eq.~(\ref{e18}) applies), in principle it {\sl may} be 
possible to write
\begin{equation}
\label{e27}
\widetilde{G}^{-1}(k_F;z) 
\equiv {1\over\hbar}\big\{z - [\varepsilon_{k_F}^0
+\hbar\widetilde{\Sigma}(k_F;z)]\big\}
\sim \alpha (z-\varepsilon_F) + \widetilde{L}(z),\;\;\mbox{\rm as}\;\; 
z\to\varepsilon_F,
\end{equation}
with $\widetilde{L}(z)$ satisfying $\lim_{z\to\varepsilon_F} 
\widetilde{L}(z)/(z-\varepsilon_F) = 0$, that is $\widetilde{L}(z)
= o(z-\varepsilon_F)$ for $z\to\varepsilon_F$.
\footnote{\label{f24}
\small
Because of $\widetilde{G}(z^*) = \widetilde{G}^{\dag}(z)$, it follows
that $\widetilde{L}(z^*) = \widetilde{L}^*(z)$. Further, unless
$\varepsilon\in (\mu_N,\mu_{N+1})$, in general $\lim_{\eta\downarrow 0}
\{\widetilde{L}(\varepsilon+i\eta) - 
\widetilde{L}(\varepsilon-i\eta)\} \not\equiv 0$. This is a 
consequence of the fact that $z=\mu_N\equiv\varepsilon_F$
and $z=\mu_{N+1}$ are branch points of $\widetilde{G}(k;z)$,
as discussed in \S~III. }
In Eq.~(\ref{e27}), $\alpha$ stands for a constant to be specified below. 
Equation (\ref{e27}) implies $\widetilde{G}^{-1}(k_F;z)$, and thus
$\widetilde{\Sigma}(k_F;z)$, to be at least once {\sl continuously
differentiable} with respect to $z$ at $z=\varepsilon_F$ (see footnote
\ref{f2b}); $m$th continuous differentiability of $\widetilde{G}^{-1}
(k_F;z)$, and thus of $\widetilde{\Sigma}(k_F;z)$, with respect to $z$ at 
$z=\varepsilon_F$ amounts to the condition $\widetilde{G}^{-1}(k_F;z)
\sim \sum_{j=0}^m\; \alpha_j (z-\varepsilon_F)^j + \widetilde{L}(z)$
with $\{\alpha_j\}$ finite constants and $\lim_{z\to\varepsilon_F} 
\widetilde{L}(z)/(z-\varepsilon_F)^m = 0$.
\footnote{\label{f24a}
\small
We note that, when a function, say $\widetilde{f}(z)$, is analytic inside 
and on the boundary of a {\sl simply-connected} region ${\cal R}$ of the 
complex $z$-plane, it is infinitely-many times differentiable in ${\cal 
R}$, so that once-differentiability of $\widetilde{f}(z)$ at a point, say 
$z_0$, interior to ${\cal R}$ implies infinitely-many times differentiability 
of $\widetilde{f}(z)$ at $z=z_0$. In the case under consideration, since 
$z=\varepsilon_F$ is a branch point, it {\sl cannot} be interior to any 
{\sl simply-connected} region of analyticity of $\widetilde{G}(k_F;z)$ (see 
\S~A.1). Consequently, $m$th differentiability of this function at 
$z=\varepsilon_F$ does {\sl not} imply its $(m+1)$th differentiability at 
this point. Consider for instance $\widetilde{f}(z) {:=} (z-z_0)^{p}$ with 
$p > 0$ a real non-integer. This function has two branch points, $z=z_0$ 
and $1/z = 0$. With $m$ an integer satisfying $m < p < m+1$, $\widetilde{f}
(z)$ is seen to be $m$ times, but {\sl not} $m+1$ times, differentiable 
at $z=z_0$. }
Consequently, $\partial\Sigma(k_F;\varepsilon)/\partial
\varepsilon\vert_{\varepsilon=\varepsilon_F}$ is well-defined and one has 
$\alpha \equiv \hbar^{-1}\big(1 -\hbar \partial\Sigma(k_F;\varepsilon)/
\partial\varepsilon\vert_{\varepsilon=\varepsilon_F}\big) \equiv 
1/(\hbar Z_{k_F})$ (see Eq.~(\ref{e22}) above).
Using the representation $z - \varepsilon_F = \varrho \exp(i\varphi)$, 
with $\varrho$ a non-vanishing constant, to be let to approach zero, from 
Eq.~(\ref{e23}) it can straightforwardly be shown that Eq.~(\ref{e24}) 
remains intact, even though $z=\varepsilon_F$ is not a {\sl pole} 
(i.e. an isolated singularity) of $\widetilde{G}(k;z)$. 

Through the Kramers-Kronig relation (see Appendices B and C) for ${\rm 
Re}\Sigma(k;\varepsilon)$ in terms of ${\rm Im}\Sigma(k;\varepsilon)$ and 
the asymptotic relation ${\rm Im}\Sigma(k;\varepsilon) \sim \mp \alpha_k 
(\varepsilon-\varepsilon_F)^2$, for $\varepsilon {\mathstrut_{\displaystyle 
{<} }^{\displaystyle >}} \varepsilon_F$, it can readily be deduced that 
(see Appendix C)
\begin{equation}
\label{e28}
{\rm Re}\Sigma(k;\varepsilon)\sim \Sigma(k;\varepsilon_F)
+ \beta_k (\varepsilon-\varepsilon_F),\;\;\;\;
\mbox{\rm with}\;\;\;\; \beta_k \leq 0\;\; \mbox{\rm for}\;\; k\to k_F.
\end{equation}
Concerning the sign assigned to $\beta_k$ for $k$ close to $k_F$, as 
is seen from Eq.~(\ref{e30}) below, a positive $\beta_{k_F}$ would imply 
$Z_{k_F} > 1$ which is impossible on account of a combination of the 
following three reasons. 

(i) By definition (see Eq.~(\ref{e23}) above) $0\leq {\sf n}(k) \leq 1$; 
this can also be traced back to the Pauli exclusion principle. 

(ii) According to Eq.~(\ref{e24}), $Z_{k_F} = {\sf n}(k_F^-) - 
{\sf n}(k_F^+)$ so that in view of (i), $\vert Z_{k_F} \vert \leq 1$. 

(iii) $Z_{k_F} < 0$ is excluded owing to ${\rm Im}\Sigma(k;
\varepsilon) {\mathstrut_{\displaystyle {>} }^{\displaystyle <}} 0$, 
for $\varepsilon {\mathstrut_{\displaystyle {<} }^{\displaystyle >}} \mu$. 

In Appendix C (see text following Eq.~(\ref{ec9})) we demonstrate 
that $\beta_k \leq 0$, for $k$ in a neighbourhood of $k_F$, directly 
follows from a general sum rule involving ${\rm Im}\Sigma(k;\varepsilon)$. 
We therefore have
\footnote{\label{f25a}
\small
Consider the analytic function $\widetilde{f}(z) \equiv {\sf u}(x,y) 
+ i {\sf v}(x,y)$, where $x+iy \equiv z$ and ${\sf u}(x,y)$ and 
${\sf v}(x,y)$ are real. Upon setting $y$ equal to zero, we obtain 
$\widetilde{f}(x) = {\sf u}(x,0) + i {\sf v}(x,0)$. Replacing $x$ in 
the argument of $\widetilde{f}$ by $z$, we obtain $\widetilde{f}(z) = 
{\sf u}(z,0) + i {\sf v}(z,0)$ (in a similar manner one obtains 
$\widetilde{f}(z) = {\sf u}(0,-iz) + i {\sf v}(0,-iz)$). This expression 
prescribes how to obtain $\widetilde{\Sigma}(k;z)$ from the knowledge of 
$\Sigma(k;\varepsilon)$. In the present case, as $z\to\varepsilon_F$,
we have: $\widetilde{\Sigma}(k;z) \sim \Sigma(k;\varepsilon_F) +\beta_k 
(z-\varepsilon_F) - i\alpha_k (z-\varepsilon_F)^2$, for $0 \leq \arg(z) 
< \pi$, and $\widetilde{\Sigma}(k;z) \sim \Sigma(k;\varepsilon_F) 
+\beta_k (z-\varepsilon_F) + i\alpha_k (z-\varepsilon_F)^2$, for 
$-\pi \leq \arg(z) < 0$. The self-energy as specified in this way
is seen to preserve the reflection property $\widetilde{\Sigma}(k;z^*) 
= \widetilde{\Sigma}^*(k;z)$ for ${\rm Im}(z)\not=0$ (see 
Eq.~(\ref{e12})). } 
\begin{equation}
\label{e29}
\Sigma(k;\varepsilon) \sim \Sigma(k;\varepsilon_F) 
+\beta_k (\varepsilon-\varepsilon_F)
\mp i\alpha_k (\varepsilon-\varepsilon_F)^2,\;\;
\varepsilon {\mathstrut_{\displaystyle {<} }^{\displaystyle >}} 
\varepsilon_F. 
\end{equation}
A simple calculation reveals that for $Z_{k_F}$ in Eq.~(\ref{e24})
(or Eq.~(\ref{e27})) 
\begin{equation}
\label{e30}
Z_{k_F} \equiv (1 -\hbar\beta_{k_F})^{-1}
\end{equation}
holds. A similar analysis based on the Kramers-Kronig relation for the 
self-energy (see Appendix C) yields that when 
\footnote{\label{f26}
\small
Recall that ${\sim'}$ indicates that the corresponding asymptotic relation 
is correct up to a multiplicative constant.}
${\rm Im}\Sigma(k;\varepsilon) {\sim'} \vert\varepsilon -\varepsilon_F
\vert^{\sigma}$ for $\varepsilon\to\varepsilon_F$, the following hold. 

(i) ${\rm Re}\Sigma(k;\varepsilon) -\Sigma(k;\varepsilon_F) {\sim'} 
(\varepsilon-\varepsilon_F)$ when $1 < \sigma\leq 2$. 

(ii) ${\rm Re}\Sigma(k;\varepsilon) -\Sigma(k;\varepsilon_F){\sim'} 
(\varepsilon -\varepsilon_F) \ln\vert \varepsilon-\varepsilon_F \vert$ 
when $\sigma=1$ (``marginal Fermi liquid'' (Varma, {\sl et al.} 1989, 
Littlewood and Varma 1991, Kotliar, {\sl et al.} 1991) (see Appendix C). 

(iii) ${\rm Re}\Sigma(k;\varepsilon) -\Sigma(k;\varepsilon_F) {\sim'} 
\vert\varepsilon -\varepsilon_F \vert^{\sigma}$ when $0 < \sigma < 1$ 
(see \S~IV.C, text following Eq.~(\ref{e22}), and identify $\sigma$ with 
$\gamma_0$; see also Appendix~D and identify $\sigma$ with $1-2\gamma_0$
in Eqs.~(\ref{ed21}) and (\ref{ed22}) and with $1-\gamma_0$ in 
Eqs.~(\ref{ed28}) and (\ref{ed29})). 

It can straightforwardly be shown (using the same strategy as in the 
case of $\sigma=2$ employed above) that, for $0 < \sigma \leq 1$, 
$Z_{k_F}$ on the right-hand side of Eq.~(\ref{e24}) is vanishing, whereas 
for $1 < \sigma \leq 2$ it takes a finite value. Vanishing of $Z_{k_F}$ 
corresponds to disappearance of the Landau QPs on the Fermi surface. 
Although for $1 < \sigma \leq 2$, $Z_{k_F}$ is non-vanishing, it is evident 
that the smaller the $\sigma$, the shorter are the life-times of the QPs 
close to the Fermi surface, if QPs can at all be meaningfully defined 
(recall condition (B) introduced in \S~I). We should emphasis that as 
$\widetilde{G} (k_F;z)$ does {\sl not} possess an isolated pole at 
$z=\varepsilon_F \equiv\mu_N$, but a branch point, it follows that even 
in the case of the Landau Fermi liquids, the QPs on the Fermi surface are 
not truly infinitely long-lived but only so in an asymptotic sense.

We conclude that what Migdal (1957) describes as `leaving of a pole from 
the interior of contour ${\cal C}$ upon changing $k$ from $k_F^-$ to 
$k_F^+$' is to be understood as follows: for $k=k_F^+$, Eq.~(\ref{e17})
(or Eq.~(\ref{e26})) {\sl cannot} be satisfied when $z=\varepsilon_F \equiv 
\mu_N$, that is $E_{k_F^+} (\mu_N) \not= \mu_N$; rather we have $E_{k_F^+}
(\mu_{N+1}) = \mu_{N+1}$ and, as can be viewed from Fig.~2, $z=\mu_{N+1}$ 
indeed does not lie in the interior of ${\cal C}$.

In closing this Section, we indicate the following observation. In \S\S~I 
and IV.C we argued that Fermi liquids are distinguished by {\sl two} 
specific aspects, namely that of condition (A), that is continuous 
differentiability of $\Sigma(k_F;\varepsilon)$ with respect to 
$\varepsilon$ in a neighbourhood of $\varepsilon =\varepsilon_F$, {\sl 
and} of condition (B), that is continuous differentiability of 
$\Sigma(k;\varepsilon_F)$ with respect to $k$ in a neighbourhood of
$k=k_F$. It is owing to these that the Landau quasi-particle dispersion 
as presented in Eq.~(\ref{e19}) can be rigorously obtained from the 
quasi-particle equation, Eq.~(\ref{e17}), which is applicable to {\sl all} 
isotropic metallic systems of spin-less fermions; conditions (A) and (B) 
are prerequisite to assigning a {\sl unique} mass as well as velocity to 
an elementary excitation at and in a close vicinity of the Fermi surface 
(see Eq.~(\ref{e20}) above). However, as we have observed in the present 
Section, in obtaining Eq.~(\ref{e24}), only Eq.~(\ref{e27}) combined with 
the condition $\lim_{z\to\varepsilon_F} \widetilde{L} (z)=0$ (which 
together embody aspect (A)) played a role. Therefore, a non-vanishing 
$Z_{k_F}$ which is universally considered as {\sl the} hallmark for Fermi 
liquids, is only a {\sl necessary} condition for a metallic system to 
qualify as a Fermi liquid, but by no means a {\sl sufficient} condition 
(see \S~I). In other words, metallic systems with $Z_{k_F}\not=0$ can 
in principle be non-Fermi liquids in the above sense.

\section{(Non-)Fermi liquids and non-differentiability of the self-energy 
with respect to momentum close to and on the Fermi `surface'}

From Eq.~(\ref{e23}), making use of Eq.~(\ref{e25}), one obtains
\begin{equation}
\label{e31}
{{\partial {\sf n}(k)}\over\partial k}
= \int_{\cal C} {{{\rm d} z}\over 2\pi i}\;
{1\over \big(z - \widetilde{E}_k(z)\big)^2}\;
{{\partial\widetilde{E}_k(z)}\over\partial k}.
\end{equation} 
In view of the singular behaviour of ${\sf n}(k)$ at $k=k_F$, in 
employing Eq.~(\ref{e31}) we approach $k_F$ from either left or 
right, i.e. by $\partial {\sf n}(k)/\partial k\vert_{k= k_F^{\mp}}$ 
we imply left and right derivatives, respectively.

For our further discussions it is convenient to introduce some notational 
conventions. We denote the circular contour circumscribing $\varepsilon_F$ 
in Fig.~2 by ${\cal C}_s$ (subscript $s$ refers to `singular', in view 
of the singularity of $\widetilde{G}(k_F;z)$ at $z=\varepsilon_F$) and 
its complement, ${\cal C}\backslash {\cal C}_s$, by ${\cal C}_r$ (subscript 
$r$ refers to `regular'). Consequently, we write ${\sf n}(k) = {\sf n}_s(k) 
+ {\sf n}_r(k)$ with ${\sf n}_s(k)$ and ${\sf n}_r(k)$ corresponding to
the $z$-integrals along ${\cal C}_s$ and ${\cal C}_r$, respectively (here 
we assume the radius $\varrho$ characterising ${\cal C}_s$ to be 
infinitesimally small). 

In arriving at the result in Eq.~(\ref{e24}), we have made use of 
${\sf n}_r(k_F^{+}) = {\sf n}_r(k_F^{-})$ so that Eq.~(\ref{e24}) 
in fact amounts to ${\sf n}_s(k_F^-)-{\sf n}_s(k_F^+) = Z_{k_F}$. A 
similar property, namely $\partial {\sf n}_r(k)/\partial k\vert_{k=k_F^-} 
=\partial {\sf n}_r(k)/\partial k\vert_{k=k_F^+}$, may not be applicable 
in general, in particular since $\partial {\sf n}_r(k)/\partial k\vert_{k
=k_F^{\mp}}$ may be unbounded. Below we demonstrate that: 
\footnote{
\small
Similar considerations can be given for the case $k=k_F^+$.}
i) for Fermi liquids $\partial {\sf n}(k)/\partial k\vert_{k=k_F^-}$ is 
{\sl either} finite, in which case it follows that $\widetilde{\Sigma}
(k;z)/\partial k\vert_{k=k_F^-}$ is finite for {\sl all} $z$, {\sl or} 
infinite, in which case $\partial \beta_k/\partial k\vert_{k=k_F^-}$ is 
also infinite and consequently $\partial \{\widetilde{\Sigma}(k;z) 
-\Sigma(k;\varepsilon_F)\}/\partial k\vert_{k=k_F^-}$ is infinite for 
{\sl all} $z$ (note that, for Fermi liquids, $\Sigma(k;\varepsilon_F)$ is 
{\sl by definition} a continuously differentiable function of $k$ in 
a neighbourhood of $k_F$); ii) for non-Fermi liquids characterised by
$Z_{k_F}=0$, $\partial {\sf n}_s(k)/\partial k\vert_{k=k_F^-} = 0$. 
Further, although for these systems ${\sf n}(k_F^-) - {\sf n}(k_F^+) 
=0$ holds, $\partial {\sf n}(k)/\partial k$ may or may not diverge as 
$k\uparrow k_F$ ($k\downarrow k_F$); in the one-dimensional Luttinger 
model (Luttinger 1963, Mattis and Lieb 1965), which is a prototype of 
one-dimensional systems of interacting spin-less fermions (Haldane 1980, 
1981), with `weak' interaction --- as characterised by a small anomalous 
dimension, 
\footnote{
\small
The non-interacting case corresponds to $\alpha\equiv 2\gamma_0=0$.
See \S~IV.C and Appendix D. }
namely $0 < \alpha < 1$ ---, $\partial {\sf n}(k)/\partial k$ does diverge 
for $k\to k_F^{\mp}$, and consequently $\partial\widetilde{\Sigma}(k;z)/
\partial k\vert_{k=k_F^-}$ is infinite for {\sl all} $z$; here, contrary 
to the case of Fermi liquids, $\partial\Sigma(k;\varepsilon_F)/\partial 
k\vert_{k=k_F^-}$ is {\sl not} necessarily finite. This aspect can be 
explicitly verified from the available explicit expression for the 
spectral function corresponding to the retarded single-particle Green 
function (see Appendix D). 

Now we proceed with demonstrating the above statements. 
In cases where Eq.~(\ref{e27}) and $\lim_{z\to\varepsilon_F}
\widetilde{L}(z) = 0$ are satisfied, making use of the Cauchy residue 
theorem, one obtains
\begin{equation}
\label{e32}
\left.{{\partial {\sf n}_s(k)}\over\partial k}\right|_{k=k_F^-}
=\hbar Z_{k_F}^2 \left. {{\partial^2\Sigma(k;\varepsilon)}\over 
\partial \varepsilon \partial k}\right|_{\varepsilon=\varepsilon_F
\atop k=k_F^-}, \;\;\;
\left.{{\partial {\sf n}_s(k)}\over\partial k}\right|_{k=k_F^+} = 0.
\end{equation}
In view of the derivative with respect to $\varepsilon$, $\Sigma(k;
\varepsilon)$ in Eq.~(\ref{e32}) may be replaced by $\Sigma(k;\varepsilon)
-\Sigma(k;\varepsilon_F)$. From this it follows that a possible divergence 
of $\partial {\sf n}_s(k)/\partial k$ for $k\uparrow k_F$ {\sl cannot} be 
due to $\Sigma(k;\varepsilon_F)$ and therefore a divergent $\partial
{\sf n}_s(k)/\partial k$ for $k\uparrow k_F$ is {\sl not} in conflict with 
the fundamental assumption of the Fermi-liquid theory, namely that of 
continuous differentiability of $\Sigma(k;\varepsilon_F)$ in a 
neighbourhood of $k=k_F$. We note in passing that, through employing 
Eqs.~(\ref{e29}) and (\ref{e30}), we can also write $\partial {\sf n}_s(k)/
\partial k\vert_{k=k_F^-} = \partial Z_k/\partial k\vert_{k=k_F}$, 
where $Z_k {:=} (1-\hbar\beta_k)^{-1}$.
\footnote{
\small
We should emphasise that here $Z_k$ is merely a generalisation of 
$Z_{k_F}$ and does not necessarily have the same physical significance 
as $Z_{k_F}$ does. To appreciate the relevance of this remark, one should 
realise that for $k\not=k_F$, Eq.~(\ref{e17}) may {\sl not} have a 
real-valued solution.} 
In Appendix C we demonstrate that divergence of $\partial\beta_k/\partial 
k$ (which in view of the latter result implies that of $\partial 
{\sf n}_s(k)/\partial k$) at any $k=k_0$, e.g. $k=k_F$, signals divergence 
of $\partial \widetilde{\Sigma}(k;z)/\partial k\vert_{k=k_0}$ for {\sl all} 
$z$, ${\rm Im}(z)\not=0$; in Appendix D we explicitly show this to be the 
case for the (one-dimensional) Luttinger model for spin-less fermions. 
With reference to Eq.~(\ref{e31}), as well as Eq.~(\ref{e25}) where 
$\widetilde{E}_k(z)$ is defined in terms of $\widetilde{\Sigma}(k;z)$, 
divergence of $\partial {\sf n}_s(k)/\partial k$ at $k=k_F^-$ is thus seen 
to imply divergence of $\partial {\sf n}(k)/\partial k$ at $k=k_F^-$. This 
completes demonstration of our point (i) above. It is relevant to mention 
that for a three-dimensional isotropic system of fermions interacting 
through a hard-core potential, qualifying as a Fermi liquid, $\partial 
{\sf n}(k)/\partial k$ has been shown to be logarithmically divergent 
for $k\to k_F^{\mp}$ (Belyakov 1961, Sartor and Mahaux 1980).
\footnote{\label{f26b}
\small
In our judgement, nothing precludes the possibility that this divergence 
is an artifact of the second-order perturbation theory employed in 
the treatment by Belyakov (1961).}
On the other hand, for a similar system of electrons, interacting 
through the long-range Coulomb interaction, no divergence is observed in 
$\partial {\sf n}(k)/\partial k$, $k\to k_F^{\mp}$, within the framework 
of the random-phase approximation (RPA) (Daniel and Vosko 1960); 
calculations of ${\sf n}(k)$ beyond the RPA are also available (Geldart, 
Houghton and Vosko 1964), however the results do not provide information 
with regard to regularity or otherwise of $\partial {\sf n}(k)/\partial 
k$ at $k=k_F^{\mp}$.
\footnote{\label{f26c}
\small
For a calculation of ${\sf n}(k)$ at large $k$ as well as some general 
discussions concerning ${\sf n}(k)$ see also Yasuhara and Kawazoe (1976).}

For non-Fermi liquids, such as the marginal Fermi liquid considered in
Appendix C, the result $\partial {\sf n}_s(k)/\partial k\vert_{k=k_F^-}
= 0$ follows exactly for the same reason that ${\sf n}_s(k_F^-) = 0$
for these systems: through parametrising ${\cal C}_s$ in terms of the 
circular-polar angle $\varphi$, one observes that despite singularity 
of the integrand at $\varepsilon_F$, the $\varphi$-integration over 
$[-\pi,\pi)$ yields an identically-vanishing contribution. As our
considerations in \S~V have made explicit, at the root of this property 
lies the non-differentiability with respect to $\varepsilon$ of 
$\Sigma(k_F;\varepsilon)$ at $\varepsilon_F$. We emphasise that this
is {\sl sufficient} for disqualifying a system to be a Fermi liquid; our 
discussions in \S~IV.C have made evident that this non-differentiability
condition stands in the way of obtaining the quasi-particle energy
dispersion in Eq.~(\ref{e19}) from Eq.~(\ref{e17}).

For a {\sl general} non-Fermi-liquid metallic system, in {\sl arbitrary} 
$d$-dimensional spatial space (if indeed such a system exists in $d > 1$), 
our knowledge with regard to behaviour of $\partial {\sf n}(k)/\partial k$ 
in some neighbourhoods of $k=k_F^{\mp}$ cannot surpass that of $\partial 
{\sf n}(k)/\partial k$ corresponding to Fermi-liquid systems.
\footnote{
\small
Here we employ the notion of `neighbourhood' in a loose sense, since, as 
we have mentioned earlier, here by $\partial {\sf n}(k)/\partial k$, $k\to 
k_F$, we express {\sl left} or {\sl right} derivative of ${\sf n}(k)$, 
thus excluding $k=k_F$ from the interior of the interval over which ${\sf 
n}(k)$ is being differentiated. }
Since, by definition, to a non-Fermi-liquid metallic system {\sl no} 
quasi-particle energy dispersion similar to that in Eq.~(\ref{e19}) can 
correspond (see above), and this can be effected through a non-continuous 
differentiable $\Sigma(k_F;\varepsilon)$ in a neighbourhood of 
$\varepsilon=\varepsilon_F$, the non-continuous-differentiability with 
respect to $k$ of $\Sigma(k; \varepsilon_F)$ pertaining to these systems 
cannot in general be ruled in but also not ruled out. In $d=1$ dimension, 
the exactly-solvable Luttinger (1963) model provides us with the 
opportunity to arrive at a rigorous statement, however. In this model, for 
$k$ close to $k_F$ one has (see, e.g., Voit 1993b) ${\sf n}(k) \sim 1/2 
- C_1 {\rm sgn}(k-k_F) \vert k-k_F \vert^{\alpha} + C_2 (k - k_F)$ where 
$C_1$ and $C_2$ are constants and $\alpha \equiv 2\gamma_0$, the `anomalous 
dimension', is not universal but depends on the nature and strength of the 
particle-particle interaction in the system under investigation; for the 
one-dimensional Hubbard model corresponding to a finite repulsive on-site 
interaction, $U$, and away from the half-filling of the Hubbard band, 
$0 <\alpha < 1/8$, with the upper limit achieved for infinitely large $U$. 
One observes that for $0<\alpha < 1$, which in view of the latter 
observation should amount to a wide range of values for $\alpha$, 
${\sf n}(k)$, although continuous, is {\sl not} continuously differentiable 
in a neighbourhood of $k=k_F$; $\partial {\sf n}(k)/\partial k$ diverges 
as $k\to k_F^{\mp}$.

\section{Summary and concluding remarks}

In this work we have presented a critical analysis of a celebrated theorem 
due to Luttinger (1961) concerning energies and life-times of low-energy 
QP excitations in interacting systems. This theorem has played the crucial 
role of classifying {\sl all} metallic systems in spatial dimensions 
larger than unity as Landau Fermi liquids, with the explicit assumption 
with regard to applicability of the many-body perturbation theory to 
all orders to these systems.
\footnote{\label{f27}
\small
This on account of the fact that the proof of this theorem has been 
based on an infinite-order perturbation expansion of the self-energy 
operator.}
In this work we have demonstrated that {\sl Luttinger's (1961) proof 
involves an implicit assumption} with regard to the dispersion of the 
QP energies, which assumption we have explicitly shown to be specific 
to Fermi-liquid systems. We have therefore shown that Luttinger's proof 
amounts to {\sl a demonstration of consistency} of the mentioned implicit 
assumption with the property ${\rm Im}\Sigma(k;\varepsilon) {\sim'}
(\varepsilon -\varepsilon_F)^2$, for $\varepsilon\to\varepsilon_F$, to 
{\sl all} orders of the many-body perturbation theory. It follows that, 
contrary to the commonly-held view, the many-body perturbation theory
does {\sl not} necessarily break down when applied to systems whose 
self-energies are of non-Fermi-liquid type. In the absence of any theorem 
to replace the Luttinger theorem, one could reasonably conjecture that 
it is most likely that, in particular in two spatial dimensions, 
non-Fermi-liquid-like metallic systems if not abundant, should not be 
rare and that these may be correctly addressed within the framework 
of the many-body perturbation theory. In this connection it is important 
to point out the following two observations. 

First, {\sl any} static and local approximation to the self-energy operator 
implies a Fermi-liquid behaviour for the QPs, no matter how ingenious such 
approximation may be. The entire body of the {\sl energy-band} methods based 
on the {\sl conventional} density-functional theory, involving a local and 
energy-independent effective potential (for a comprehensive review see 
Dreizler and Gross 1990), pertains to this category of approximation 
frameworks. The (self-consistent) Hartree-Fock scheme which involves the 
non-local static exchange self-energy, {\sl may} describe non-Fermi liquid 
metals, however only those whose $Z_{k_F}=1$. 
\footnote{
\small
For uniform systems of electrons interacting through the Coulomb potential, 
the derivative with respect to $k$ of the Hartree-Fock self- energy is 
logarithmically divergent at $k=k_F$ (Ashcroft and Mermin 1981, p.~334) 
--- see Appendix~B. Since $\Sigma^{HF}$ is independent of $\varepsilon$ 
(Appendix~B), it is therefore analytic over the entire $z$-plane and thus
the associated $Z_{k_F}=1$ (see \S~V). Therefore a metal within the 
Hartree-Fock scheme may be non-Fermi liquid solely on account of violation 
of condition (B) introduced in \S~I. } 
Second, in evaluating the self-energy operator for a Coulomb system, within 
the framework of the many-body perturbation theory, it has to be realised 
that in metallic systems the long range of the {\sl bare} Coulomb interaction 
gives rise to divergent self-energy contributions (from the second order 
in the interaction onwards). Thus, for such systems, perturbation expansion 
has to be in terms of the dynamically-screened electron-electron interaction 
function. In calculating this function, at least {\sl all} polarisation 
diagrams of the random-phase approximation have to be taken into account, 
for all these diagrams have unbounded contributions and only in combination 
give rise to a finite result. Our considerations with regard to the problem 
of unbounded self-energy diagrams have led us to draw the additional 
conclusion that, irrespective of our above-indicated finding, Luttinger's 
(1961) proof {\sl cannot} have bearing on metallic systems of electrons 
interacting through the long-range Coulomb interaction function. This 
follows from the fact that Luttinger's proof has been based on a 
perturbation expansion of the self-energy operator in terms of the {\sl 
bare} electron-electron interaction. In this connection we point out that 
analytic properties of a series of functions of complex variable do not 
coincide with those of the constituent terms, unless the series be {\sl 
uniformly} convergent (Whittaker and Watson 1927, pp. 91 and 92, Titchmarsh 
1939, pp. 95-98). Further, the sum of a series which is not {\sl absolutely 
convergent} depends on the order of summation of terms in the series 
(Whittaker and Watson 1927, pp. 18 and 25). It is evident that, in 
particular, perturbation series which involve singular terms {\sl cannot} 
be absolutely or uniformly convergent.
 
We have further elaborated upon the analytic properties of the 
single-particle Green function and the self-energy operator as functions 
of a complex-valued energy parameter, $z$. In particular we have pointed 
out that the commonly-used equation for energies of the QPs in an 
interacting system can yield, if any, only real-valued solutions. To 
obtain complex-valued solutions (concerning systems in the thermodynamic 
limit), the self-energy operator has to be analytically continued into a 
{\sl non-physical} RS of the complex energy plane.

We have paid careful attention to the precise nature of the singular points 
of the single-particle Green function and the self-energy operator. For 
instance we have pointed out that the Fermi energy is not an {\sl isolated 
singularity} of the Green function and therefore {\sl cannot} be a pole. 
In the light of this, we have considered a celebrated theorem due to Migdal 
(1957) and discussed how its proof is independent of the commonly-made 
assumption, that on the Fermi surface the Fermi energy were a {\sl pole} 
of the single-particle Green function. We have explicitly shown that for 
the momentum distribution function ${\sf n}(k)$ pertaining to a metallic 
system to be discontinuous at $k=k_F$ by a finite amount --- the magnitude 
of this discontinuity being, according to the Migdal theorem, equal to 
$Z_{k_F}$ ---, it is only necessary that the self-energy $\Sigma(k_F;
\varepsilon)$ be a continuously differentiable function of $\varepsilon$ 
in a neighbourhood of $\varepsilon=\varepsilon_F$. Consequently metallic 
systems with $Z_{k_F}\not= 0$ are {\sl not} necessarily Fermi liquids. 
Fermi liquids, we have shown, have in addition the property that their 
corresponding $\Sigma(k;\varepsilon_F)$ is a continuously differentiable 
function of $k$ in a neighbourhood of $k=k_F$. In spite of this, $\Sigma
(k;\varepsilon)$ pertaining to Fermi liquids may not be continuously 
differentiable function of $k$ in a neighbourhood of $k=k_F$ when 
$\varepsilon\not=\varepsilon_F$. These observations, which we have 
explicitly examined on the self-energies of Fermi-, marginal Fermi- 
and Luttinger-liquids, clearly demonstrate how uncritical Taylor 
expansions of various functions of $k$ and $\varepsilon$ associated 
with the self-energy can lead to incorrect description of the physical 
properties of the systems under consideration at low energies.

\vskip 10mm
\section*{Acknowledgements}
With appreciation I acknowledge support by the Max-Planck-Gesellschaft, 
Germany. 
It is a pleasure for me to thank Professor Peter B. Littlewood and members 
of the Theory of Condensed Matter Group for their kind hospitality at 
Cavendish Laboratory where this work was completed. 
I record my indebtedness to Girton College, Cambridge, for support in the 
course of my present association with Cavendish Laboratory. 
I take this opportunity and extend my heartfelt appreciations to Dr Martin 
W. Ennis for generously and patiently sharing his insight especially into 
the music of J.S. Bach and kind hospitality. 

\appendix
\section{Mathematical preliminaries}

Since in the present work we repeatedly encounter a number of specific 
mathematical notions, we devote this Appendix to a brief exposition of 
these.

\subsection{On the types of singularity}

A point at which a function $g(z)$ of complex variable $z$ is not 
{\sl analytic}
\footnote{\label{f29}
\small
{\sl Analytic}, {\sl regular} and {\sl holomorphic} are alternative
but equivalent designations.}
is called a {\sl singular} point (for details of what follows see, for
example, Whittaker and Watson 1927, pp. 102 and 104, Titchmarsh 1939, 
pp. 89-95). Such a point is either {\sl isolated} or {\sl non-isolated}; 
$z_0$ is isolated if there exists a $\delta > 0$ such that inside the 
region $\vert z - z_0\vert < \delta$ there exists no other singular point 
of $g(z)$ than $z_0$. Otherwise $z_0$ is not isolated.

A singularity may be {\sl removable}; such singularity corresponds to 
a point $z_0$ at which $g(z)$ is not defined, but $\lim_{z\to z_0} 
g(z)$ exists. Thus $z=0$ is a removable singularity of $g(z) {:=} 
\sin(z)/z$.

{\sl Limiting} (or {\sl accumulation}) point of an infinite sequence 
of poles, is {\sl not} classified as a pole and thus is considered as an
{\sl essential} singularity, and here, a non-isolated essential singularity. 
For instance, the sequence of poles of $g(z) {:=} \sum_{n=0}^{\infty} 
1/(n![1+a^{2n}z^2])$, with $a > 1$, have $z=0$ as their limiting point 
which is not isolated. This function has {\sl no} Taylor or Laurent 
series expansion (see \S~A.2) over any domain of the $z$-plane which has 
$z=0$ as its {\sl interior}.

Let $g(z)$ be a single-valued function throughout a domain ${\cal D}$ 
at whose interior point $z_0$, $g(z)$ is singular. Suppose that the {\sl 
principal part} of the Laurent series expansion (see \S~A.2) of $g(z)$ 
around $z=z_0$ terminates with the term $a_{-n}/(z - z_0)^n$, with $a_{-n}$ 
a non-vanishing constant. In such case, $z_0$ is called a {\sl pole} 
of order $n$. Poles are thus by definition {\sl isolated} singularities.

If the principal part of the Laurent expansion of $g(z)$ around $z=z_0$ 
does not terminate (i.e., if there exists {\sl no} $n_0$ such that $a_{-n} 
=0$ for all $n > n_0$), $z_0$ is an {\sl isolated} essential singularity 
of $g(z)$. The function $g(z) {:=} \exp(z)$ has one such singularity at 
the point of infinity, i.e. at $z=1/\zeta$ where $\zeta=0$. 

{\sl Branch points} belong to the class of singular points and concern 
multi-valued functions. Let $g(z)$ be one such function. By traversing 
a closed contour which circumscribes only one branch point of $g(z)$, 
one obtains, upon arriving at the starting point $z_1$, a different 
value for $g(z_1)$, indicating change of the initial branch of $g(z)$ 
into a different branch; for a branch point of order $p$, 
the original branch is recovered after completion of $p$ revolutions 
along the mentioned contour. Thus $(z-z_0)^{1/3}$ has a third-order 
branch point at $z=z_0$. Functions can also possess branch points of 
infinite order; for $g(z){:=}\ln(z)$, $z=0$ and $1/z=0$ are such points.

\subsection{Taylor, Laurent and asymptotic series}

In the Abstract of this work we have referred to the expression 
${\rm Im}\Sigma(k;\varepsilon) \sim \mp \alpha_k (\varepsilon 
-\varepsilon_F)^2$, $\varepsilon {\mathstrut_{\displaystyle {<} 
}^{\displaystyle >}} \varepsilon_F$, for $\varepsilon\to\varepsilon_F$, 
as an {\sl asymptotic} relation. Here we specify what asymptotic relations 
and asymptotic series are and in what essential respects these series 
differ from the Taylor and the Laurent series (Whittaker and Watson 
1927, pp. 93, 94 and 100). We also indicate the interrelation between 
branch points (see \S~A.1) and divergent asymptotic series. 

When $g(z)$ is analytic at $z=z_0$, then, by definition, there exists 
an {\sl open} domain ${\cal D}$ of which $z_0$ is an interior and over 
which $g(z)$ is analytic. Within a circle around $z_0$ embedded within 
${\cal D}$, $g(z)$ can be represented in terms of a Taylor series: $g(z) 
= \sum_{n=0}^{\infty} a_n (z - z_0)^n$, with {\sl unique} coefficients 
$a_n$. Thus $\{(z-z_0)^n \vert n=0,1,\dots\}$ can be considered as a 
{\sl complete} basis in a (finite) neighbourhood of $z_0$ for {\sl any} 
function that is analytic at $z=z_0$ (note that this set does not contain 
such term as, for instance, $(z-z_0)^{\sigma}$ with $\sigma$ non-integer). 
This basis can be extended to $\{(z-z_0)^n \vert n=-m,-m+1,\dots, 0,1,
\dots\}$ in order to form a {\sl complete} basis in a (finite)
neighbourhood of $z_0$ for representing {\sl any} function which has a 
pole of order $m$ or lower at $z_0$ and is analytic in a neighbourhood of 
$z=z_0$; the resulting representation is the well-known Laurent series 
expansion (\S~A.1). It {\sl cannot} be extended, through increasing $m$ 
in $\{(z-z_0)^n \vert n=-m,-m+1,\dots, 0,1, \dots\}$, for representing 
functions that possess a branch point at $z=z_0$ (see \S~A.3). It is 
important to point out that the Taylor and the Laurent series expansions 
are very special in that they provide {\sl uniform} representations of an 
analytic function; when these series are convergent for any $\varrho$ and 
$\varphi$ in $z-z_0 \equiv \varrho \exp(i\varphi)$, they are {\sl uniformly} 
convergent for {\sl all} values of $\varphi$, $\varphi\in [0,2\pi)$. Branch 
points are peculiar in that they do not allow for {\sl any} uniform 
representation (note the `$\mp$' in the above expression for ${\rm Im}
\Sigma(k;\varepsilon)$). We shall return to this point further on in this 
Section.

Asymptotic series (in the sense of Poincar\'e) (Whittaker and Watson 
1927, Ch. VIII, Copson 1965, Dingle 1973, Lauwerier 1977) are more 
general than the
Taylor and the Laurent series in that they are in terms of ``basis 
functions'' that are not necessarily of the form $(z-z_0)^n$, with $n$ 
possibly negative, zero and positive integer. To an asymptotic series 
corresponds a so-called {\sl asymptotic sequence}: The set $\{\phi_n(z) 
\vert n=0,1,\dots\}$ is an asymptotic sequence with respect to $z_0$ 
provided it possesses the property $\phi_{n+1}(z)/ \phi_n(z)\to 0$, 
denoted by $\phi_{n+1}(z) = o\big(\phi_n(z)\big)$, for $z\to z_0$. Thus 
$\{(z-z_0)^n \vert n=-m,-m+1,\dots, 0,1,\dots\}$ is an asymptotic sequence 
with respect to $z_0$. We point out that some authors, for example 
Whittaker and Watson (1927, Ch.~VIII), reserve the designation `asymptotic 
series' for those series that are both asymptotic (in the above sense) 
{\sl and} divergent. We do {\sl not} follow this restrictive convention.

As we have mentioned above, an asymptotic series expansion of a function 
may not be convergent. It may 
not also be {\sl uniform}; functions with non-uniform asymptotic series 
expansions for $z\to z_0$ are characterised by possessing different 
asymptotic series for different sectors of the $z$-plane around $z=z_0$. 
Thus, for instance, $g(z) {:=} \exp(z) + \exp(-z) \tanh (1/z)$ has the 
following asymptotic forms for $z\to 0$: $g(z)\sim 2 \cosh(z) \sim 2 
+ z^2 + \dots$, when ${\rm Re}(z) > 0$, and $g(z)\sim 2 \sinh(z) 
\sim 2 z + z^3/3 +\dots$, when ${\rm Re}(z) < 0$ (Lauwerier 1977, p. 
11). The fact that in different sectors around a point in the complex 
$z$-plane a function can have different asymptotic expansions with 
respect to the same asymptotic sequence, is referred to as the {\sl 
Stokes phenomenon} (Watson 1952, Dingle 1973); `Stokes lines' are
thus the branch cuts in our discussions.

The Taylor series of analytic functions based on point $z_0$ are {\sl
convergent} uniform asymptotic series for $z\to z_0$ (similarly for the 
Laurent series). It can be shown (Lauwerier 1977, pp. 12-14) that an
analytic function can be associated with {\sl even} a divergent asymptotic 
series corresponding to the asymptotic sequence $\{ (z-z_0)^n \vert n=0,1,
\dots\}$; the given (divergent) series is its asymptotic expansion. Thus, 
for instance, after Borel (or Euler) transformation (Whittaker and Watson 
1927, pp. 154 and 155, Dingle 1973, pp. 405-408, Lauwerier 1977, pp. 45-50) 
of the divergent series $f(z) {:=} \sum_{n=0}^{\infty} (-1)^n n! z^n$, one 
formally obtains $f_B(z) {:=} \sum_{n=0}^{\infty} (-1)^n z^n = (1 + z)^{-1}$. 
Through the Borel back-transformation of $f_B(z)$, $\overline{f}_B(z) {:=} 
\int_0^{\infty} {\rm d} x\; \exp(-x) f_B(x z)$, one obtains a function, 
i.e. $\overline{f}_B(z)$, which is analytic in the sector $-\pi < \arg(z) < 
\pi$ of the $z$-plane. 
\footnote{\label{f30}
\small
It can be shown that $\overline{f}_B(z)\equiv \exp(1/z) E_1(1/z)/z$ 
where $E_1(z)$ stands for the exponential-integral function (see, e.g.,
Abramowitz and Stegun 1972, p. 228). We note that $\exp(1/z)$ has
an isolated essential singularity at $z=0$ (see \S~A.1) and that
$E_1(1/z)$ has branch points at $z=0$ and $1/z=0$. }
Through replacing $f_B(x z)$ by its {\sl formal} geometric expansion
$1 - xz + (xz)^2 -\dots$, the term-by-term evaluation of the latter 
integral yields the original divergent series. We observe that the 
divergence of the original asymptotic series is closely associated with 
the restricted sector of the $z$-plane around $z=0$ over which 
$\overline{f}_B(z)$ is analytic.

\subsection{Many-valued functions: Physical and non-physical Riemann 
sheets}

Many-valuedness (Whittaker and Watson 1927, pp. 96-98, Titchmarsh 1939,
pp. 138-164) is a generic property of correlation functions of complex 
energy pertaining to systems in the thermodynamic limit (see \S~A.4). 
In particular, the single-particle Green function and the self-energy 
operator are many-valued (see \S~II). 

An $n$-valued function of complex variable $z$ over domain ${\cal D}$ 
may be thought of as embodying $n$ branches of a single-valued function 
over the extended domain consisting of the union of $n$ replicas of 
${\cal D}$; since these domains signify the same region of the complex 
plane, they are to be distinguished by considering them as being located 
on different sheets, namely the {\sl Riemann sheets} (RSs), of the complex 
plane. In the present paper we single out one particular branch of the 
many-valued functions that we encounter, such as the single-particle 
Green function, because of its {\sl physical} significance. To this 
particular branch we refer as the function in question on the {\sl 
physical} RS or the `physical' function (compare with $G(\varepsilon)$
in Eq.~(\ref{e1}) which we have designated as the `physical' 
single-particle Green function). 

Throughout this paper we employ the following notational conventions. 
Let $f(\varepsilon)$ be a function of real energy variable $\varepsilon$, 
which we consider to describe some dynamical physical property of the
many-particle system under consideration. By $\widetilde{f}(z)$ we denote 
the analytic function that {\sl uniformly} approaches $f(\varepsilon)$ 
when $z\to\varepsilon$; the so-called `edge-of-the-wedge' theorem 
(Streater and Wightman 1964) asserts uniqueness of $\widetilde{f}(z)$. 
\footnote{
\small
The condition of {\sl uniformity} can be shown, {\sl a posteriori}, to be 
redundant: the existence of the limit $\widetilde{f}(z) \to f(\varepsilon)$, 
as $z\to\varepsilon$, with $\varepsilon$ varying over an {\sl open} interval 
where $f(\varepsilon)$ is {\sl continuous}, implies uniformity of 
$\widetilde{f}(z) \to f(\varepsilon)$; see p.~75 in Streater and Wightman 
(1964). }
We refer to $\widetilde{f}(z)$ as $\widetilde{\widetilde{f}}(z)$ on the 
physical RS; $\widetilde{\widetilde{f}}(z)$ denotes the above-mentioned 
single-valued function over the union of all the RSs. We recall that in 
the main text we have referred to, for example, $\widetilde{G}(z)$ in 
Eq.~(\ref{e4}) as the single-particle Green function on the physical RS.

An example will clarify this. Let $f(\varepsilon){:=}\ln(\varepsilon)$, 
for $\varepsilon > 0$. We can identify $\ln(z)$ by $\widetilde{f}(z)$ 
since for $z\to\varepsilon$, with $\varepsilon > 0$, indeed $\ln(z)
\to\ln(\varepsilon)$. One can easily verify that ${\rm Ln}_n(z) {:=} 
\ln\vert z\vert + i\{\arg(z) + 2\pi n\}$, with $n$ an integer (positive, 
zero and negative) and $-\pi \leq \arg(z) < \pi$, is {\sl the} analytic 
function (i.e. $\widetilde{\widetilde{f}}(z)$) of which $\widetilde{f}(z) 
{:=} \ln(z)$ is the physical branch; thus this branch corresponds to 
$n=0$. The physical branch $\widetilde{f}(z)$, like any other branch of 
$\widetilde{\widetilde{f}}(z)$, is a many-valued function (branch points 
are {\sl not} removable [see \S~A.1] singularities). It has two branch 
points, one at $z=0$ and the other at $1/z=0$. Our above convention $-\pi
\leq\arg(z) < \pi$ specifies the branch cut of $\widetilde{f}(z)$, which 
connects the two branch points of $\widetilde{f}(z)$, to be along the 
negative $\varepsilon$-axis. One can determine {\sl all} branches of 
$\widetilde{\widetilde{f}}(z)$ from the knowledge of any of its branches. 
Two of infinitely many branches of $\widetilde{\widetilde{f}}(z)$, 
$\widetilde{f}_1(z)$ and $\widetilde{f}_2(z)$ say, that can be obtained 
by means of a {\sl direct} analytic continuation of $\widetilde{f}(z)$, 
through its branch cut, into two different non-physical RSs are {\sl 
uniquely} determined from the following requirements: $\lim_{\eta\downarrow 
0}\{\widetilde{f}_1 (\varepsilon+i\eta)-\widetilde{f}(\varepsilon-i\eta)\} 
= 0$ and $\lim_{\eta\downarrow 0}\{\widetilde{f}_2(\varepsilon-i\eta)
-\widetilde{f}(\varepsilon+i\eta) \}=0$, for $\varepsilon < 0$. It is 
not difficult to verify that $\widetilde{f}_1(z)\equiv {\rm Ln}_{-1}(z)$ 
and $\widetilde{f}_2(z) \equiv {\rm Ln}_{+1}(z)$.

\subsection{A physically-motivated example}

Here we apply the above concepts to a simple model function which 
accommodates a number of salient features of the physical functions 
which we deal with in the main part of this paper. We shall particularly 
emphasise the role played by the process of evaluating the thermodynamic
limit in modifying the nature of the singular points of dynamic 
correlation functions. 

Consider $f(\varepsilon;N,\Omega) {:=} {1\over\Omega} \sum_{\ell}[\Theta(
\varepsilon_{\ell} - e_0)-\Theta(\varepsilon_{\ell} -e_1)]
/(\varepsilon_{\ell} -\varepsilon)$, with $\varepsilon_{\ell+1} > 
\varepsilon_{\ell}$. Here $e_0$ and $e_1$ are finite constants for 
which we assume $e_0 < \varepsilon_{\ell} < e_1$ for {\sl some} values 
of $\ell$; $N$ and $\Omega$ indicate that $f$ is a function of the number 
$N$ of particles as well as the volume $\Omega$ of the system. In the 
`thermodynamic limit' ($N\to\infty$, $\Omega\to\infty$ and finite 
concentration $C {:=} N/\Omega$), $f$ is only a function of $C$. A brief 
glance at the contents of \S~II should clarify our present choice for $f$.

Suppose that in the thermodynamic limit $(\varepsilon_{\ell+M} 
-\varepsilon_{\ell})\to 0$ for {\sl any} finite value of $M$. This 
condition implies that in taking the limit, the function $f(\varepsilon;
N,\Omega)$ becomes ill-defined for {\sl any} real $\varepsilon$ in the 
interval $[e_0,e_1]$; an $\varepsilon$ in this interval will be ``pinched'' 
(Itzykson and Zuber 1988, pp. 302 and 303) by poles of $f(\varepsilon;
N,\Omega)$. To avoid this problem, the thermodynamic limit has to be 
effected {\sl after} replacing the real energy variable $\varepsilon$ by 
a complex energy variable $z$, since a complex $z$ cannot be ``pinched'' 
by the real poles of $f(z;N,\Omega)$ as $N$ and $\Omega$ approach infinity.

To be specific, let us now assume that in the thermodynamic limit poles 
of $f(z;N,\Omega)$ populate the interval $[e_0,e_1]$ with a constant 
density equal to $A \times \Omega$. We write
\begin{eqnarray}
\label{ea1}
f(z;N,\Omega) &=& {1\over\Omega} \sum_{\ell} 
{{\Theta(\varepsilon_{\ell} - e_0)
-\Theta(\varepsilon_{\ell} - e_1)} \over \varepsilon_{\ell} - z}
\nonumber\\
&\cong& A \int_{e_0}^{e_1}
{{{\rm d}\varepsilon'}\over \varepsilon' - z} {=:} 
\widetilde{f}(z;C),\;\;\;
\mbox{\rm for}\;\;\; N\to\infty,\;\Omega\to\infty.
\end{eqnarray}
One trivially obtains
\begin{equation}
\label{ea2}
\widetilde{f}(z;C) = A\{\ln(z-e_1)-\ln(z-e_0)\}.
\end{equation}
Evidently, the process of evaluating the thermodynamic limit has led 
to a dramatic change in $f(z;N,\Omega)$, transforming it into 
$\widetilde{f}(z;C)$ which analytically is distinct from $f(z;N,\Omega)$. 
For instance, $\widetilde{f}(z;C)$ has {\sl no} poles, rather it 
possesses two infinite-order branch points, at $z=e_0$ and $z=e_1$. 
Consequently, contrary to $f(z;N,\Omega)$, which is single-valued, 
$\widetilde{f}(z;C)$ is a many-valued function of $z$. 

We shall now explicitly demonstrate that $\widetilde{f}(z;C)$ as
presented in Eq.~(\ref{ea2}) is the `physical' branch of 
$\widetilde{\widetilde{f}}(z)$ (see \S~A.3). To this end, consider 
$\widetilde{f}(\varepsilon\pm i\eta;C)\equiv A \int_{e_0}^{e_1} 
{\rm d}\varepsilon'/ (\varepsilon' -\varepsilon\mp i\eta)$. 
For $\eta\downarrow 0$ we have $1/(\varepsilon' -\varepsilon \mp i\eta) 
= {\cal P}\{1/(\varepsilon'-\varepsilon)\}\pm i\pi\delta(\varepsilon'
-\varepsilon)$, where ${\cal P}$ denotes the Cauchy `principal value'. It 
is then trivially seen that $\widetilde{f}(\varepsilon\pm i\eta;C) = 
A\{\ln\vert e_1 -\varepsilon\vert -\ln\vert e_0-\varepsilon \vert\} 
\pm i\pi A\Theta(\varepsilon- e_0)\Theta(e_1 -\varepsilon)$ for $\eta
\downarrow 0$. Hence, for $\varepsilon < e_0$ and $\varepsilon > e_1$ 
the `physical' branch $\widetilde{f}(\varepsilon\pm i\eta;C)$ must be 
real-valued, while for $\varepsilon \in (e_0,e_1)$ it must satisfy 
$\lim_{\eta\downarrow 0} \{\widetilde{f}(\varepsilon+i\eta;C) 
-\widetilde{f}(\varepsilon-i\eta;C)\} = 2\pi i A$; thus the interval 
$[e_0,e_1]$ constitutes the branch cut of $\widetilde{f}(z;C)$. These 
conditions are {\sl exactly} fulfilled by the expression in Eq.~(\ref{ea2});
note that $\ln(z)$, as distinct from ${\rm Ln}_n(z)$ for $n\not=0$, 
stands for the principal-value logarithm (see, e.g., Abramowitz and
Stegun 1972, p. 67). Thus $\widetilde{f}(z;C)$ in Eq.~(\ref{ea2}) is indeed 
the analytic continuation of $f(\varepsilon;C)$ into the {\sl physical} RS.

It is instructive to consider two examples concerning continuations of 
$\widetilde{f}(z;C)$ into non-physical RSs. Let $\widetilde{g}(z) 
{:=} A \{ {\rm Ln}_1 (z-e_1) -\ln(z-e_0) \}$. It is easily verified that 
for $e_0 < \varepsilon < e_1$, $\lim_{\eta\downarrow 0} \{\widetilde{f}
(\varepsilon+i\eta;C)-\widetilde{g}(\varepsilon-i\eta)\}=0$, which 
implies that $\widetilde{g}(z)$ is {\sl the} analytic continuation of 
$\widetilde{f}(z;C)$ from the upper-half plane of the physical RS through 
the branch cut $[e_0,e_1]$ into the lower-half plane of a non-physical RS. 
That is $\widetilde{g}(z)$ is 
a branch of $\widetilde{\widetilde{f}}(z;C)$ on a non-physical RS. As 
for the second example, consider $\widetilde{h}(z){:=} A\{ {\rm Ln}_{-1}
(z -e_1)-\ln(z -e_0)\}$. For $\varepsilon\in (e_0,e_1)$ we have 
$\lim_{\eta\downarrow 0} \{\widetilde{f}(\varepsilon-i\eta;C) 
-\widetilde{h}(\varepsilon +i\eta)\}=0$ so that $\widetilde{h}(z)$ is 
the analytic continuation of $\widetilde{f}(z;C)$ from the lower-half 
plane through the branch cut $[e_0,e_1]$ into the upper-half plane of 
a non-physical RS.

\section{Asymptotic behaviour of the self-energy at large energies }

Here we demonstrate that, for $\vert z\vert\to \infty$, $\widetilde{\Sigma}
(z) \sim \Sigma^{HF}$, the Hartree-Fock self-energy,  
\begin{equation}
\label{eb1}
\Sigma^{HF}({\bf r},{\bf r}') \equiv \Sigma^H({\bf r},{\bf r}')
+ \Sigma^F({\bf r},{\bf r}') \equiv {1\over\hbar}
v_H({\bf r};[n])\delta({\bf r}-{\bf r}') 
- {1\over 2\hbar} v({\bf r}-{\bf r}')\varrho({\bf r},{\bf r}'),
\end{equation}
where $v_H({\bf r};[n]) {:=} \int {\rm d}^d r'\; v({\bf r}-{\bf r}')
n({\bf r}')$ stands for the Hartree potential, which is a functional of 
the electronic number density $n$ in the ground state, and $\varrho$ for 
the reduced single-particle density matrix. We note that $\varrho$ in the 
definition for $\Sigma^F$ pertains to the fully interacting system and
is distinct from the Slater-Fock reduced density matrix $\varrho_0$; contrary 
to the former, the latter {\sl is} idempotent, i.e. $\varrho_0 \varrho_0 
\equiv \varrho_0$, whereas $\varrho \varrho \not\equiv \varrho$. This 
difference may have some far-reaching consequences. For instance, in a 
uniform isotropic system of electrons interacting through the long-range 
Coulomb interaction function, for $\Sigma^F(k)$ evaluated in terms of 
$\varrho_0$, which we denote by $\Sigma^F_{\sf s}(k)$, one has (Ashcroft 
and Mermin 1981, p. 334)
\footnote{\label{fb1}
\small
For this system, $\Sigma^H(k)$ is divergent but cancels an equally-divergent 
contribution due to the field of the positively-charged uniform background.}
\begin{equation}
\label{eb2}
\Sigma^F_{\sf s}(k) = -{{2 e^2}\over \pi\hbar} k_F\; {\cal F}(k/k_F),
\;\;\; 
{\cal F}(x) {:=} {1\over 2} + {{1-x^2}\over 4x} 
\ln\left| {{1+x}\over 1-x}\right|;
\end{equation}
the first derivative with respect to $k$ of $\Sigma^F_{\sf s}(k)$ is seen 
to be logarithmically divergent at $k=k_F$, an aspect which {\sl may} be an 
artifact of $\Sigma^F_{\sf s}(k)$ and may not be shared by $\Sigma^F(k)$.

From the Dyson equation we have $\Sigma(\varepsilon) = G_0^{-1}
(\varepsilon) - G^{-1}(\varepsilon)$. Therefore to deduce the asymptotic 
expansion of $\Sigma(\varepsilon)$ for $\vert\varepsilon\vert\to\infty$, 
we need to determine those for $G_0^{-1}(\varepsilon)$ and $G^{-1}
(\varepsilon)$. The asymptotic series for $G_0^{-1}(\varepsilon)$ and
$G^{-1}(\varepsilon)$ can be deduced from those pertaining to 
$G_0(\varepsilon)$ and $G(\varepsilon)$, respectively, through reliance 
on the following result from the theory of asymptotic analysis (see Copson 
1965, pp. 8 and 9): Let ${\tilde f}(z)$ have the following asymptotic 
expansion for $\vert z\vert\to\infty$, where $z$ is a complex variable: 
${\tilde f}(z) \sim f_0 + f_1/z + f_2/z^2 +\dots$, with $f_0$, $f_1$, 
$\dots$ constants, independent of $z$. Then, provided that $f_0\not=0$,
$1/{\tilde f}(z) \sim 1/f_0 + {\bar f}_1/z + {\bar f}_2/z^2 + \dots$ holds 
for $\vert z\vert\to\infty$, where ${\bar f}_1 = - f_1/f_0^2$, ${\bar f}_2 
= (f_1^2 - f_0 f_2)/f_0^3$, etc.
\footnote{\label{fb2}
\small
When $f_0 = 0$ and $f_1\not=0$, then one should apply this
result to ${\tilde g}(z) {:=} z {\tilde f}(z)$; from the
asymptotic series for $1/{\tilde g}(z)$ that for $1/{\tilde f}(z)$
is obtained through a multiplication by $z$.}

From the Lehmann representation for $G(\varepsilon)$ in Eq.~(\ref{e1})
it follows that
\begin{equation}
\label{eb3}
G(\varepsilon) \sim {{G_{\infty_1}}\over \varepsilon}
+ {{G_{\infty_2}}\over\varepsilon^2} + \dots,
\;\;\;\; \mbox{\rm for}\;\; \vert\varepsilon\vert\to\infty,
\end{equation}
where
\begin{eqnarray}
\label{eb4}
G_{\infty_1}({\bf r},{\bf r}') &=& \hbar\sum_s 
\Lambda_s({\bf r}) \Lambda_s^*({\bf r}') 
\equiv \hbar\delta({\bf r}-{\bf r}'),\\
\label{eb5}
G_{\infty_2}({\bf r},{\bf r}') &=& \hbar\sum_s \varepsilon_s
\Lambda_s({\bf r}) \Lambda_s^*({\bf r}')
\equiv \hbar \Big(\Xi_{<}({\bf r},{\bf r}') 
+ \Xi_{>}({\bf r},{\bf r}')\Big);
\end{eqnarray}
for $\Xi_{<}$ and $\Xi_{>}$ see Eqs.~(\ref{eb9}) and (\ref{eb10}) below.
A similar expression to Eq.~(\ref{eb3}) holds for $G_0(\varepsilon)$.
Therefore from our above considerations with regard to the asymptotic
series for $1/{\tilde f}(z)$ in terms of the coefficients of that for 
${\tilde f}(z)$, it follows that $\Sigma(\varepsilon)$ possesses the 
following asymptotic series expansion
\begin{equation}
\label{eb6}
\Sigma(\varepsilon) \sim \Sigma_{\infty_0} + {{\Sigma_{\infty_1}}\over
\varepsilon} + {{\Sigma_{\infty_2}}\over \varepsilon^2} + \dots,
\;\;\;\; \mbox{\rm for}\;\; \vert\varepsilon\vert\to\infty.
\end{equation}
Here we are interested in the leading asymptotic term $\Sigma_{\infty_0}$ 
which our above considerations indicate to be determined from $G_{0;
\infty_2}$ and $G_{\infty_2}$ as follows
\begin{equation}
\label{eb7}
\Sigma_{\infty_0} = {1\over\hbar^2} \big\{
- G_{0;\infty_2} + G_{\infty_2} \big\}.
\end{equation}

For the many-body Hamiltonian of the form
\begin{equation}
\label{eb8}
\widehat{H} = \int {\rm d}^d r\; \widehat{\psi}^{\dag}({\bf r})
\left[ {{-\hbar^2}\over 2 m_e}\nabla^2 + u({\bf r}) \right]
\widehat{\psi}({\bf r})
+ {1\over 2} \int {\rm d}^d r {\rm d}^d r'\;
\widehat{\psi}^{\dag}({\bf r}) \widehat{\psi}^{\dag}({\bf r}')
v({\bf r}-{\bf r}')\widehat{\psi}({\bf r}')
\widehat{\psi}({\bf r}),
\end{equation}
with $u({\bf r})$ the local external potential, making use of the 
definition for the Lehmann amplitudes and energies as given in 
Eq.~(\ref{e3}) and the fact that $E_{M,s}\vert\Psi_{M,s}\rangle = 
\widehat{H}\vert\Psi_{M,s}\rangle$, one readily obtains
\begin{eqnarray}
\label{eb9}
\Xi_{<}({\bf r},{\bf r}') &{:=}& \sum_s \theta(\mu -\varepsilon_s)
\varepsilon_s \Lambda_s({\bf r}) \Lambda_s^*({\bf r}')\nonumber\\
&=& -\langle\Psi_{N,0}\vert \widehat{\psi}^{\dag}({\bf r}')
\left[\widehat{H},\widehat{\psi}({\bf r})\right]_{-}
\vert\Psi_{N,0}\rangle,\\
\label{eb10}
\Xi_{>}({\bf r},{\bf r}') &{:=}& \sum_s \theta(\varepsilon_s -\mu)
\varepsilon_s \Lambda_s({\bf r}) \Lambda_s^*({\bf r}')\nonumber\\
&=& -\langle\Psi_{N,0}\vert
\left[\widehat{H},\widehat{\psi}({\bf r})\right]_{-}
\widehat{\psi}^{\dag}({\bf r}')
\vert\Psi_{N,0}\rangle.
\end{eqnarray}
By applying the canonical anti-commutation relations for the field 
operators in the Schr\"odinger picture one arrives at the following 
result
\begin{equation}
\label{eb11}
G_{\infty_2}({\bf r},{\bf r}') = \hbar
\left\{\left[{{-\hbar^2}\over 2 m_e}\nabla^2 + u({\bf r})
+ v_H({\bf r};[n])\right] \delta({\bf r}-{\bf r}')
- {1\over 2} v({\bf r}-{\bf r}') \varrho({\bf r},{\bf r}')\right\}.
\end{equation}
The corresponding expression for the non-interacting Green function 
follows from this expression by setting the coupling constant of the 
electron-electron interaction equal to zero,
\begin{equation}
\label{eb12}
G_{0;\infty_2}({\bf r},{\bf r}') = \hbar
\left[{{-\hbar^2}\over 2 m_e}\nabla^2 + u({\bf r})\right]
\delta({\bf r}-{\bf r}').
\end{equation}

From Eqs.~(\ref{eb7}), (\ref{eb11}), (\ref{eb12}) and (\ref{eb1})
we obtain
the following result
\begin{equation}
\label{eb13}
\Sigma_{\infty_0} \equiv \Sigma^{HF},
\end{equation}
which completes our demonstration. We note that Eqs.~(\ref{eb3}) and 
(\ref{eb6}) hold equally for $\widetilde{G}(z)$ and $\widetilde{\Sigma}
(z)$, respectively, with merely $\varepsilon$ replaced by $z$.

\section{The Kramers-Kronig relations for the self-energy; applications
to isotropic Fermi and marginal-Fermi liquids}

The analyticity of a {\sl function} of the complex variable $z = x + iy$ 
in a domain ${\cal D}$ of the $z$-plane implies that in ${\cal D}$
its real (imaginary) part is up to a constant {\sl uniquely} determined 
by its imaginary (real) part (Morse and Feshbach 1953, pp. 356-358). For 
functions, such as ${\tilde f}(z)$, whose singularities are located along 
the $x$-axis and whose real and imaginary parts, ${\sf u}(x,y)$ and 
${\sf v}(x,y)$, approach zero not slower than $1/\varrho^p$, with $p > 0$, 
as $\varrho {:=} (x^2 + y^2)^{1/2} \to\infty$, the Kramers-Kronig relations 
establish the mentioned unique relationships between ${\sf u}(x,0^{\pm})$ 
and ${\sf v}(x,0^{\pm})$.

The self-energy $\widetilde{\Sigma}(z)$ is analytic over the entire
$z$-plane with the exception of the real axis.
\footnote{
\small
This follows from the Dyson equation $\widetilde{\Sigma}(z) =
\widetilde{G}_0^{-1}(z) -\widetilde{G}^{-1}(z)$ and the fact that neither 
$\widetilde{G}_0(z)$ nor $\widetilde{G}(z)$ possesses zero eigenvalues 
for ${\rm Im}(z)\not=0$ (see last paragraph in \S~II). }
Two aspects should be addressed before a Kramers-Kronig pair of relations 
can be set up for the self-energy. First, as the considerations in Appendix 
B have demonstrated, $\widetilde{\Sigma}(z)\sim \Sigma^{HF}$ for $\vert 
z\vert\to\infty$, so that we need first to introduce
\begin{equation}
\label{ec1}
\widetilde{\Sigma}_{\star}(z) {:=} \widetilde{\Sigma}(z) - \Sigma^{HF},
\end{equation}
which, following Eq.~(\ref{eb6}), approaches zero according to $\sim 
\Sigma_{\infty_1}/z$. Second, as $\widetilde{\Sigma}(z)$ is an operator, 
it cannot be assigned real and imaginary parts; moreover, as
$\widetilde{\Sigma}(z)$ is not gauge invariant, real and imaginary 
parts of the matrix elements $\{\langle\alpha\vert \widetilde{\Sigma}(z)
\vert\beta\rangle\}$, for a given representation, are dependent upon the 
choice of the gauge. We need therefore generalise for operators the 
notions `real part' and `imaginary part' before the corresponding 
Kramers-Kronig relations can be meaningfully defined. We introduce
\begin{equation}
\label{ec2}
\widetilde{\Sigma}'(z) {:=} {1\over 2} \big\{
\widetilde{\Sigma}(z) + \widetilde{\Sigma}^{\dag}(z) \big\},
\;\;\;
\widetilde{\Sigma}''(z) {:=} {1\over 2 i} \big\{
\widetilde{\Sigma}(z) - \widetilde{\Sigma}^{\dag}(z) \big\}.
\end{equation}
We similarly introduce $\widetilde{\Sigma}_{\star}'(z)$ and 
$\widetilde{\Sigma}_{\star}''(z)$; $\widetilde{\Sigma}'(z)$ and 
$i \widetilde{\Sigma}''(z)$ ($\widetilde{\Sigma}_{\star}'(z)$ and 
$i \widetilde{\Sigma}_{\star}''(z)$) are Hermitian and anti-Hermitian 
components of $\widetilde{\Sigma}(z)$ ($\widetilde{\Sigma}_{\star}(z)$)
and are unique. This {\sl uniqueness} enables us to deduce from 
\begin{equation}
\label{ec3}
\widetilde{\Sigma}_{\star}(\varepsilon\pm i\eta)
= \pm {1\over \pi i} {\cal P}\int_{-\infty}^{\infty}
{\rm d}\varepsilon'\; {{\widetilde{\Sigma}_{\star}(\varepsilon'\pm 
i\eta)}\over \varepsilon' -\varepsilon},
\end{equation}
which follows from the application of the Cauchy theorem together with 
the above-indicated analytic and asymptotic properties of 
$\widetilde{\Sigma}_{\star}(z)$, the following pair of representation-free 
Kramers-Kronig relations
\begin{eqnarray}
\label{ec4}
\Sigma_{\star}'(\varepsilon) &=& - {1\over\pi}
{\cal P}\int_{-\infty}^{\infty} {\rm d}\varepsilon'\;
{{\mbox{\rm sgn}(\mu -\varepsilon') \Sigma_{\star}''(\varepsilon')}
\over \varepsilon' -\varepsilon},\\  
\label{ec5}
\Sigma_{\star}''(\varepsilon) &=& + {1\over\pi}
{\cal P}\int_{-\infty}^{\infty} {\rm d}\varepsilon'\;
{{\mbox{\rm sgn}(\mu -\varepsilon) \Sigma_{\star}'(\varepsilon')}
\over \varepsilon' -\varepsilon}.
\end{eqnarray}
In obtaining these relations we have made use of our convention
in Eq.~(\ref{e5}). We note that $\Sigma'(\varepsilon)
\equiv \Sigma^{HF} + \Sigma_{\star}'(\varepsilon)$ and
$\Sigma''(\varepsilon)\equiv \Sigma_{\star}''(\varepsilon)$.

\subsection{Fermi liquids}

Now we employ Eq.~(\ref{ec4}) and obtain from
\footnote{\label{fc1}
\small
Note that $\varepsilon_F \equiv \mu_N$, $\alpha_k \geq 0$ and that 
this asymptotic relation is valid for $k$ in a neighbourhood of $k_F$, 
say for $0\leq k {\mathstrut^{\displaystyle {<} }_{\displaystyle 
{\sim} }} 3 k_F$ --- see footnote \ref{f11}.}
$\Sigma''(k;\varepsilon) \equiv {\rm Im}\Sigma(k;\varepsilon) \sim \mp 
\alpha_k (\varepsilon -\varepsilon_F)^2$, $\varepsilon {\mathstrut_{
\displaystyle {<} }^{\displaystyle >}} \varepsilon_F$, the associated 
$\Sigma'(k;\varepsilon)$; in this way we obtain Eq.~(\ref{e28}) as 
well as an explicit expression for $\beta_k$ in terms of $\alpha_k$ 
and $\Sigma''(k;\varepsilon)$. To this end, we subdivide the interval of 
the $\varepsilon'$-integration in Eq.~(\ref{ec4}) into three subintervals, 
$(-\infty,-\Delta]$, $(-\Delta,+\Delta]$ and $(+\Delta,+\infty)$, where we 
assume $\Delta > \vert\delta\varepsilon\vert$ with $\delta\varepsilon 
{:=} \varepsilon -\varepsilon_F$. Upon change of variables and transforming 
integrals over the negative $\varepsilon'$-axis into those over the positive 
$\varepsilon'$-axis we obtain
\begin{equation}
\label{ec6}
\Sigma_{\star}'(k;\varepsilon) \sim {{-2\alpha_k \delta\varepsilon}
\over \pi}\; {\cal P} \int_{0}^{\Delta} {\rm d}\varepsilon'\;
{{ {\varepsilon'}^2}\over {\varepsilon'}^2 - \delta\varepsilon^2}
+ {1\over\pi} \int_{\Delta}^{\infty} {\rm d}\varepsilon'\;
\Big\{ {{\Sigma''(k;\varepsilon'+\varepsilon_F)}\over \varepsilon'
-\delta\varepsilon} + {{\Sigma''(k;-\varepsilon'+\varepsilon_F)}\over
\varepsilon'+\delta\varepsilon} \Big\}.
\end{equation}
The first integral on the right-hand side of Eq.~(\ref{ec6}) is standard 
(Gradshteyn and Ryzhik 1965, p.~59) and one has ${\cal P}\int_0^{\Delta} 
{\rm d}\varepsilon'\; {\varepsilon'}^2/({\varepsilon'}^2 - 
\delta\varepsilon^2) = \Delta - (\delta\varepsilon/2) \ln\left| 
(\Delta +\delta\varepsilon)/(\Delta-\delta\varepsilon)\right|$. Since, 
in the second integral on the right-hand side of Eq.~(\ref{ec6}), 
$\varepsilon'\ge\Delta$ and since $\Delta > \vert\delta\varepsilon\vert$, 
we replace $1/(\varepsilon'-\delta\varepsilon)$ and $1/(\varepsilon'
+\delta\varepsilon)$ by their respective geometric series expansions,
in powers of $\delta\varepsilon/\varepsilon'$, and thus obtain a powers 
series in $\delta\varepsilon$ for the second integral. Collecting terms 
of the zeroth and the first order in $\delta\varepsilon$, while making 
use of Eq.~(\ref{e28}), we obtain
\begin{equation}
\label{ec7}
\Sigma(k;\varepsilon_F) \equiv \Sigma'(k;\varepsilon_F)
\cong \Sigma^{HF}(k) + S_0^{(\Delta)}(k),\;\;\;\;\;\;
\beta_k \cong S_1^{(\Delta)}(k) - {2\Delta\alpha_k\over\pi},
\end{equation}
where
\begin{equation}
\label{ec8}
S_m^{(\Delta)}(k) {:=} {1\over\pi} \int_{\Delta}^{\infty}
{ {{\rm d}\varepsilon'}\over {\varepsilon'}^{m+1}}\;
\Big\{ \Sigma''(k;\varepsilon'+\varepsilon_F)
+ (-1)^m\; \Sigma''(k;-\varepsilon'+\varepsilon_F)\Big\}.
\end{equation}
In Eq.~(\ref{ec7}), `$\cong$' signifies the approximate nature of the 
results in so far as $\Delta$ is finite and, moreover, in evaluating the 
$\varepsilon'$-integral over $(-\Delta,+\Delta]$ we have employed a 
truncated asymptotic expansion for $\Sigma''(k;\varepsilon)$. As should 
be evident, our choice for a finite $\Delta$ has it root in the requirement 
$\Delta > \vert\delta\varepsilon\vert$. Since Eqs.~(\ref{ec7}) and 
(\ref{ec8}) are valid independently of the value of $\delta\varepsilon$, 
it is possible to let $\Delta$ in these equations approach zero; because 
$\Sigma''(k;\pm\varepsilon'+\varepsilon_F) {\sim'} {\varepsilon'}^2$ for 
$\varepsilon'\to 0$, $S_{m}^{(\Delta\downarrow 0)}(k)$, which involves 
$1/{\varepsilon'}^{m+1}$, exists for $m=0,1$, and is equal to 
$S_{m}^{(0)}(k)$; on the other hand, for $m \geq 2$, $S_{m}^{(\Delta)}(k)$ 
does {\sl not} have a limit for $\Delta\downarrow 0$. Thus we can write
\begin{equation}
\label{ec9}
\Sigma(k;\varepsilon_F) \equiv \Sigma'(k;\varepsilon_F) =
\Sigma^{HF}(k) + S_0^{(0)}(k),\;\;\;\;\;\;
\beta_k = S_1^{(0)}(k).
\end{equation}
We point out that, since, for $\varepsilon' \geq 0$, $\Sigma''(k;\varepsilon'
+\varepsilon_F) \leq 0$ and $\Sigma''(k;-\varepsilon'+\varepsilon_F)
\geq 0$,
\footnote{\label{fc2}
\small
Violation of these inequalities implies breakdown of causality
and instability of the ground state of the system (see Eq.~(\ref{e25})
in comparison with Eq.~(\ref{e1})).}
it follows that $S_m^{(\Delta)}(k) \leq 0$ for {\sl all} odd values
of $m$. Specifically, following Eq.~(\ref{ec9}), we have $\beta_k \leq 0$
(concerning the range of $k$ over which this result can be relied upon,
see footnotes \ref{f11} and \ref{fc1}). As we have indicated in \S~V (see 
text following Eq.~(\ref{e28})), for $k\to k_F$ this inequality guarantees 
$Z_{k_F} \leq 1$.

It is interesting to note that $S_{m}^{(\Delta)}(k)$, $m=0,1$, can be 
written in the following appealing form
\begin{eqnarray}
\label{ec10}
S_0^{(\Delta)}(k) &{:=}& {1\over\pi}\; {\rm Im} 
\Big[ \int_{\mu+\Delta}^{i\infty}
{ {{\rm d} z}\over (z-\mu)}\;
\big\{ \widetilde{\Sigma}(k;z) + \widetilde{\Sigma}(k;2\mu-z) 
- 2 \Sigma^{HF}(k)\big\} \Big],\\
\label{ec11}
S_1^{(\Delta)}(k) &{:=}& {1\over\pi}\; {\rm Im} 
\Big[ \int_{\mu+\Delta}^{i\infty}
{ {{\rm d} z}\over (z-\mu)^2}\;
\big\{ \widetilde{\Sigma}(k;z) 
-\widetilde{\Sigma}(k;2\mu-z)\big\} \Big],
\end{eqnarray}
where the term $2\Sigma^{HF}(k)$, which is real-valued, has been 
introduced in order to render the corresponding integral existent (see 
Eq.~(\ref{eb13})). In Eqs.~(\ref{ec10}) and (\ref{ec11}) we have further 
restored $\mu$ which originates from Eq.~(\ref{ec4}) and which is {\sl 
not} exactly equal to $\varepsilon_F$ (see \S~IV.A). We point out that 
in these expressions, contrary to that in Eq.~(\ref{ec8}), $\Delta$ {\sl 
cannot} be set equal to zero, since for $\Delta=0$ the integrals enclosed 
by square brackets (logarithmically) diverge as a consequence of the 
integrands being proportional to $1/(z-\mu)$ for $z\to\mu$; since, 
however, the singular contributions are real-valued, $\Delta$ can be 
made as small as desired, provided that it is kept non-zero; thus 
$S_m^{(0^+)}(k)$, $m=0,1$, is well-defined. 

The relevance of Eqs.~(\ref{ec10}) and (\ref{ec11}) rests in the following: 
The fact that $\widetilde{\Sigma}(k;z)$ is {\sl analytic} over the entire 
$z$-plane (excluding the real axis outside $(\mu_N,\mu_{N+1})$), implies 
that if $\partial^{\ell} S_m^{(0^+)}(k)/\partial k^{\ell}$, $m=0,1$ 
(including $\ell=0$), is singular at some $k$, this singularity must be 
shared by the integrand of $\partial^{\ell} S_m^{(0^+)}(k)/\partial 
k^{\ell}$, $m=0,1$, for {\sl all} $z$, ${\rm Im}(z)\not=0$. In other 
words, Eqs.~(\ref{ec10}) and (\ref{ec11}) make explicit that possible 
singularities in $\partial^{\ell} S_m^{(0^+)}(k)/\partial k^{\ell}$, 
$m=0,1$, are {\sl not} specific to $\widetilde{\Sigma}(k;z)$ for some 
$z=z_0$, but to $\widetilde{\Sigma}(k;z)$ for {\sl all} $z$, 
${\rm Im}(z)\not=0$.
\footnote{\label{fc3}
\small
Consider $g(k) {:=} \int_{\cal C} {\rm d} z\; \widetilde{f}(k;z)$, where 
the contour of integration ${\cal C}$ lies inside an open domain ${\cal D}$ 
of the complex $z$-plane where $\widetilde{f}(k;z)$ is {\sl analytic}. 
Suppose that the $\ell$th-order derivative (including $\ell = 0$) with 
respect to $k$ of $g(k)$ is divergent at $k=k_0$. This divergence {\sl 
cannot} be due to some {\sl isolated} singularity (singularities) of 
$\partial^{\ell}\widetilde{f}(k;z)/ \partial k^{\ell}\vert_{k=k_0}$ 
along ${\cal C}$, for {\sl analyticity} of $\widetilde{f}(k;z)$ enables 
one to deform ${\cal C}$ inside ${\cal D}$, thus avoiding the mentioned 
singularity (singularities), without changing the value of the integral 
(Cauchy's theorem); exceptions to this concern the cases where 
$\partial^{\ell} \widetilde{f}(k;z)/\partial k^{\ell}\vert_{k=k_0}$ is 
divergent either at the end-points of ${\cal C}$, which cannot be 
dislocated, or at two neighbouring points which `pinch' ${\cal C}$. Note 
that in Eqs.~(\ref{ec10}) and (\ref{ec11}) the contours of integration 
are arbitrary, as long as they connect $\mu+\Delta$ with the point of 
infinity (the `end-points') and are located on the upper-half of the 
$z$-plane (signifying ${\cal D}$). Therefore, excluding end-point and 
`pinch' singularities, divergence of $\partial^{\ell} g(k)/\partial 
k^{\ell}$ at $k=k_0$ implies divergence of $\partial^{\ell} 
\widetilde{f}(k;z)/\partial k^{\ell}$ at $k=k_0$ for {\sl all} $z$ inside 
${\cal D}$. In Eqs.~(\ref{ec10}) and (\ref{ec11}) one of the end-points, 
namely the point of infinity, is harmless, as the integrands in both 
expressions are vanishing at this point; by choosing $\mu=(\mu_N 
+\mu_{N+1})/2$ and $\Delta=(\mu_{N+1}-\mu_N)/2$, owing to the assumption 
of continuous-differentiability with respect to $k$ of $\Sigma(k;\mu_N)$ 
and $\Sigma(k;\mu_{N+1})$ in a neighbourhood of $k=k_F$, a possible 
divergence of $S_m^{(0^+)}(k)$ or $\partial S_m^{(0^+)} (k)/\partial k$, 
$m=0,1$, at $k=k_F$ {\sl cannot} be ascribed to an end-point singularity.}

In order to demonstrate the utility of our above considerations, let us 
assume $\Sigma^{HF}(k)$, similar to $\Sigma_{\sf s}^{HF}(k)$ (see
Eq.~(\ref{eb2})), has a divergent first-order derivative with respect 
to $k$ at $k=k_F$ (see Appendix B). For the system under consideration 
to be a Fermi liquid, $\Sigma(k;\varepsilon_F)$ must, by definition, be a 
continuously differentiable function of $k$ at $k=k_F$ (see \S\S~I and 
IV.C). From Eq.~(\ref{ec9}) it follows that the assumed singularity due 
to $\partial\Sigma^{HF}(k)/\partial k$ at $k=k_F$ must therefore be 
compensated by a counter contribution arising from $S_0^{(0^+)}(k)$. With 
reference to our above discussions, in such an event $\partial
\{\widetilde{\Sigma}(k;z) + \widetilde{\Sigma}(k;2\mu-z) - 
2\Sigma^{HF}(k)\}/\partial k$ must be singular at $k=k_F$ for {\sl all} 
$z$, ${\rm Im}(z) \not=0$ (see specifically footnote \ref{fc3}). For the 
special choice of $z=\mu +\Delta$, with $\mu=(\mu_N+\mu_{N+1})/2$ and 
$\Delta =(\mu_{N+1}-\mu_N) /2$, $\widetilde{\Sigma}(k;z)$ and 
$\widetilde{\Sigma}(k;2\mu - z)$ are continuously differentiable functions 
of $k$ in a neighbourhood of $k=k_F$ (by the Fermi-liquid assumption); 
nonetheless, in accordance with our expectation $\partial
\{\widetilde{\Sigma}(k;z) + \widetilde{\Sigma}(k;2\mu-z) 
-2\Sigma^{HF}(k)\}/\partial k$ remains divergent at $k=k_F$ owing to the 
assumed divergence of $\partial \Sigma^{HF}(k)/\partial k$ at $k=k_F$. 
On the other hand, for an arbitrary $z$ the continuous differentiability 
with respect to $k$ of $\widetilde{\Sigma}(k;z)$ and $\widetilde{\Sigma}
(z;2\mu-z)$ is not guaranteed, so that divergence of $\partial 
\{\widetilde{\Sigma}(k;z) + \widetilde{\Sigma}(k;2\mu-z) - 
2\Sigma^{HF}(k)\}/\partial k$ at $k=k_F$ can be in part due to that of 
$\partial\{\widetilde{\Sigma}(k;z) + \widetilde{\Sigma}(k;2\mu-z) \}/
\partial k$ at $k=k_F$. In view of the asymptotic relation in 
Eq.~(\ref{e29}) (see footnote \ref{f25a}), it follows that $\beta_k$ and 
$\alpha_k$ may {\sl not} be necessarily continuously differentiable at 
$k=k_F$. The divergence of $\partial \beta_k/\partial k$ as $k\to 
k_F^{\mp}$ gives rise to a divergent $\partial {\sf n}_s(k)/\partial k$, 
as well as a divergent $\partial {\sf n}(k)/\partial k$, at $k=k_F^{\mp}$ 
(for a discussion of this case see \S~VI). Since for Fermi liquids (here 
as characterised by $Z_{k_F}\not= 0$), a divergent $\partial {\sf n}(k)/
\partial k$, at $k=k_F^{\mp}$, has proved possible (Belyakov 1961, Sartor 
and Mahaux 1980), we observe that in general even for Fermi liquids 
$\Sigma(k;\varepsilon)$ is {\sl not} a continuously differentiable
function of $k$ (in a neighbourhood of $k=k_F$) for $\varepsilon \not=
\varepsilon_F^{\mp}$; at $\varepsilon= \varepsilon_F^{\mp}$, $\beta_k$ 
and $\alpha_k$ are excluded from contributing to $\Sigma(k;\varepsilon)$ 
(see Eq.~(\ref{e29})).

\subsection{Marginal Fermi liquids}

Having considered the case of isotropic Fermi liquids in considerable
detail, below we briefly deal with the case of the isotropic marginal 
Fermi liquids. 

For marginal Fermi liquids we have $\Sigma''(k;\varepsilon) \equiv {\rm 
Im}\Sigma(k;\varepsilon) \sim -\alpha_k (\varepsilon -\varepsilon_F)$, as
$\varepsilon\to \varepsilon_F$, with $\alpha_k \geq 0$ (see \S~V). For the 
$\Sigma_{\star}'(k;\varepsilon)$ corresponding to this $\Sigma''(k;
\varepsilon)$ one obtains an expression which except for the first term 
is identical with that presented in Eq.~(\ref{ec6}). For this first term, 
which we denote by $\Sigma_{\star;1}'(k;\varepsilon)$, we have (as in the 
Fermi-liquid case, below $\Delta > \vert\delta\varepsilon\vert$)
\begin{equation}
\label{ec12}
\Sigma_{\star;1}'(k;\varepsilon) {:=} {{-2\alpha_k\delta\varepsilon}
\over\pi} {\cal P}\int_0^{\Delta} {\rm d}\varepsilon'\;
{{\varepsilon'}\over {\varepsilon'}^2 - \delta\varepsilon^2}
\equiv {{-2\alpha_k\delta\varepsilon}
\over\pi} \Big\{ \ln(\Delta) - \ln\vert\delta\varepsilon\vert
- {1\over 2\Delta^2}\;\delta\varepsilon^2\Big\};
\end{equation}
(see Gradshteyn and Ryzhik 1965, p.~59).
One observes that here, contrary to the Fermi-liquid case, the limit 
$\Delta \downarrow 0$ cannot be taken. This is related to the fact that 
here $\Sigma''(k;\pm\varepsilon'+\varepsilon_F) {\sim'} \varepsilon'$ for
$\varepsilon'\to 0$, so that {\sl only} $S_0^{(\Delta)}(k)$ has a limit 
for $\Delta \downarrow 0$. Therefore, in the present case, only the first 
expression in Eq.~(\ref{ec9}) is meaningful. Further, from Eq.~(\ref{ec12}) 
we observe that the {\sl leading} asymptotic contribution to 
$\Sigma_{\star}'(k;\varepsilon)$, as $\varepsilon\to\varepsilon_F$, 
involves $\ln\vert\delta\varepsilon\vert$ which is singular for 
$\delta\varepsilon=0$. Combining
the above results, we have $\Sigma(k;\varepsilon) \sim \Sigma(k;
\varepsilon_F) + \beta_k (\varepsilon -\varepsilon_F) \ln\vert\varepsilon 
-\varepsilon_F\vert + [\gamma_k + i\alpha_k]\; (\varepsilon-\varepsilon_F)$, 
as $\varepsilon\to\varepsilon_F$, where
\begin{eqnarray}
\label{ec13}
\beta_k &\equiv& {{2\alpha_k}\over\pi},\\
\label{ec14}
\gamma_k &\cong&  -{{2\alpha_k}\over\pi} \ln(\Delta)
+ S_1^{(\Delta)}(k).
\end{eqnarray}
The first term on the right-hand side of Eq.~(\ref{ec14}) cancels the 
logarithmically divergent contribution arising from the second term as 
$\Delta\downarrow 0$. It can be readily verified that to the order in 
which $-\alpha_k (\varepsilon -\varepsilon_F)$ is an exact representation 
of $\Sigma''(k;\varepsilon)$ for $\varepsilon\to\varepsilon_F$, $\partial 
\gamma_k/\partial \Delta = 0$. This invariance property correctly reflects 
the fact that $\gamma_k$ must not depend on the cut-off energy $\Delta$.

\section{Single-particle Green function and Self-energy of the Luttinger 
model; some asymptotic expressions}

In this Appendix we consider the single-particle Green function and the 
self-energy pertaining to the Luttinger model (Luttinger 1963, Mattis 
and Lieb 1965) for spin-less fermions in some detail. 
\footnote{\label{f26a}
\small
We note that an explicit expression for the single-particle Green function 
for the Luttinger model with $\delta$-function interaction in real space 
(this model is equivalent with the $1+1$ Thirring (1958) model, dealt with 
by Johnson (1961)) has been evaluated by Theumann (1967). The corresponding 
expression in Eq.~(2.19)$_T$ (here subscript $T$ denotes an equation from 
Theumann's work) turns out to be identically vanishing for the case of 
non-interacting Luttinger model. To verify this, note that the coupling 
constant $\lambda$ of interaction as presented in Eq.~(1.4)$_T$, determines 
$\Omega$ in Eq.~(2.3)$_T$ which in turn fixes $\beta$ in Eq.~(2.14)$_T$. 
It is seen that for non-interacting particles, corresponding to $\lambda=0$
and thus $\beta = 1$, $G_1(k,\omega)$ is identically vanishing, owing to 
$\sin\pi\beta$ in Eq.~(2.19)$_T$. We have not attempted to identify the 
origin of this shortcoming of $G_1(k,\omega)$ in Eq.~(2.19)$_T$. } 
From the expression for the former function we can {\sl explicitly} test 
the validity of one of our findings in Appendix C, namely that when $\partial
\widetilde{\Sigma}(k;z)/\partial k$ is divergent at $k=k_0$ and $z=z_0$, 
it is divergent at $k=k_0$ for {\sl all} $z$. Further, we consider the 
asymptotic behaviour of both $G(k;\varepsilon)$ and $\Sigma(k;\varepsilon)$ 
for $\varepsilon$ in the close vicinity of $\varepsilon_F$. In doing so we 
separately deal with the case where $k=k_F$ and where $k\not=k_F$. In 
order to remain close to the available literature concerning the Luttinger 
model, in this Appendix we adopt the commonly-used notations and units 
and therefore in these deviate from the other parts of the present work.
We specify the new notations as we proceed through this Appendix.
In this Appendix $\hbar=1$. 

\subsection{The single-particle Green function and its 
derivative with respect to momentum}

We first construct the single-particle Green function $G$ from the 
spectral function $\rho$ corresponding to the retarded part $G^R$ of the 
Green function. For this spectral function we have (Voit 1993b, 1993a)
\begin{eqnarray}
\label{ed1}
\rho_r(q,\omega) &=& {\Lambda\over 2 v_0 \Gamma^2(\gamma_0)}
\Theta(\omega+r v_0 q) \Theta(\omega-r v_0 q)
\gamma\Big(\gamma_0,{\Lambda\over 2 v_0} (\omega+ v_0 r q)\Big) 
\nonumber\\
& &\;\;\;\times \Big({\Lambda\over 2 v_0} (\omega- v_0 
r q)\Big)^{\gamma_0 -1} \exp\Big(-{\Lambda\over 2 v_0}
(\omega - v_0 r q)\Big) 
\nonumber\\
& &\;\;\;\;\;\;\; + \big(\omega \to -\omega, q \to -q\big),
\;\;\; \mbox{\rm for}\;\; \gamma_0 \not= 0,\\
\label{ed2}
\rho_r(q,\omega) &\equiv& \rho_{0;r}(q,\omega)
{:=} \delta(\omega - v_0 r q),
\;\;\; \mbox{\rm for}\;\; \gamma_0 = 0.
\end{eqnarray}
Here $r = \mp$ specifies the left and right branches respectively
of the single-particle spectrum in the Luttinger model, $\omega \equiv 
\varepsilon -\mu$ and $k \equiv q + r k_F$. It is important to note that 
contrary to the 2D and 3D cases, {\sl here $k$ and $q$ take on both 
positive and negative values. Moreover $q$ is measured with respect to 
\footnote{
\small
Thus a better notation for $q$ would be $q_r$ in order to emphasise this 
$r$-dependence of the origin.}
$\, r k_F$; $k$, on the other hand, is measured with respect to the origin.} 
Further, $\Lambda$ stands for the (finite) cut-off on the range of the 
interaction in the momentum space --- the precise value of $\Lambda$ is not 
of relevance here (S\'olyom 1979; see \S\S~9 and 2 herein); $\Gamma(z)$ 
stands for the Gamma function (Abramowitz and Stegun 1972, p.~255), 
$\gamma_0 \equiv \alpha/2$, with $\alpha$ the `anomalous dimension' (see 
\S~IV.C) and $\gamma(a,z)$ denotes the incomplete Gamma function 
(Abramowitz and Stegun 1972, p.~260). Unless we explicitly indicate 
otherwise, below $0 < \alpha < 1$ and thus $0 < \gamma_0 < 1/2$. With 
\begin{equation}
\label{ed3}
\rho_r(q,\omega) {:=} -{1\over\pi}\; {\rm Im}G_r^R(q,\omega),
\end{equation}
where the $G_r^R$ denotes the retarded Green function, making use of the 
fact that ${\rm Im}G_r(q,\omega) = {\rm sgn}(\omega)\; {\rm Im}G_r^R(q,
\omega)$, we have
\begin{equation}
\label{ed4}
G_r(q,\omega) = \int_{-\infty}^{\infty}
{\rm d}\omega'\; {{\rho_r(q,\omega')}
\over \omega -\omega' +i \eta\; {\rm sgn}(\omega') }, \;\;\;
\eta\downarrow 0.
\end{equation}
From Eq.~(\ref{e25}), introducing $\omega_{r;q}^0 {:=} r v_0 q$, we 
obtain (compare with Eq.~(\ref{e32}))
\footnote{\label{f27c}
\small
It is readily verified that $\widetilde{G}_r(q,z) \equiv
\int_{-\infty}^{\infty} {\rm d}\omega'\; \rho_r(q,\omega')/(z - \omega')$, 
for ${\rm Im}(z)\not= 0$; in view of Eq.~(\ref{e5}), $G_r(q,\omega) = 
\lim_{\eta\downarrow 0} \widetilde{G}_r(q,\omega \pm i \eta)$, for 
$\omega {\mathstrut_{\displaystyle {<} }^{\displaystyle >}} 0$, which can 
be easily shown to coincide with that presented in Eq.~(\ref{ed4}). 
We point out that, with $\rho_{0;r}(q,\omega) = \delta(\omega - 
\omega_{r;q}^0)$ (see Eq.~(\ref{ed2})), one obtains $G_{0;r}(q,\omega) 
= 1/[\omega-\omega_{r;q}^0 + i\eta\, {\rm sgn}(\omega_{r;q}^0)]$ and 
consequently $\widetilde{G}_{0;r}(q,z) = 1/[z - \omega_{r;q}^0]$, for 
${\rm Im}(z)\not=0$. }
\begin{equation}
\label{ed5}
{{\partial\widetilde{G}_r(q,z)}\over\partial q}
= \widetilde{G}_r^2(q,z)\;
\Big\{ {{\partial \omega_{r;q}^0}\over\partial q} 
+ {{\partial\widetilde{\Sigma}_r(q,z)}\over\partial q} \Big\}.
\end{equation}
Since $\widetilde{G}_r(q,z)$ is an entire function of $z$ for all $z$, 
${\rm Im}(z)\not=0$ (see \S~II) and, since $\partial \omega_{r;q}^0/
\partial q \equiv r v_0$ is finite, any possible divergence of $\partial
\widetilde{G}_r (q,z)/\partial q$, as function of $q$ and $z$ (${\rm Im}
(z)\not=0$), must be due to a divergence of $\partial \widetilde{\Sigma}_r
(q,z)/\partial q$ on the right-hand side of Eq.~(\ref{ed5}). In order to 
be able to evaluate $\partial \widetilde{G}_r(q,z)/\partial q$, which 
according to Eq.~(\ref{ed4}) involves $\partial\rho_r(q,\omega)/\partial 
q$, some changes in the expression on the right-hand side of Eq.~(\ref{ed1}) 
are necessary. To appreciate this, note that $\partial\Theta (\omega - r 
v_0 q)/\partial q = -r v_0 \delta(\omega - r v_0 q)$, from which, on 
account of the term $(\omega -r v_0 q)^{\gamma_0-1}$, one immediately 
observes that independent of the value of $q$, $\partial \rho_r(q,\omega)/
\partial q = \infty$. This would imply a peculiar situation where $\rho_r
(q,\omega)$, which is continuous almost everywhere, would be {\sl nowhere} 
on the $q$-axis differentiable. As we shall see, this is wholly attributable 
to the fact that the expressions on the right-hand side of Eq.~(\ref{ed1}) 
which are multiplied by {\sl distributions} $\Theta(\omega \mp r v_0 q)$ 
and $\Theta(-\omega \pm r v_0 q)$ do {\sl not} qualify as {\sl test 
functions} (Gelfand and Shilov 1964).  For simplicity, in the following 
we shall explicitly deal with the case corresponding to $r=+$. We write 
\begin{eqnarray}
\label{ed6}
\rho_+(q,\omega) &\equiv & \Theta(\omega + v_0 q) \Theta(\omega - v_0 q)
\phi(q,\omega) + (\omega \to -\omega, q\to -q),\\
\label{ed7}
\phi(q,\omega) &{:=}& {\cal A}_1\, \phi_1(\omega+v_0 q)\, 
\phi_2(\omega- v_0 q),\\
\label{ed8}
{\cal A}_{\sigma} &{:=}& {1\over \Gamma^2(\gamma_0)} 
\Big({\Lambda\over 2 v_0}\Big)^{\sigma},\\
\label{ed9}
\phi_1(x) &{:=}& \gamma\Big(\gamma_0,{\Lambda\over 2 v_0} x\Big),\\
\label{ed10}
\phi_2(x) &{:=}& \Big({\Lambda\over 2 v_0} x\Big)^{\gamma_0 -1}
\exp\Big(-{\Lambda\over 2 v_0} x\Big).
\end{eqnarray}
Since $0 <\gamma_0 < 1/2$, it is evident that the problem at hand (i.e.
$\partial\rho_+(q,\omega)/\partial q =\infty$) has its origin in the 
singularity of $\phi_2(x)$ at $x=0$. Through application of the 
prescription as specified in Fig.~3 (see caption to this Figure), we 
define $\widetilde{G}_{+,\nu}(q,z)$ and consider $\widetilde{G}_+(q,z)$ as 
being obtained through $\lim_{\nu\downarrow 0}\widetilde{G}_{+,\nu}(q,z)$. 
We have
\begin{eqnarray}
\label{ed11}
\widetilde{G}_{+,\nu}(q,z) &\equiv&
\int_{-\infty}^{-v_0 q} {\rm d}\omega'\;
{\phi(-q,-\omega')\over z -\omega'} 
+\int_{v_0 q + \nu}^{\infty} {\rm d}\omega'\;
{\phi(q,\omega')\over z -\omega'}\nonumber\\
& &+\int_{v_0 q -\nu}^{v_0 q} {\rm d}\omega'\;
{\phi(q,v_0 q +\nu) \big[1 + (\omega' - v_0 q)/\nu\big]
\over z - \omega'}
+ \int_{v_0 q}^{v_0 q +\nu} {\rm d}\omega'\;
{\phi(q, v_0 q +\nu)\over z -\omega'}.
\end{eqnarray} 
It is from this expression that the derivative with respect to $q$ has 
to be taken; we then have $\partial\widetilde{G}_+(q,\omega)/\partial q 
{:=} \lim_{\nu\downarrow 0}\partial\widetilde{G}_{+,\nu}(q,\omega)/
\partial q$. After some algebra, one obtains the following final result
\begin{eqnarray}
\label{ed12}
{\partial\widetilde{G}_+(q,z)\over\partial q} =
&-& v_0 {\phi(-q,v_0 q)\over z + v_0 q}
+ \int_{-\infty}^{-v_0 q} {\rm d}\omega'\;
{\partial\phi(-q,-\omega')/\partial q\over z -\omega'}
\nonumber\\
&+& {\cal A}_1 v_0 \int_{v_0 q}^{\infty} {\rm d}\omega'\;
{ \{\partial\phi_1(\omega'+v_0 q)/\partial\omega'\}\, 
\phi_2(\omega'-v_0 q)\over z - \omega'}\nonumber\\
&+& {\cal A}_1 v_0 \int_{v_0 q}^{\infty} {\rm d}\omega'\;
\phi_2(\omega'- v_0 q)\, {\partial\over\partial\omega'}
\Big\{ {\phi_1(\omega'+v_0 q)\over z -\omega'}\Big\}.
\end{eqnarray}
In arriving at this expression, we have made use of the properties 
$\partial\phi_{1,2}(\omega'\pm v_0 q)/\partial q = \pm v_0 
\partial\phi_{1,2}(\omega'\pm v_0 q)/\partial\omega'$.
It is evident that the integrands of the last two integrals on the 
right-hand side of Eq.~(\ref{ed12}) are singular (owing to $\phi_2
(\omega'-v_0 q)$) at the lower limits of the integration boundaries; 
however since $0 < \gamma_0 < 1/2$, these singularities are integrable 
and therefore $\partial\widetilde{G}_+(q,z)/\partial q$ as 
presented in Eq.~(\ref{ed12}) is a well-defined function of $q$ and $z$. 

We are now in a position to draw the important conclusion that since 
$\phi(-q,v_0 q) = {\cal A}_1 \phi_1(0)\phi_2(2 v_0 q)\propto q^{\gamma_0 
-1}$, $\partial\widetilde{G}_+(q,z)/\partial q$ diverges as $q\downarrow
0$, for {\sl all} $z$. This is in full conformity with our statement 
in Appendix C that, when $\partial\widetilde{G}_+(q,z)/\partial q$ is 
unbounded at a particular value of $q$, it is unbounded at that $q$ for 
{\sl all} $z$.

\subsection{Self-energy and its asymptotic forms close to the Fermi 
points and the Fermi energy}

Now we proceed with evaluating the asymptotic behaviour of $\Sigma_r(q,
\omega)$ for $\omega\to 0$. We consider two cases, corresponding to $q=0$ 
and $q\not=0$. We demonstrate that, as $\gamma_0$ approaches zero, 
the self-energy of the Luttinger (1963) model becomes vanishingly small
in a manner that is specific to non-interacting Fermi systems.

Since we are interested in cases where $0 < \gamma_0 < 1/2$, for the 
determination of the {\sl leading} asymptotic contribution 
to $G_r(q,\omega)$, and thus to $\Sigma_r(q,\omega)$, for $q\to 0$ (i.e. 
$k\to k_F$) and $\omega\to 0$ (i.e. $\varepsilon\to\varepsilon_F$) we need 
to merely take account of the $\omega'$-integration in Eq.~(\ref{ed4}) 
restricted to the interval $[-\Delta,\Delta]$, where $\Delta$ is some 
finite constant satisfying $\vert \omega\vert < \Delta$ (see Appendix C). 
Thus we define ({\sl cf.} Eq.~(\ref{ed4}))
\begin{equation}
\label{ed13}
\widetilde{G}_r^{(\Delta)}(q,z) {:=}
\int_{-\Delta}^{\Delta} {\rm d}\omega'\;
{\rho_r(q,\omega')\over z -\omega'}.
\end{equation}
Below we shall deal with $G_r^{(\Delta)}(q,\omega)$. 

\subsubsection{The $q=0$ case} 

From the expression in Eq.~(\ref{ed1}) it can readily be deduced
that (see Eqs.~(\ref{ed6})-(\ref{ed10}))
\begin{equation}
\label{ed14}
\rho_r(q=0,\omega) \sim {\cal A}_{\gamma_0+1}\, 
\vert\omega\vert^{2\gamma_0-1},\;\;\;
\mbox{\rm for}\;\; \omega\to 0.
\end{equation}
Making use of the general result ($\vert\omega\vert < \Delta$)
\begin{eqnarray}
\label{ed15}
{\cal P}\int_0^{\Delta} {\rm d}\omega'\; 
{ {{\omega'}^{\sigma}}\over {\omega'}^2 - \omega^2}
&=& {\pi\over 2} \tan\left({{\pi\sigma}\over 2}\right)\;
\vert\omega\vert^{\sigma -1} 
+ {{\Delta^{\sigma-1}}\over \sigma -1}\;
{}_2F_1\left(1,{{1-\sigma}\over 2};{{3-\sigma}\over 2};
{{\omega^2}\over\Delta^2}\right)\nonumber\\
&=& {\pi\over 2} \tan\left({{\pi\sigma}\over 2}\right)\;
\omega^{\sigma -1}
+ {\Delta^{\sigma-1}\over \sigma -1}
+ {\Delta^{\sigma-3}\over \sigma-3}\; \omega^2
+ {\cal O}(\omega^4),\nonumber\\
& &\;\;\;\;\;\;\;\;\;\;\;\;\;\;\;\;\;\;\;\;\;\;\;\;\;\;\;\;\;\;\;\;\; 
\sigma > -1,\;\sigma\not= 1,3,\dots, 
\end{eqnarray}
where ${}_2F_1(a,b;c;z)$ stands for the Gauss Hypergeometric
function (Abramowitz and Stegun 1972, p.~556), for $0 <\gamma_0 < 1$ 
and $\vert\omega\vert < \Delta$ we obtain 
\begin{equation}
\label{ed16}
G_r^{(\Delta)}(q=0,\omega)
\sim -\pi {\cal A}_{\gamma_0+1}
\Big\{ \tan\Big({\pi (2\gamma_0-1)\over 2}\Big)
+ i\Big\}\,{\rm sgn}(\omega) \vert\omega\vert^{2\gamma_0-1}.
\end{equation}
Evidently, for $0 < \gamma_0 < 1/2$, the right-hand side of this 
expression diverges as $\omega\to 0$. As a consequence, for $0 <\gamma_0 
< 1/2$, the possibly non-vanishing constant $\{G_r(q=0,\omega) 
-G_r^{(\Delta)}(q=0,\omega)\}$ is asymptotically irrelevant for $\omega
\to 0$ and we can write
\begin{equation}
\label{ed17} 
G_r(q=0,\omega) 
\sim \pi {\cal A}_{\gamma_0+1}
\Big\{ \cot(\pi\gamma_0)
- i\Big\}\,{\rm sgn}(\omega) \vert\omega\vert^{2\gamma_0-1},
\;\; \omega\to 0\;\;\; (0 < \gamma_0 < 1/2).
\end{equation}

From the Dyson equation we have
\begin{equation}
\label{ed18}
\Sigma_r(q,\omega) = G_{0;r}^{-1}(q,\omega) - G_r^{-1}(q,\omega).
\end{equation}
Since $G_{0;r}(q,\omega) = 1/[\omega - v_0 r q + i\, {\rm sgn}
(v_0 r q)]$ (see footnote \ref{f27c}), it follows that
\begin{equation}
\label{ed19}
G_{0;r}^{-1}(q,\omega) = \omega - v_0 r q,
\end{equation}
and thus $G_{0;r}^{-1}(q=0,\omega) = \omega$. For $0 <\gamma_0 < 1$, the 
leading-order contribution to $\Sigma_r(q=0,\omega)$ is therefore entirely 
due to $G_r^{-1}(q=0,\omega)$ on the right-hand side of Eq.~(\ref{ed18}). 
From Eq.~(\ref{ed16}) we thus obtain ($0 < \gamma_0 < 1/2$)
\begin{equation}
\label{ed20}
G_r^{-1}(q=0,\omega) \sim
{1\over \pi {\cal A}_{\gamma_0+1} \big(\cot^2(\pi\gamma_0)+1\big)}
\Big\{ \cot(\pi\gamma_0) + i\Big\}\, {\rm sgn}(\omega)
\vert\omega\vert^{1-2\gamma_0}.
\end{equation}
As a consequence of our above arguments, for $0 < \gamma_0 < 1/2$ we 
have (for $\gamma_0$ in this range, $\cot(\pi\gamma_0) > 0$)  
\begin{eqnarray}
\label{ed21}
\Sigma_r'(q=0,\omega) &\sim& {-\cot(\pi\gamma_0) {\rm sgn}(\omega)\over
\pi {\cal A}_{\gamma_0+1} \big(\cot^2(\pi\gamma_0) +1\big)}
\vert\omega\vert^{1-2\gamma_0},\\
\label{ed22}
\Sigma_r''(q=0,\omega) &\sim& {- {\rm sgn}(\omega) \over
\pi {\cal A}_{\gamma_0+1} \big(\cot^2(\pi\gamma_0) +1\big)}
\vert\omega\vert^{1-2\gamma_0}.
\end{eqnarray}
We note that $\Sigma_r(q=0,\omega=0) = 0$ (see Eq.~(\ref{ed17}) and text 
preceding it), which in view of our considerations in Appendix~B amounts 
to the fact that in the Luttinger model $\Sigma_r^{HF}(q=0)$ is entirely 
cancelled by correlation effects. It is further interesting to note that
\begin{equation}
\label{ed23}
{\Sigma_r''(q=0,\omega)\over 
\Sigma_r'(q=0,\omega) -\Sigma_r(q=0,\omega=0)} \sim 
\tan(\pi\gamma_0),\;\;\; \omega\to 0,
\end{equation}
which is seen to approach zero (like $\sim \pi\gamma_0$) for $\gamma_0 
\downarrow 0$. As we have mentioned in \S~IV.C, $\gamma_0 = 0$ corresponds 
to the non-interacting system, so that the expressions presented in 
Eqs.~(\ref{ed21}) and (\ref{ed22}) reproduce the non-interacting limit 
in a continuous way.

\subsubsection{The $q\not=0$ case} 

Here we confine our considerations to the case corresponding to $r=+$. 
From Eq.~(\ref{ed11}) (taking the limit $\nu\downarrow 0$) it readily 
follows that for $0 < \gamma_0 < 1$ the leading asymptotic contribution 
to $G_+(q,\omega)$, as $\omega\to 0$, is due to 
\begin{equation}
\label{ed24}
\widetilde{{\cal G}}_+^{(\Delta)}(q,z) {:=}
\int_{v_0 q}^{v_0 q +\Delta} {\rm d}\omega'\;
{\phi(q,\omega')\over z -\omega'}.
\end{equation}
For $z = v_0 q + \omega + i\eta$, $\eta\downarrow 0$,
with $0 < \omega < \Delta$, making use of 
$\gamma\big(\gamma_0,\Lambda (\omega'+2 v_0 q)/[2 v_0]\big)
= \gamma\big(\gamma_0,\Lambda q\big) + {\cal O}(\omega')$,
$\exp(-\Lambda\omega'/[2 v_0]) = 1 + {\cal O}({\omega'})$,
while employing the standard result ($0 < \omega < \Delta$)
\begin{eqnarray}
\label{ed25}
{\cal P}\int_0^{\Delta} {\rm d}\omega'\;
{{\omega'}^{\sigma-1}\over \omega' -\omega}
&=& -\pi \cot(\pi\sigma) \omega^{\sigma -1} 
+ {\Delta^{\sigma -1}\over \sigma -1}\,
{}_2F_1\big(1,1-\sigma;2-\sigma;{\omega\over\Delta}\big)\nonumber\\
&=& -\pi \cot(\pi\sigma) \omega^{\sigma -1} 
- {\Delta^{\sigma-1}\over 1-\sigma} - {\Delta^{\sigma-2}
\over 2-\sigma} \omega + {\cal O}\big(\omega^2\big),\nonumber\\
& &\;\;\;\;\;\;\;\;\;\;\;\;\;\;\;\;\;\;\;\;\;\;\;\;\;\;\;\;\;\;\;
\sigma > 0,\;\sigma \not=1,2,\dots,
\end{eqnarray}
we obtain ($0 < \omega < \Delta$, $0 < \gamma_0 < 1$ and $q\not=0$)
\begin{equation}
\label{ed26}
{\cal G}_+^{(\Delta)}(q,\omega) \sim G_+(q,\omega) 
\sim \pi\,\gamma\big(\gamma_0,\Lambda q\big)\, {\cal A}_{\gamma_0}
\Big\{ \cot(\pi\gamma_0) - i\Big\}\, \omega^{\gamma_0 -1},
\;\;\; \omega\downarrow 0.
\end{equation}
Note that the exponent $\gamma_0 - 1$, as compared with $2\gamma_0 -1$ 
for the case corresponding $q=0$, implies that even for $\gamma_0$ in the 
{\sl open} interval $(0,1)$ (as opposed to $(0,1/2)$), $G_+(q\not=0,\omega)$ 
is to leading order asymptotically identical with the corresponding auxiliary 
function ${\cal G}_+^{(\Delta)}(q\not=0,\omega)$ as $\omega\downarrow 0$; 
recall that the leading asymptotic terms of $G_r^{(\Delta)}(q=0,\omega)$ 
and $G_r(q=0,\omega)$ for $\omega\to 0$ (see Eqs.~(\ref{ed16}) and 
(\ref{ed17})) are identical only when $0 <\gamma_0 < 1/2$. Further, similar 
to the case corresponding to $q=0$, the divergence of $G_+(q,\omega)$ for 
$0 <\gamma_0 < 1$, as $\omega\downarrow 0$, implies that the possibly 
non-vanishing constant $\{G_+(q\not=0,\omega) -{\cal G}_+^{(\Delta)}
(q\not=0,\omega)\}$ is asymptotically irrelevant.

Proceeding as in the case of $q=0$, for $\omega\downarrow 0$ we obtain 
($0 < \gamma_0 < 1$)
\begin{equation}
\label{ed27}
G_+^{-1}(q,\omega) \sim
{1\over \pi \gamma\big(\gamma_0,\Lambda q\big)
{\cal A}_{\gamma_0} \big(\cot^2(\pi\gamma_0)+1\big)}
\Big\{ \cot(\pi\gamma_0) + i \Big\}
\omega^{1-\gamma_0}.
\end{equation}
As a consequence of our above arguments, for $0 < \gamma_0 < 1$ we have 
(for $\gamma_0$ in this range, $\cot(\pi\gamma_0)$ takes on both positive 
as well as negative values)  
\begin{eqnarray}
\label{ed28}
\Sigma_+'(q,\omega) &\sim& {-\cot(\pi\gamma_0)\over
\pi \gamma\big(\gamma_0,\Lambda q\big)
{\cal A}_{\gamma_0} \big(\cot^2(\pi\gamma_0) +1\big)}
\omega^{1-\gamma_0},\\
\label{ed29}
\Sigma_+''(q,\omega) &\sim& {-1\over
\pi \gamma\big(\gamma_0,\Lambda q\big)
{\cal A}_{\gamma_0} \big(\cot^2(\pi\gamma_0) +1\big)}
\omega^{1-\gamma_0}.
\end{eqnarray}
We point out that $\Sigma_+(q,\omega=0) = 0$ (see Eq.~(\ref{ed17}) and 
text preceding it), which in view of our considerations in Appendix~B leads 
to the conclusion that, in the Luttinger model $\Sigma_+^{HF}(q)$ is 
entirely cancelled by correlation effects. As in the case corresponding
to $q=0$, here we have (compare with Eq.~(\ref{ed23}))
\begin{equation}
\label{ed30}
{\Sigma_+''(q,\omega)\over 
\Sigma_+'(q,\omega) -\Sigma_+(q,\omega=0)} \sim 
\tan(\pi\gamma_0),\;\;\; \omega\downarrow 0,
\end{equation}
which approaches zero (like $\sim \pi \gamma_0$) for $\gamma_0 
\downarrow 0$; this result is consistent with the fact that $\gamma_0 
= 0$ corresponds to the non-interacting system. 

We further point out that since $\partial\gamma(a,z)/\partial z = \exp(-z) 
z^{a-1}$, from Eqs.~(\ref{ed28}) and (\ref{ed29}) we obtain that $\partial
\Sigma_+(q, \omega)/\partial q \propto \vert q\vert^{\gamma_0-1}$ which 
diverges as $q\to 0$ for $\gamma_0 < 1$. This result is in full conformity 
with our finding above (see the last paragraph in \S~D.1, following 
Eq.~(\ref{ed12})).

In closing this Appendix, we point out that the apparent difference 
in the asymptotic behaviours of $\Sigma_r(q,\omega)$ (specifically  
concerning $r=+$ explicitly considered here) for $q\to 0$ and $q=0$,
is {\sl not} specific to the Luttinger model. It is well known (Hodges, 
Smith and Wilkins 1971, Bloom 1975, Fujimoto 1990, Fukuyama, Narikiyo 
and Hasegawa 1991, Fukuyama, Hasegawa and Narikiyo 1991) that for 
isotropic systems of interacting electrons in two spatial dimensions, 
under the {\sl assumption} that they are Fermi liquids, while 
$\Sigma''(k;\varepsilon) {\sim'} (\varepsilon -\varepsilon_F)^2$ 
for $k\not=k_F$, we have $\Sigma''(k_F;\varepsilon) {\sim'} (\varepsilon
-\varepsilon_F)^2 \ln\vert \varepsilon -\varepsilon_F\vert$.
$\hfill\Box$


\pagebreak

\begin{figure}[t!]
\protect
\label{fi1}
\centerline{
\psfig{figure=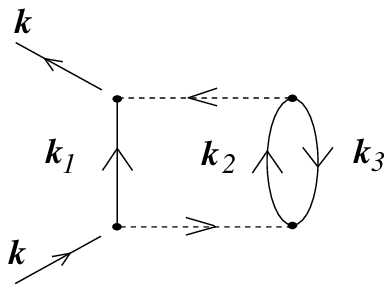,width=3.00in} }
\vskip 20pt
\caption{\rm
A second-order skeleton self-energy diagram in terms of the bare
particle-particle interaction $v$ (broken lines) employed by
Luttinger (1961). The solid lines stand for either $G_0$ or 
$G$, the single-particle Green function corresponding to the 
non-interacting and interacting systems, respectively. The external 
and internal wave-vectors, ${\bf k}$ and ${\bf k}_1$, ${\bf k}_2$ and 
${\bf k}_3$, respectively, are shown. Because of conservation of 
momentum, one of the internal wave-vectors, say ${\bf k}_3$, can 
be eliminated; ${\bf k}_3 = {\bf k}_1 + {\bf k}_2 - {\bf k}$.  }
\end{figure}

\begin{figure}[t!]
\protect
\label{fi2}
\centerline{
\psfig{figure=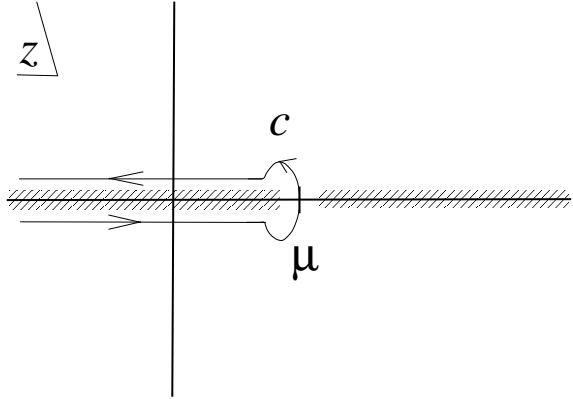,width=3.00in} }
\vskip 20pt
\caption{\rm
Contour ${\cal C}$ of integration employed in the proof of the Migdal 
theorem concerning the discontinuity of the momentum distribution 
function ${\sf n}(k)$ at $k=k_F$ and its relation to the quasi-particle
weight on the Fermi surface, $Z_{k_F}$. The shaded parts of the real 
axis indicate branch cuts of $\widetilde{G}(k;z)$; the unshaded 
part of this axis centred around the `chemical potential' $\mu$ represents 
$(\mu_N,\mu_{N+1})$ with $\mu_N$ and $\mu_{N+1}$ two branch points of 
$\widetilde{G}(k;z)$ as well as $\widetilde{\Sigma}(k;z)$. }
\end{figure}

\pagebreak
\begin{figure}[t!]
\protect
\label{fi3}
\centerline{
\psfig{figure=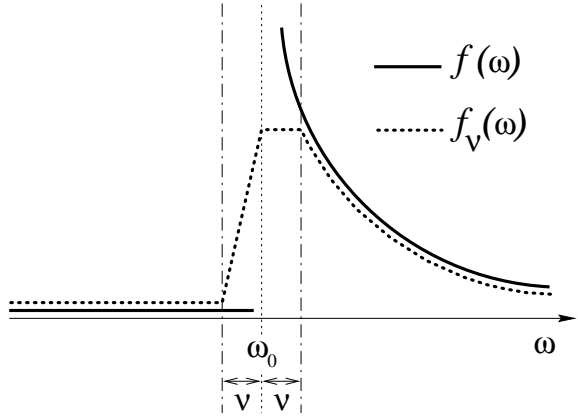,width=3.00in} }
\vskip 20pt
\caption{\rm
The {\sl distribution} (as opposed to {\sl function}) $f(\omega) {:=}
\Theta(\omega-\omega_0) g(\omega)$ and a specific function 
$f_{\nu}(\omega)$, corresponding to a finite positive value 
of $\nu$, in terms of which we choose $f(\omega)$ to be defined (this 
choice is not unique); 
$f(\omega) {:=} \lim_{\nu\downarrow 0} f_{\nu}(\omega)$.
We have the following
$f_{\nu}(\omega) \equiv 0$, for $\omega \leq \omega_0 -\nu$;
$f_{\nu}(\omega) \equiv g(\omega_0+\nu) [1 + (\omega -\omega_0)/\nu]$, 
for $\omega_0 -\nu \leq \omega \leq \omega_0$;
$f_{\nu}(\omega) \equiv g(\omega_0+\nu)$, for $\omega_0 \leq \omega 
\leq \omega_0 +\nu$ and
$f_{\nu}(\omega) \equiv g(\omega)$, for $\omega\geq \omega_0 +\nu$. The 
form of $f(\omega)$ is reminiscent of $\rho_+(q,\omega)$ pertaining 
to the Luttinger model for which $\omega_0$ coincides with $v_0 q$ and 
$\rho_+(q,\omega) \propto (\omega - v_0 q)^{\gamma_0 -1}$ for 
$\omega\downarrow v_0 q$ (here $0 < \gamma_0 < 1/2$). }
\end{figure}

\end{document}